\documentclass{aastex62}
\usepackage{booktabs}
\usepackage{graphicx}

\received{January 1, 2019}
\revised{}
\accepted{}
\submitjournal{ApJ}

\shorttitle{JVLA overview of the bursting source G25.65+1.05}
\shortauthors{Bayandina et al.}
\begin{document}

\title{JVLA OVERVIEW OF THE BURSTING H$_2$O MASER SOURCE G25.65+1.05}

\correspondingauthor{O. S. Bayandina}
\email{bayandina@asc.rssi.ru}

\author{O. S. Bayandina}
\affil{Astro Space Center, Lebedev Physical Institute, Russian Academy of Sciences, Leninskiy Prospekt 53, Moscow 119333, Russia}

\author{R. A. Burns}
\affil{Joint Institute for VLBI ERIC, Oude Hoogeveensedijk 4, 7991 PD Dwingeloo, The Netherlands}
\affil{Mizusawa VLBI Observatory, National Astronomical Observatory of Japan, 2-21-1 Osawa, Mitaka, Tokyo 181-8588, Japan}
\affil{Korea Astronomy and Space Science Institute, 776 Daedeokdae-ro, Yuseong-gu, Daejeon 34055, Republic of Korea}

\author{S. E. Kurtz}
\affiliation{Instituto de Radioastronom{\'\i}a y Astrof{\'\i}sica, Universidad Nacional Aut$\acute{o}$noma de M$\acute{e}$xico, Apartado Postal 3-72, Morelia 58089, M$\acute{e}$xico}

\author{N. N. Shakhvorostova}
\affiliation{Astro Space Center, Lebedev Physical Institute, Russian Academy of Sciences, Leninskiy Prospekt 53, Moscow 119333, Russia}
\affiliation{Astronomical Observatory, Institute for Natural Sciences and Mathematics, Ural Federal University, Kuybyshev st. 48, Ekaterinburg 620026, Russia}

\author{I. E. Val'tts}
\affiliation{Astro Space Center, Lebedev Physical Institute, Russian Academy of Sciences, Leninskiy Prospekt 53, Moscow 119333, Russia}



\begin{abstract}

The source G25.65+1.05 (RAFGL7009S, IRAS 18316-0602) is the least studied of the three regions of massive star formation known to show exceptionally powerful H$_2$O maser bursts.  
We report spectral line observations of the H$_2$O maser at 22 GHz, the methanol maser transitions at 6.7, 12.2 and 44 GHz, and the continuum in these same frequency bands with The Karl G. Jansky Very Large Array (JVLA) at the post-burst epoch of 2017.
For the first time, maps of 22 GHz H$_2$O and 44 GHz CH$_3$OH maser spots are obtained and the absolute position of the 22 GHz H$_2$O bursting feature is determined with milliarcsecond precision. We detected four continuum components, three of which are closely spaced in a linear orientation, suggesting a physical link between them.
\end{abstract}



\section{Introduction} 

Cosmic masers were discovered nearly 50 years ago, and although great
strides have been made in understanding the nature of these intriguing objects,
they continue to present behaviors that we cannot yet fully explain.
One such phenomenon is the ``super-bursts'' of water
maser emission.  Water masers are known to be variable
on a variety of time scales, but only three Galactic water masers
are known to flare to the level of 10$^5$--10$^6$~Jy (T$_B \sim 10^{17}$~K): Orion~KL, W49N,
and the recently discovered G25.65+1.05.

Because unsaturated maser amplification is exponentially related to path length, the observed maser flux depends on the viewpoint of the observer. However, changes in maser flux also pertain to activity at, or in the vicinity of, the maser region. The super-burst phenomenon may be caused by an enhanced maser pumping mechanism due to a change to more favorable physical conditions; an increase in the incident seed photons entering the maser region; or due to an increase in maser path length on the line of sight to the observer. The latter case can be achieved by a chance overlap -- when two masers of similar velocity move into superposition in the sky-plane, as described by \cite{2005ApJ..634..459S}. A comparison of flaring in different maser species and monitoring of free-free continuum levels can provide important clues to understand the nature of the flaring process.

G25.65+1.05 (associated with IRAS 18316$-$0602 and also known as RAFGL7009S) was included in the H$_2$O maser surveys of \cite{1991AA..246..249P} and \cite{1994AAS..103..541B} -- both of which reported a powerful  H$_2$O maser with flux density $>$700 Jy and V$_{peak}$ = $\sim$45 km~s$^{-1}$. Long-term monitoring of H$_2$O masers with the RT-22 of the Pushchino Radio Astronomy Observatory (Moscow region) showed flares in 2002, 2010, and 2016 with flux densities of 3400 (V$_{peak}$ = 41.05 km~s$^{-1}$), 19000 (V$_{peak}$ = 41.29 km~s$^{-1}$), and 46000 (V$_{peak}$ = 41.80 km~s$^{-1}$) Jy, respectively \citep{2018ARep...62..213L}. The next powerful burst of 65000 Jy (42.5 km~s$^{-1}$) was detected in September 2017 \citep{2017ATel10728....1V, 2019MNRAS.482L..90V} with the RT-22 of the Crimean Astrophysical Observatory. In October 2017, shortly after this burst, the source showed yet another increase of flux density -- see \cite{2017ATel10853....1V, 2019MNRAS.482L..90V}. The most recent burst was found to be short-lived: the peak flux density rose from $\sim$20 kJy to 76 kJy within half a day on November 20; the source then faded to $\sim$20 kJy on November 22 \citep{2017ATel11042....1A}. Although G25.65+1.05 is the target of intense monitoring, the transient nature of burst events may lead to some events going undetected. It is therefore likely that the super-burst phenomenon is more common than is currently thought.

The source is maser-rich, with both class I and II methanol\footnote{Hereafter we will refer to CH$_3$OH maser classes with the abbreviations proposed in \citep{2012IAUS..287..506M} cIMM and cIIMM, respectively.}, OH, and H$_2$O masers reported (e.g., \cite{1991AA..246..249P, 2018ApJS..00..00B, 2010AA..517..56F}). 
The 6.7 GHz class II methanol maser in G25.65+1.05 has been extensively observed. At HartRAO, with
the 26-m, it was detected with a peak of 113 Jy \citep{1995AAS...110..81V}. \cite{1997MNRAS.291..261W} used the Parkes 64-m and found a 105 Jy flux with V$_{peak}$ = +42 km~s$^{-1}$. The maser was detected with 181 Jy at V$_{peak}$ = +41.8 km~s$^{-1}$ with the Medicina 32-m \citep{1999AAS..134..115S} and $\sim$100 Jy at V$_{peak}$ = +42 km~s$^{-1}$ with Effelsberg 100-m \cite{2010AA..517..56F}. The Torun 32-m observations showed a flux density of 200 Jy at V$_{peak}$ = +42 km~s$^{-1}$ \citep{2000AAS.143..269S}. It was noted that the peak flux density of the 6.7 GHz maser reported by \cite{1995AAS...110..81V} and \cite{1997MNRAS.291..261W} is less than half of the value reported by \cite{1999AAS..134..115S} and \cite{2000AAS.143..269S}; i.e., the maser is variable. The presence of a circumstellar disk was suggested in \citep{2002AA.394..225Z} and \citep{2015AA..578..102S} based on 6.7 GHz maser presence and structure.

It is interesting to note that \cite{2010AA..517..56F} show three spectra of methanol: one of class II methanol at 6.7 GHz ($\sim$100 Jy, Effelsberg, 100-m) and two of class I methanol lines at 44 GHz ($\sim$10 Jy) and at 95 GHz ($\sim$2 Jy); both class I spectra were obtained with the Nobeyama 45-m telescope. 
The class I emission suggests the object is at a very early evolutionary stage, while the
class II emission suggests a later stage, after the appearance of an ultracompact (UC) HII region. 
Established methanol maser classification \citep{1987Natur.326...49B, 1991ASPC..16..119M} suggests that cIMM and cIIMM do
not coexist because of distinct pumping conditions. One of the main cIMM properties is that they
do not spatially coincide with cIIMM or with H$_2$O and OH masers. OH maser lines in G25.65+1.05,
according to recent JVLA data, show the strongest peak at 1665 MHz in the left polarization
at V$_{LSR}$ = +41.36 km~s$^{-1}$ ($\sim$9 Jy), 1 Jy at 1667 GHz, and no emission in satellite lines \citep{2018ApJS..00..00B}. Thus, a likely scenario is that this region hosts spatially separated molecular cores at different stages of evolution. 

The 6.7 GHz class II methanol maser showed a $>$40\% flux increase several months in advance of the H$_2$O maser burst \citep{2017ATel10757....1S} possibly indicating an enhancement in the local IR radiation field. 
Maser flaring, albeit at lower levels, has been reported in other sources as well. For example, \cite{2017ApJ...837L..29H} report strong maser flaring in H$_2$O, CH$_3$OH, and OH in NGC6334I, accompanied by a 4-fold increase in the dust continuum emission. Correlated flaring of different maser species has been reported by several groups, for example, formaldehyde and methanol \citep{2010ApJ..717..133A} and in water and methanol \citep{2016MNRAS.459L..56S}. 

The distance to the source is an open question -- VLBI measurements of the trigonometric parallax have never been performed for the source. Molecular line observations in most cases
argue in favor of the near kinematic distance and yield values of 3.17 kpc \citep{1996AA..308..573M}, 3.3 kpc \citep{1997MNRAS.291..261W}, and 2.7 kpc \citep{2007PASJ..59..1185S}. In contrast, HI self-absorption toward the source suggests a far kinematic distance of 12.5 kpc \citep{2011MNRAS.417.2500G}.
A probability density function for source distance calculated on the basis of  The Bar and Spiral Structure Legacy (BeSSeL, \url{http://bessel.vlbi-astrometry.org}) Survey data \citep{2016ApJ...823...77R} indicates a distance of 2.08$\pm$0.37 kpc with a probability of 64\%. In our calculations below we will use this value.

There is no optical identification with an HII region within 2$^{\prime}$ \citep{1996AA..308..573M}; i.e., this region is deeply embedded and/or is at a very early evolutionary state. The probable young stellar object of the region is a BIV star \citep{2002AA.394..225Z}. According to \cite{1995AJ....110.1762M}, the source is weak in the radio continuum with 2.7 mJy flux density at 6~cm and classified as ``probably pre-main-sequence''. \cite{1994ApJS..91..659K} presented 3.6~cm map and reported an ``irregular'' morphology for the UCHII region. The IRAS source is separated from the central continuum radio source by 18$^{\prime\prime}$ (a projected linear distance of 0.18~pc at 2.08~kpc). \cite{1995AJ....110.1762M} also reported JCMT observations at 1100, 800 and 450 microns.  They detected a dusty core, and using their data, along with the IRAS fluxes, they fit a two-component model to the spectrum.  Their study suggests a larger, cooler core (0.3 pc and 35 K) an a smaller, hotter core (0.073 pc and 123 K).  For the cold component they find a total dust mass of 11.0 M$_\odot$ with a molecular hydrogen column density of $9.8 \times 10^{23}{\rm~cm}^{-2}$. Although the JCMT and IRAS spectra suggest a two-component model, the JCMT images of both \cite{1995AJ....110.1762M} and \cite{2006AA..453..1003T} lack the angular resolution to distinguish internal structure.

Thermal lines, such as NH$_3$ and CS, detected in the region show a peak at the velocity of 
V$_{LSR}$=+42.2 km~s$^{-1}$ \citep{1996AA..308..573M, 1996AAS..115..81B}. Dense gas tracers such as N$_2$H$^+$ and C$_2$H have similar LSR
velocities of 42.6 and 42.4 km~s$^{-1}$, respectively \citep{2013AA..557..94S}.
A bipolar flow was detected in CO \citep{1996ApJ..457..267S} and CH$_3$CN \citep{2000AA..361..1095D}. The outflow was confirmed in the SiO (5-4) and HCO$^+$ (1-0) lines, but not its bipolar structure \citep{2011AA..526L..2L, 2013AA..557..94S}. \cite{2000AA..361..1095D} report multiwavelength observations with the IRAM 30-m, the Plateau
de Bure interferometer, UKIRT and ISO. Observations were carried out in CH$_3$CCH and CH$_3$OH lines, including deuterated species, and 3~mm continuum. Their maps can be interpreted as either a single, episodically
driven outflow from a single source, or as multiple outflows from several distinct, but unresolved,
sources. Observation of the $\nu$ = 1-0 S(1) H$_2$ line at 2.166 $\mu$m did not allow to determine unambiguously whether there are single or multiple outflows and driving sources in the region \citep{2006MNRAS..367..238T}.

In response to the recent super-burst discovery, we initiated a multi-part investigation of G25.65+1.05 to study the burst mechanism in the context of massive star formation. This paper reports the first results of the investigation and aims to provide a brief overview of G25.65+1.05, in addition to introducing new data that impact our understanding of G25.65+1.05 and help establish a robust contextual basis for forthcoming work on this maser burst object.

\section{Observations} 

Observations were carried out in two sessions on 2017 November 2 and December 9 with the Jansky Very Large Array (JVLA, NRAO) in the B-configuration as a Target of Opportunity program (project code 17B-408). The total observing time was 2~hours, comprising a
first session of 45 minutes and a second of 75 minutes. Four frequency bands were used: C (6.0 GHz), Ku (15 GHz), K (22 GHz) and Q (45 GHz). 

These JVLA observations of G25.65+1.05 were conducted during the post-burst period: Q-band observations were performed on 2017 November 2, i.e. after the powerful H$_2$O maser burst detected in September 2017 \citep{2017ATel10728....1V, 2019MNRAS.482L..90V}, but before the short-lived burst of November 20 \citep{2017ATel11042....1A}. C-, Ku-, and K-band observations were performed on 2017 December 9, i.e. after both recent bursts.

The pointing coordinates used were RA(J2000)=18$^h$34$^m$21$^s$, DEC(J2000)=-05$^{\circ}$59$\arcmin$42$\arcsec$ and the central velocity was 42 km~s$^{-1}$.

Table \ref{tab:obs} reports the main observational parameters. 3C~48 was used as a flux density, bandpass, and delay calibrator; J1832-1035 as a complex gain calibrator (the angular separation from the target source is 4.6$^{\circ}$). About an hour was spent on J1832-1035 scans and $\sim$20 minutes on 3C~48. 

Spectral lines were observed in windows of 1024 channels. The channel width was 0.98 kHz in C-band, 1.95 kHz in Ku-band, 7.81 kHz in K- and Q-bands. For the corresponding velocity resolutions, see Table \ref{tab:obs}. 

For the continuum observations, 31 spectral windows were used in C-, K-, and Q-bands and 23 windows in Ku-band. In all cases, each spectral window contained 128 channels of 1 MHz width.

\section{Data Reduction} 

The data were reduced with the NRAO software package CASA (Common Astronomy Software Applications, \url{http://casa.nrao.edu}). The main stages of the data calibration were performed with the CASA package ``VLA calibration pipeline''. Subsequent self-calibration was performed on the
strongest maser channel for each frequency data set.

An unresolved issue in the calibration is the flux density scale for
the maser emission.  The flux densities of the 6.7 GHz, 22 GHz and 44
GHz masers, as detected with the VLA, are 2 to 3$\times$ higher than
the corresponding single-dish fluxes.  For the 6.7 GHz cIMM, data
taken with the Torun 32-m (RT-32 of the Toru\'n Centre for Astronomy of Nicolaus Copernicus University) at about the same time reported 166 Jy while
our VLA data showed 386 Jy (factor of 2.3).  For the 22 GHz H$_2$O
maser, nearly coeval data from the Torun 32-m and the Crimean 22-m
telescope show $\sim$9.8~kJy while the VLA flux density was 29.5~kJy
(factor of 3.0).  For the 44 GHz cIMM, \cite{2010AA..517..56F} reported
10 Jy, while our VLA data show 25 Jy (factor of 2.5).  In what
follows, we have scaled the VLA flux densities by the corresponding
factors.  We continue to investigate the nature of this discrepancy.
Accordingly, the present work concentrates on astrometric results only, thus we note that
it does not affect our analysis or conclusions,
nor does the flux discrepancy occur for the continuum data.

Calibrated data were imaged using the Clark CLEAN algorithm \citep{1980AA....89..377C} with Robust weighting = 0. A two-dimensional Gaussian brightness distribution was fitted to the map in every channel with flux density above 5$\sigma$ level (see Table \ref{tab:obs} for $\sigma$-level at each frequency) to determine the position, flux density (integrated and peak) and LSR velocity of the maser spots. Here after the term ``spot'' refers to maser emission in a single channel map. 

Spectra were obtained with the CASA Spectral Profile Tool. Detected flux densities of maser sources at each observed frequency differs in the right and left polarizations by less than 2\%, ergo we present the Stokes I data.

\section{Results}

\subsection{Continuum Emission} 

\begin{figure*}
\gridline{\fig{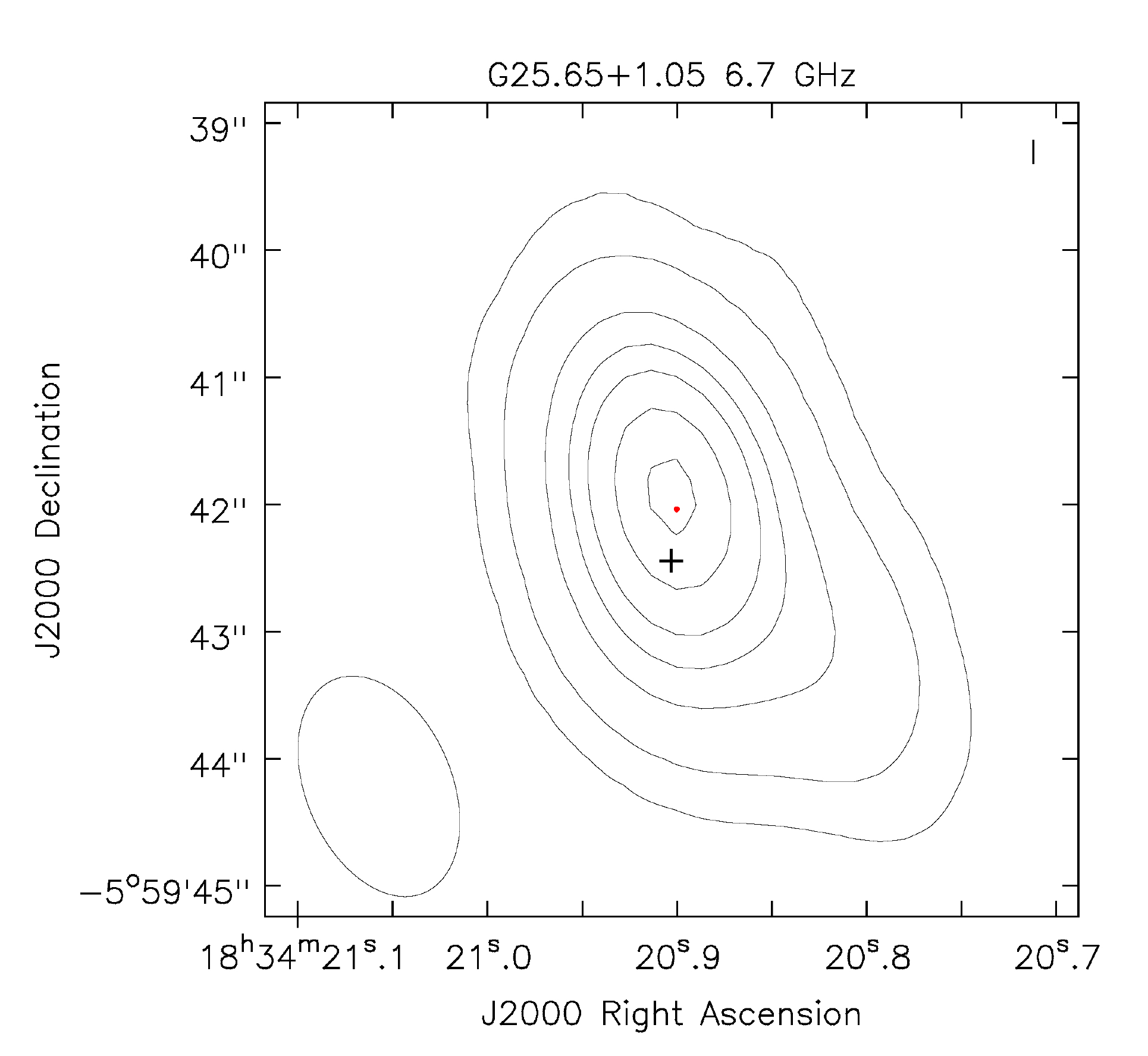}{0.45\textwidth}{(a)}
          \fig{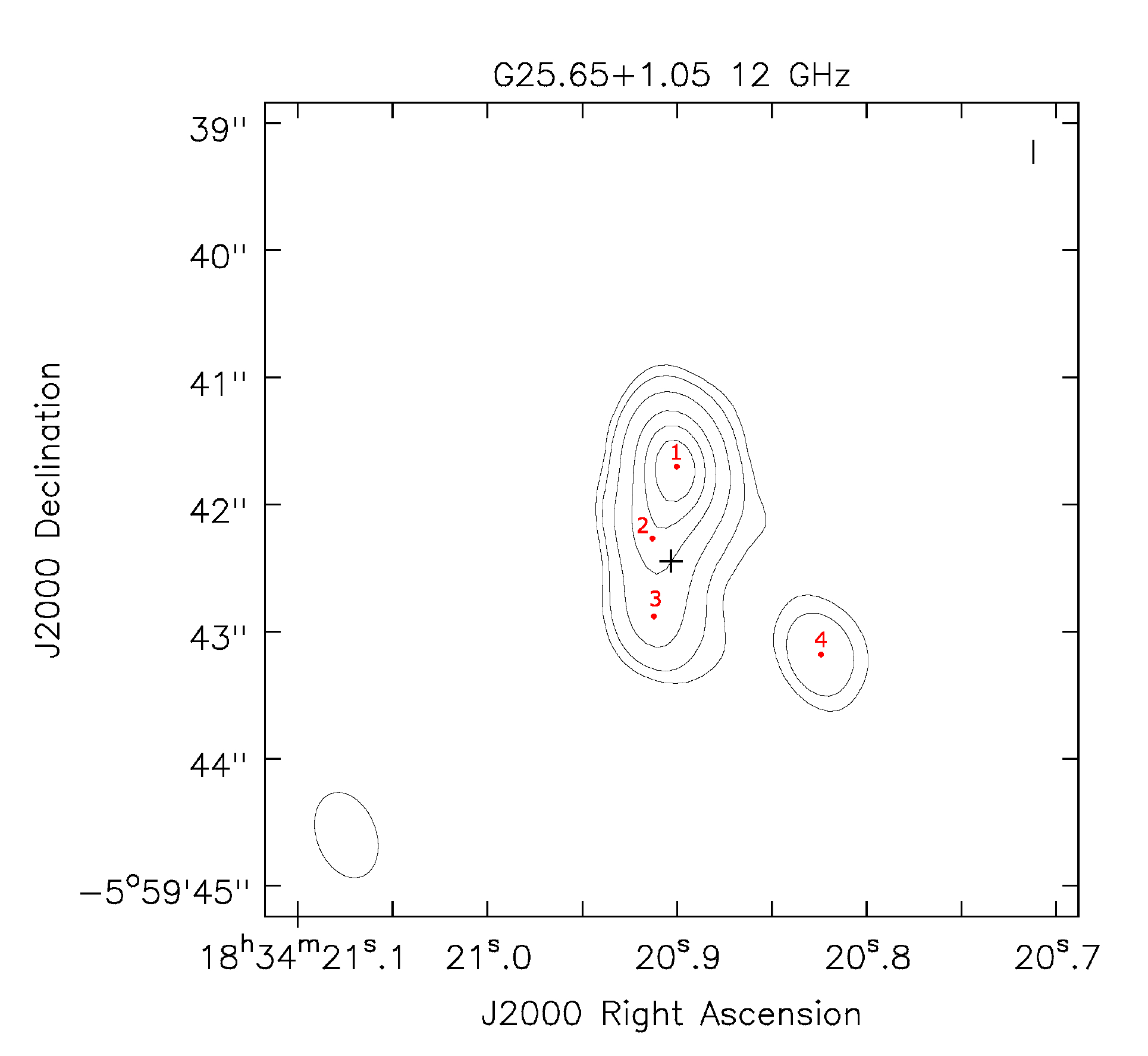}{0.45\textwidth}{(b)}
          }
\gridline{\fig{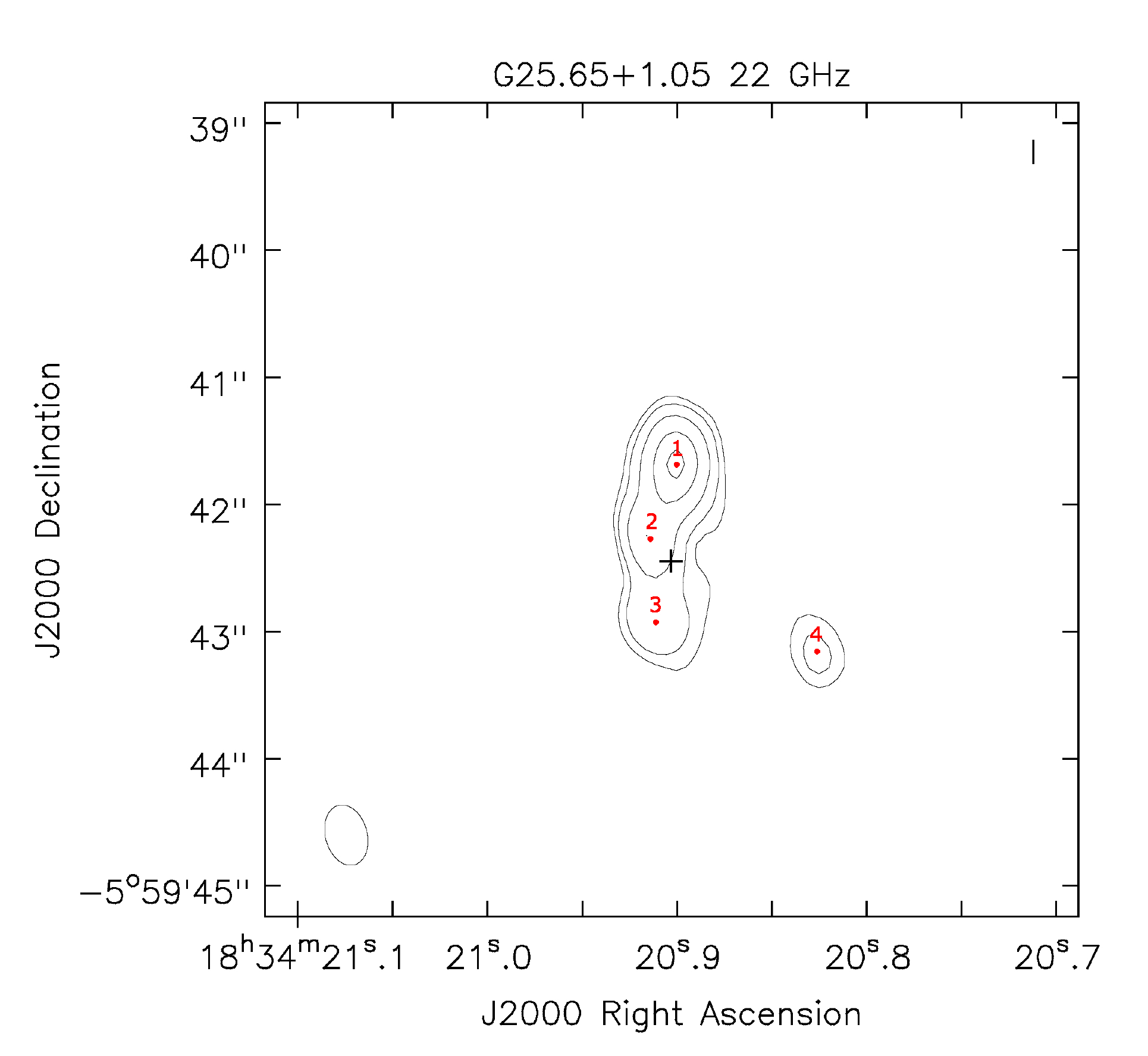}{0.45\textwidth}{(c)}
          \fig{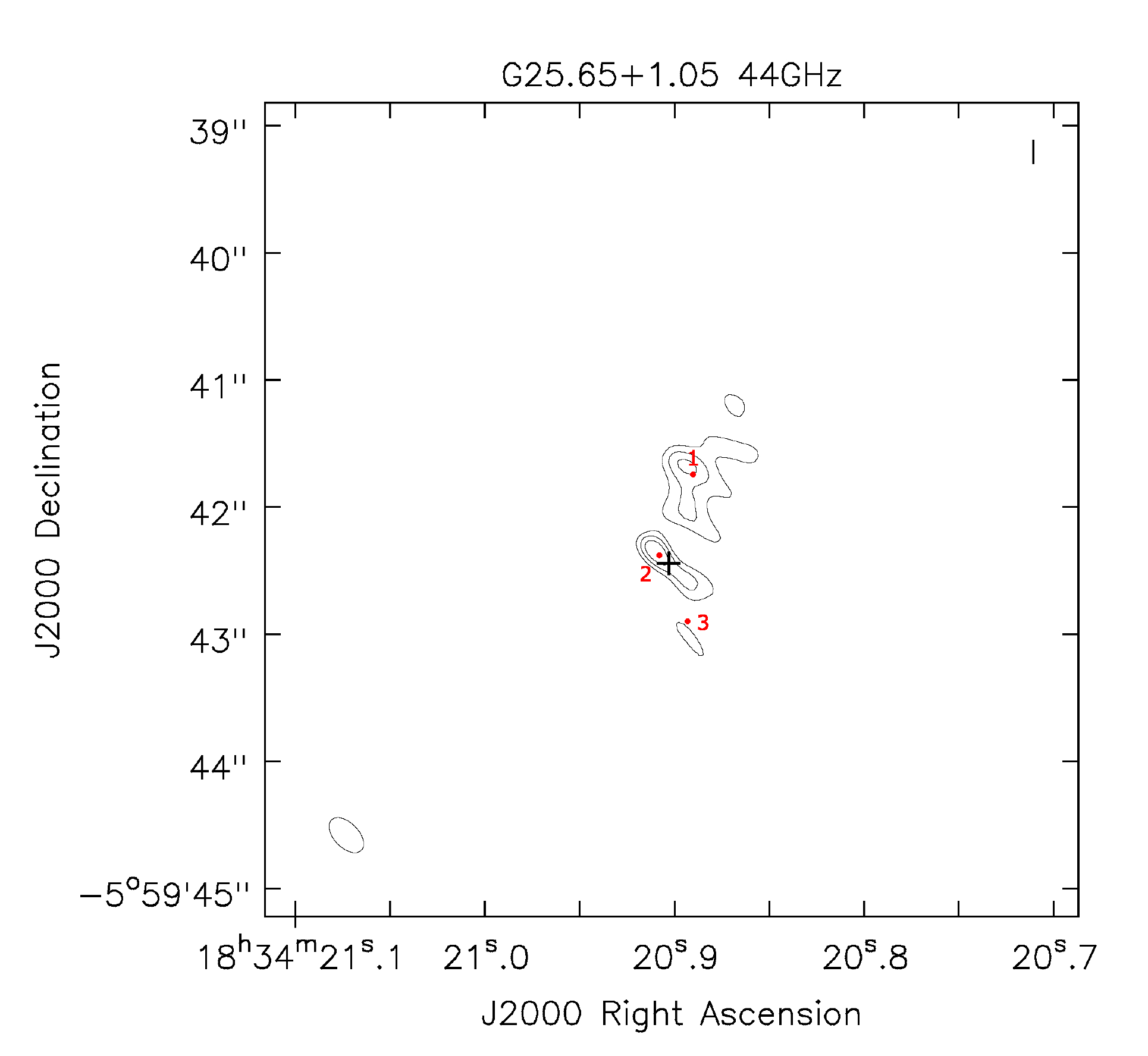}{0.45\textwidth}{(d)}
				  }
\caption{Continuum images of G25.65+1.05 at (a) 6.7 GHz (levels [3, 10, 30, 50, 70, 100, 130] $\times$ 20~$\mu$Jy/beam), (b) 12 GHz (levels [3, 5, 10, 20, 30, 40] $\times$ 40~$\mu$Jy/beam), (c) 22 GHz (levels [3, 5, 10, 20, 30] $\times$ 50~$\mu$Jy/beam), and (d) 44 GHz (levels [3, 5, 7, 10] $\times$ 70~$\mu$Jy/beam). Red points indicate detected peaks of continuum emission at a certain frequency -- see Table \ref{tab:cont}. Indices correspond to source labeling from Table \ref{tab:cont}. The black cross marks the position of the UCHII region from \citep{1994ApJS..91..659K}. \label{fig:continuum}}
\end{figure*}

Continuum emission above the 5$\sigma$ detection level was found in all four frequency bands. Four distinct continuum sources were detected in the G25.65+1.05 region, which we index as JVLA 1, 2, 3, and 4. Parameters of the detected continuum sources are listed in the Table \ref{tab:cont} and the JVLA continuum images are shown in Figure~\ref{fig:continuum}. 

\begin{figure*}
\gridline{\fig{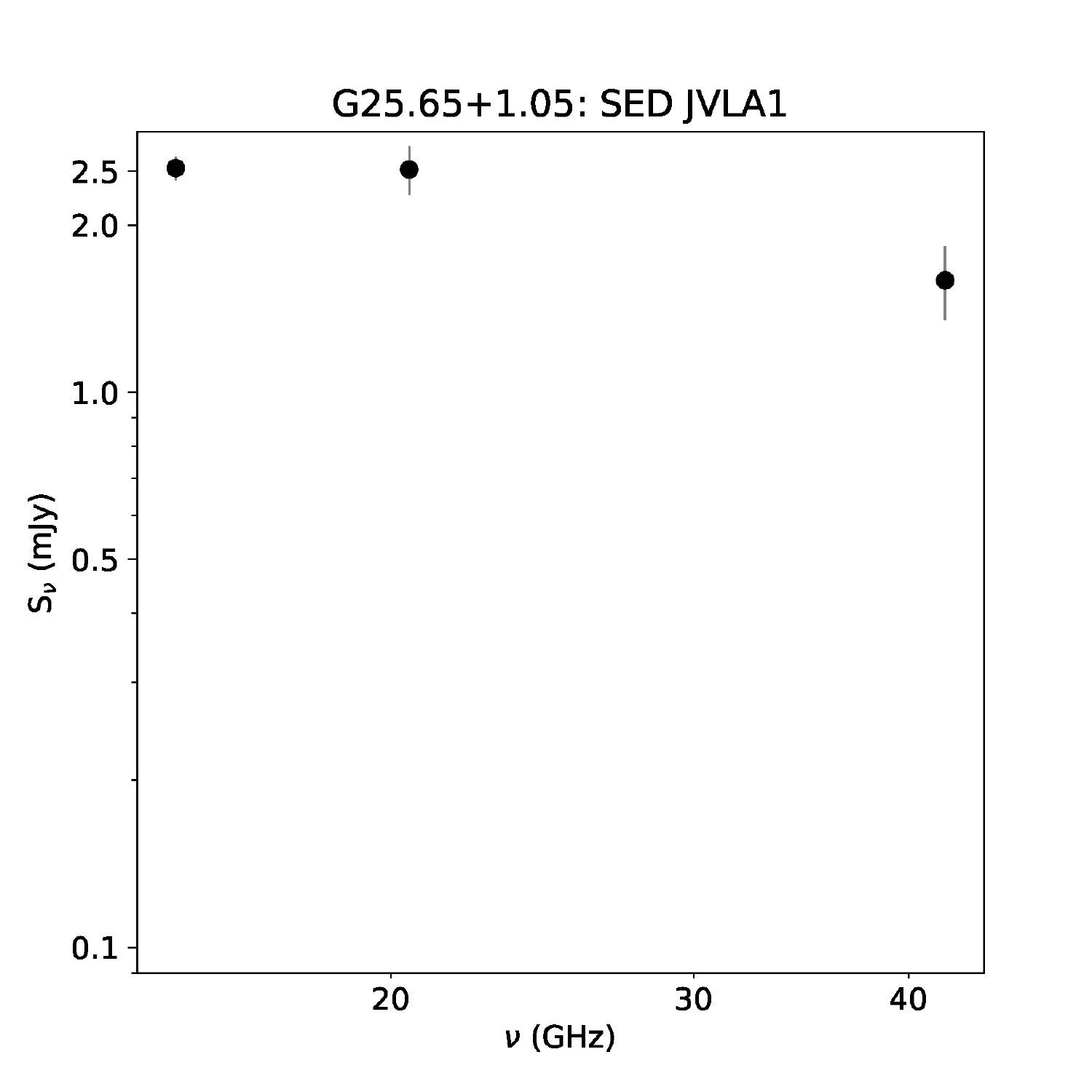}{0.45\textwidth}{(a)}
          \fig{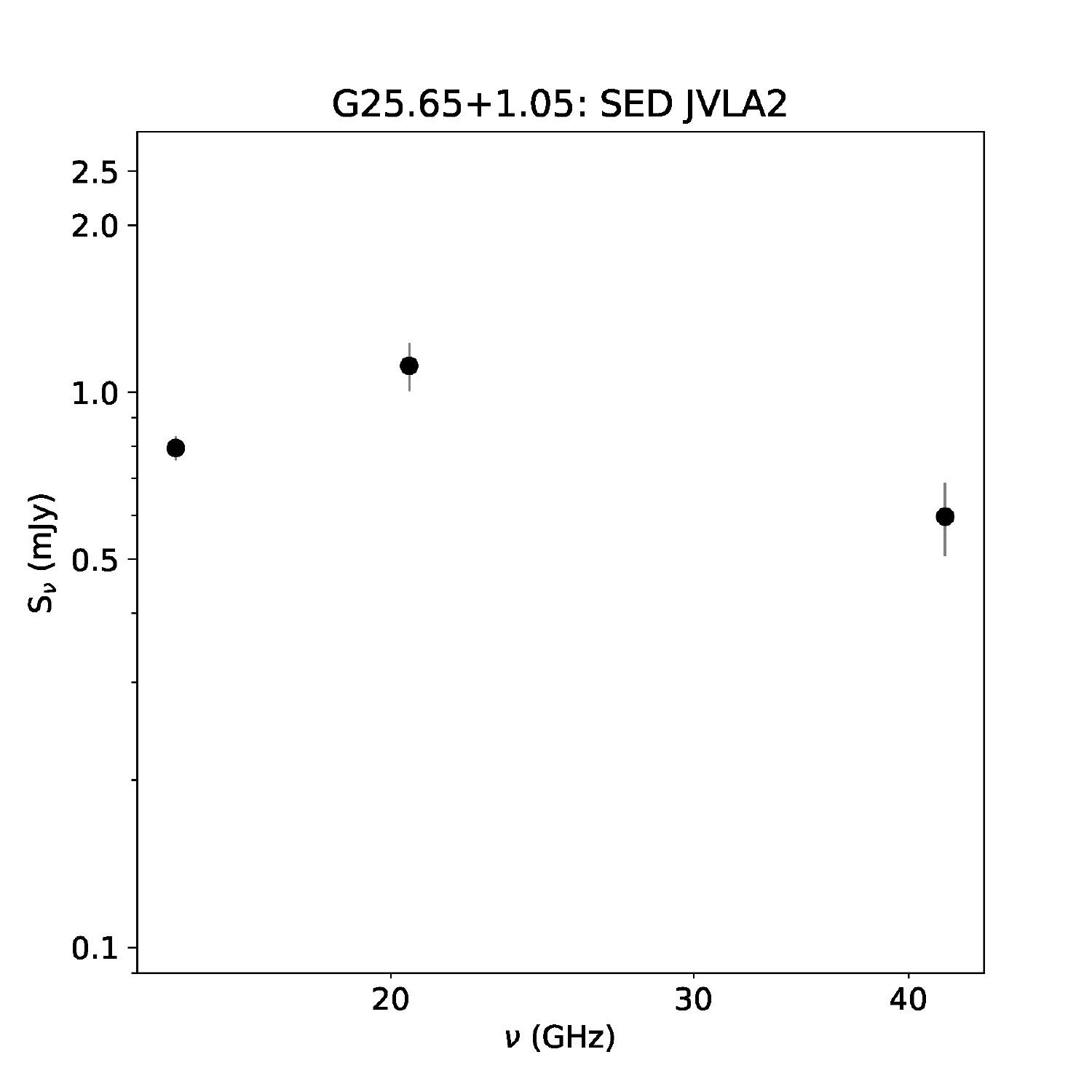}{0.45\textwidth}{(b)}
          }
\gridline{\fig{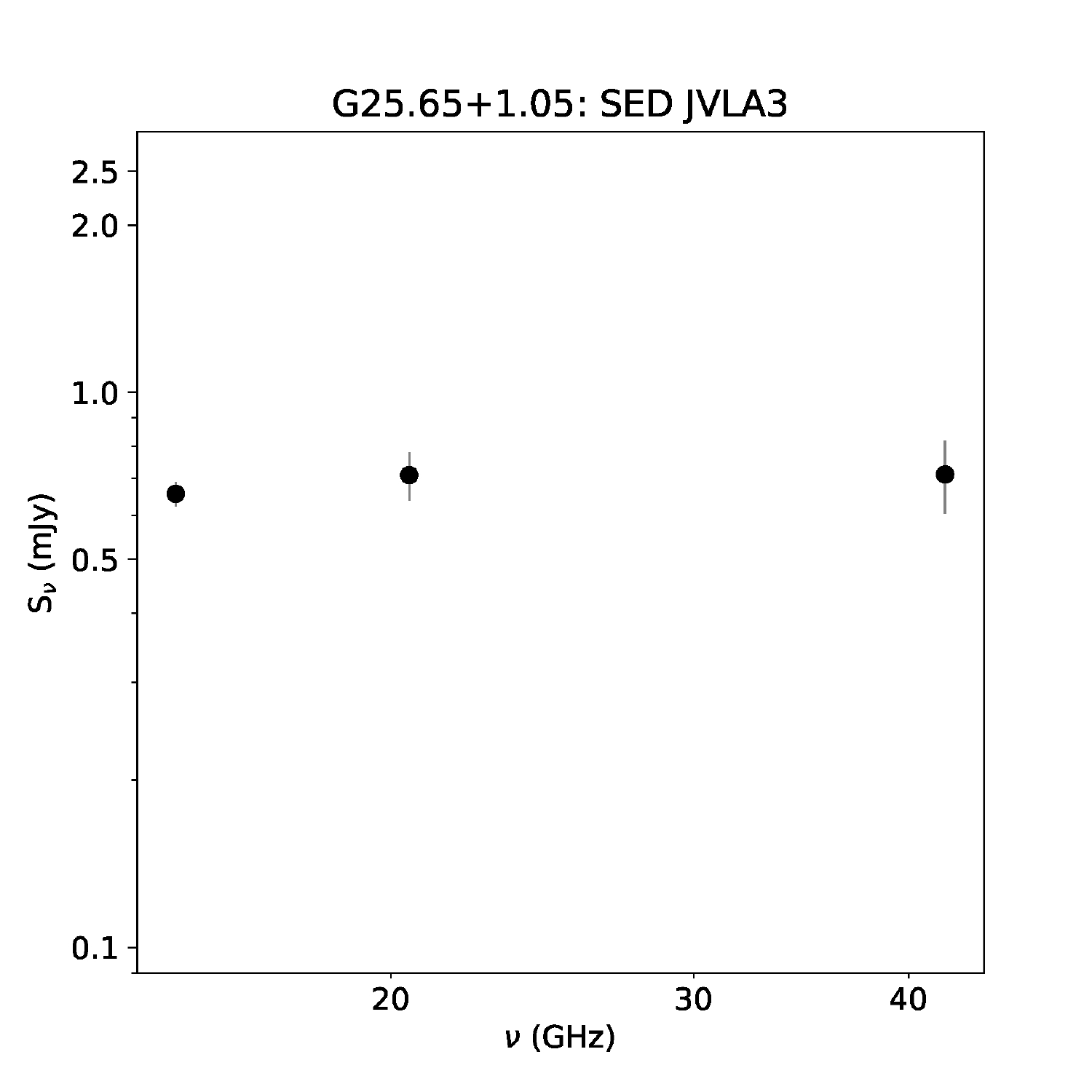}{0.45\textwidth}{(c)}
          \fig{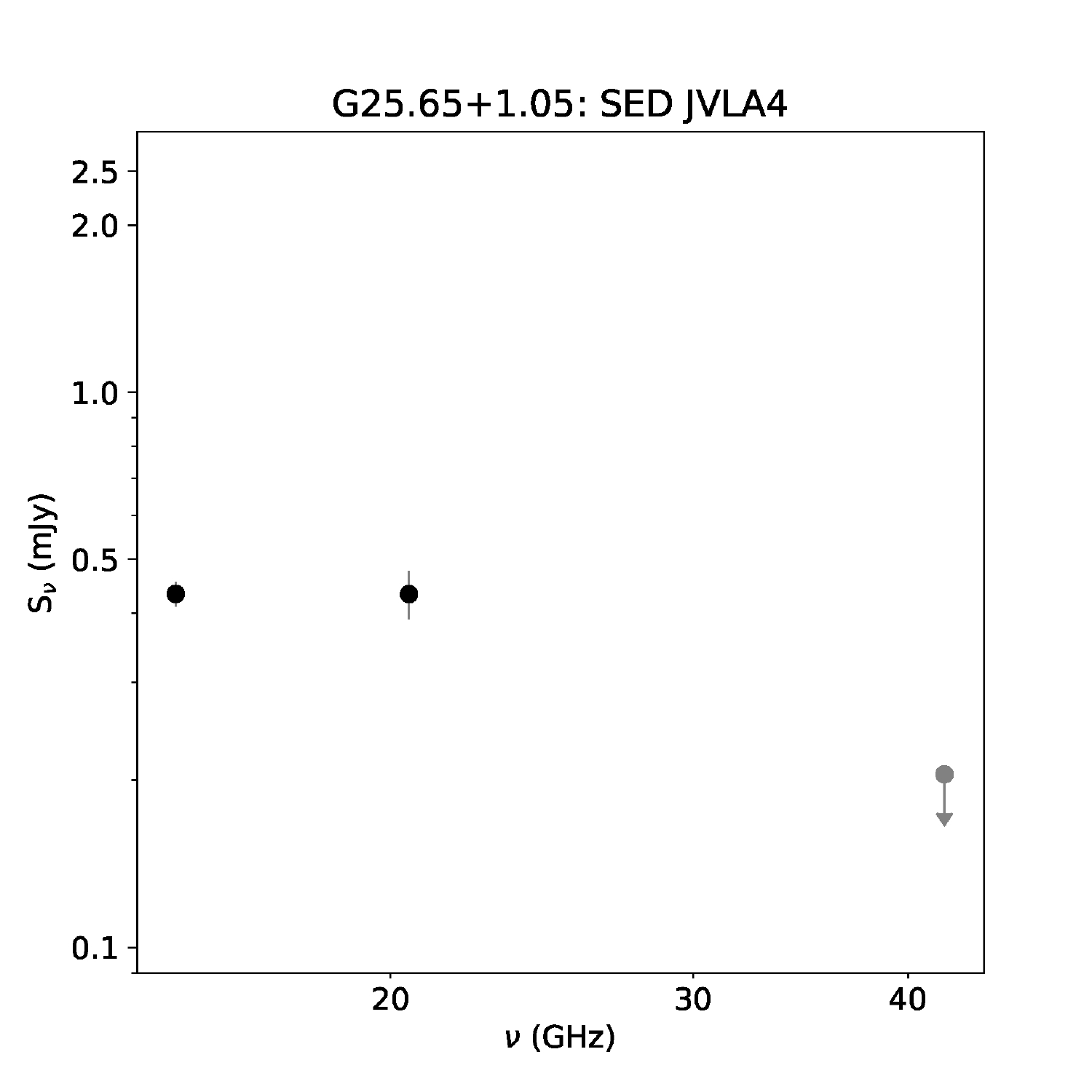}{0.45\textwidth}{(d)}
				  }
\caption{Spectral energy distribution (SED) of the detected in G25.65+1.05 continuum sources (a) JVLA1, (b) JVLA2, (c) JVLA3, and (d) JVLA4. Filled circles represent the integrated flux densities with error bars. The error bars indicate adopted in our data analysis a 5\% error at 12~GHz, a 10\% error at 22~GHz, and a 15\% error at 45 GHz.  Grey filled circle on JVLA 4 SED is an upper limit (3$\sigma$ level) at 45 GHz.  \label{fig:sed}}
\end{figure*}

Emission from the four sources is heavily blended at 6 GHz, appearing as
a single region elongated in the NE-SW direction, with JVLA 4
seen as a protrusion to the SW.   
At 12 GHz, JVLA 4 is clearly separated from the other three sources, which remain heavily blended.
At 22 GHz all four sources can be spatially distinguished and well-characterized using a 4-component Gaussian fit. 
At 44 GHz, only JVLA 1 and 2 are detected.
The statistical errors of the Gaussian fit flux densities are listed in Table \ref{tab:cont}, while the error bars in Figure~\ref{fig:sed} represent a more realistic 5\% error at 12~GHz, a 10\% error at 22~GHz, and a 15\% error at 44 GHz, which we adopted in our data analysis.

We show the flux density distributions for JVLA 1-4 in Figure \ref{fig:sed}. 
No values are shown for 6 GHz, owing to insufficient angular resolution
to distinguish one source from another.  
JVLA 1 and 2 have the best-determined spectral indices, as they are
detected (and reasonably well-resolved) at three distinct frequencies.
Both sources have declining spectra, with $\alpha$ equal to $-0.49$ and
$-0.38$, respectively, based on a linear least-squares fit to the 15, 22
and 45 GHz flux densities.  JVLA 3 and 4 have less-robust spectral index
measurements, owing to the relatively large uncertainties in their 45 GHz
flux densities.  Their spectral indices are +0.06 and $-0.78$, respectively.

\subsection{Spectral Maser Line Emission} 

Spectral line emission is detected in all frequency bands except the Ku-band (15 GHz). The position, velocity, integrated and peak flux density of each spot of 22 GHz H$_2$O maser and 6.7 and 44 GHz CH$_3$OH masers are listed in Tables \ref{tab:T22GHZ}, \ref{tab:T67GHZ}, \ref{tab:T44GHZ}, respectively.  Spectra of these three maser species and their spatial distribution are presented in Figure \ref{fig:maps}. 
Even though the structures seen in the maser maps are small (about tens of mas) compared to the angular size of the synthesized beam ($\sim$0.3$^{\prime\prime}$ at 22~GHz) the high signal-to-noise achieved on the bright maser emission is such that the centroids of Gaussian fits to the data reliably trace the general structure of emission at these scales.
An overview of maser activity in the source is presented in Figure~\ref{fig:summap}; positions of all detected continuum  and maser sources are combined in a single map. 

\subsubsection{22~GHz H$_2$O Maser}

22~GHz H$_2$O maser emission is detected in the velocity range $\sim$35-55 km~s$^{-1}$ -- the widest velocity range among all maser species detected in the region --- with the peak at 42.92 km~s$^{-1}$ (Figure \ref{fig:maps}). There are four spectral components: the line at 42.92 km~s$^{-1}$ is associated with the burst of September 2017 \citep{2017ATel10728....1V, 2019MNRAS.482L..90V} and is the closest to the systemic velocity of the source of V$_{LSR}$=+42.41 \citep{2013AA..557..94S}. 
The bursting spectral line shows an asymmetric profile with the excess in the blue wing and a line width of $\sim$1.5 km~s$^{-1}$. 
Three other lines -- the blue line at $\sim$37 km~s$^{-1}$ and two red lines at $\sim$51.5 km~s$^{-1}$ and $\sim$52.5 km~s$^{-1}$ -- have significantly lower flux densities and are separated from the systemic velocity by more than 5 km~s$^{-1}$. 

The morphological distribution of H$_2$O maser components can be divided into four groups. Three groups, which we denote as G1, G2 and G3, reside close to JVLA 1 and comprise the full range of maser velocities detected (35 to 55 km~s$^{-1}$). Another group, G4, is close to JVLA 2 and only comprises masers blueshifted with respect to the systemic velocity. No water maser emission was found near JVLA 3 or 4 (see Table \ref{tab:T22GHZ} and Figure \ref{fig:j1+j2}).

\begin{figure}
\plotone{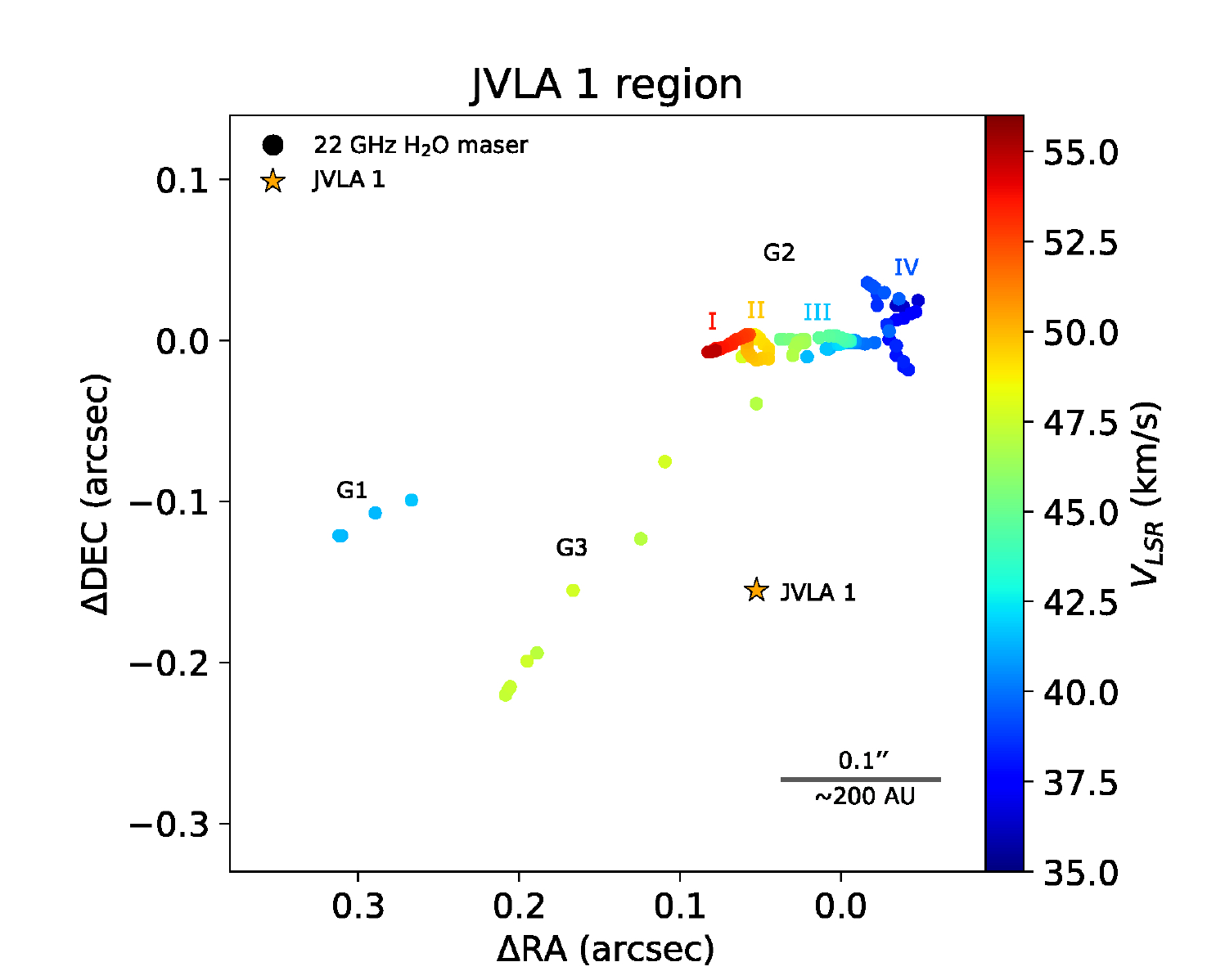}
\caption{Distribution of 22 GHz H$_2$O maser spots detected in JVLA 1 region. For 22 GHz H$_2$O maser groups and clusters labeling see Table \ref{tab:T22GHZ}.  Plot is color-coded by radial velocity (see colorbar for color scale). Positional offsets are relative to the strongest 22 GHz H$_2$O maser spot. The physical scale label (in AU) assumes the distance to the source of 2.08 kpc (the BeSSeL Survey Bayesian Distance Calculator). \label{fig:h2ojvla1}}
\end{figure}

Within the JVLA 1 maser groups, G2 masers reside $\sim$150 mas to the North of JVLA 1 and contain masers covering a wide range of velocities.
Both the G1 and G3 groups exhibit simple elongated distributions, extending toward the G2 group, and only moderate deviations from the source systemic velocity. The G1-G2-G3 groups generally form a large ($\sim$400 mas) lateral V-shape, with G2 at its apex.
The bursting maser is located within the G2 group, which exhibits the most complex spatial and velocity structure (see Figure \ref{fig:h2ojvla1}) of all groups. Particularly notable are substructures which we denote (from East to West) as group G2 clusters I, II, III and IV.
The continuous nature of clusters I-IV in position-velocity diagrams indicates that they are real physical structures.

\begin{figure*}
\gridline{\fig{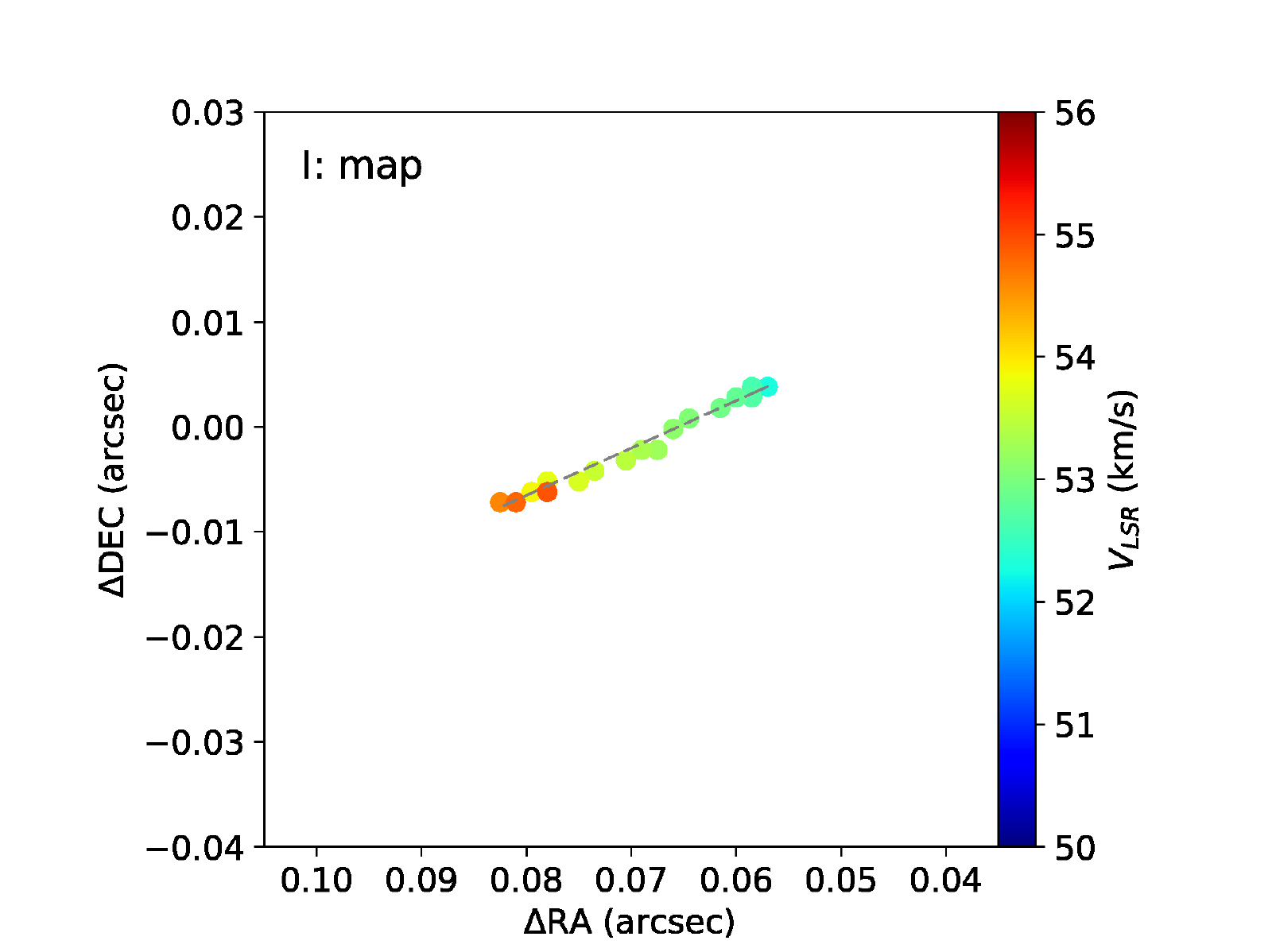}{0.35\textwidth}{}
					\fig{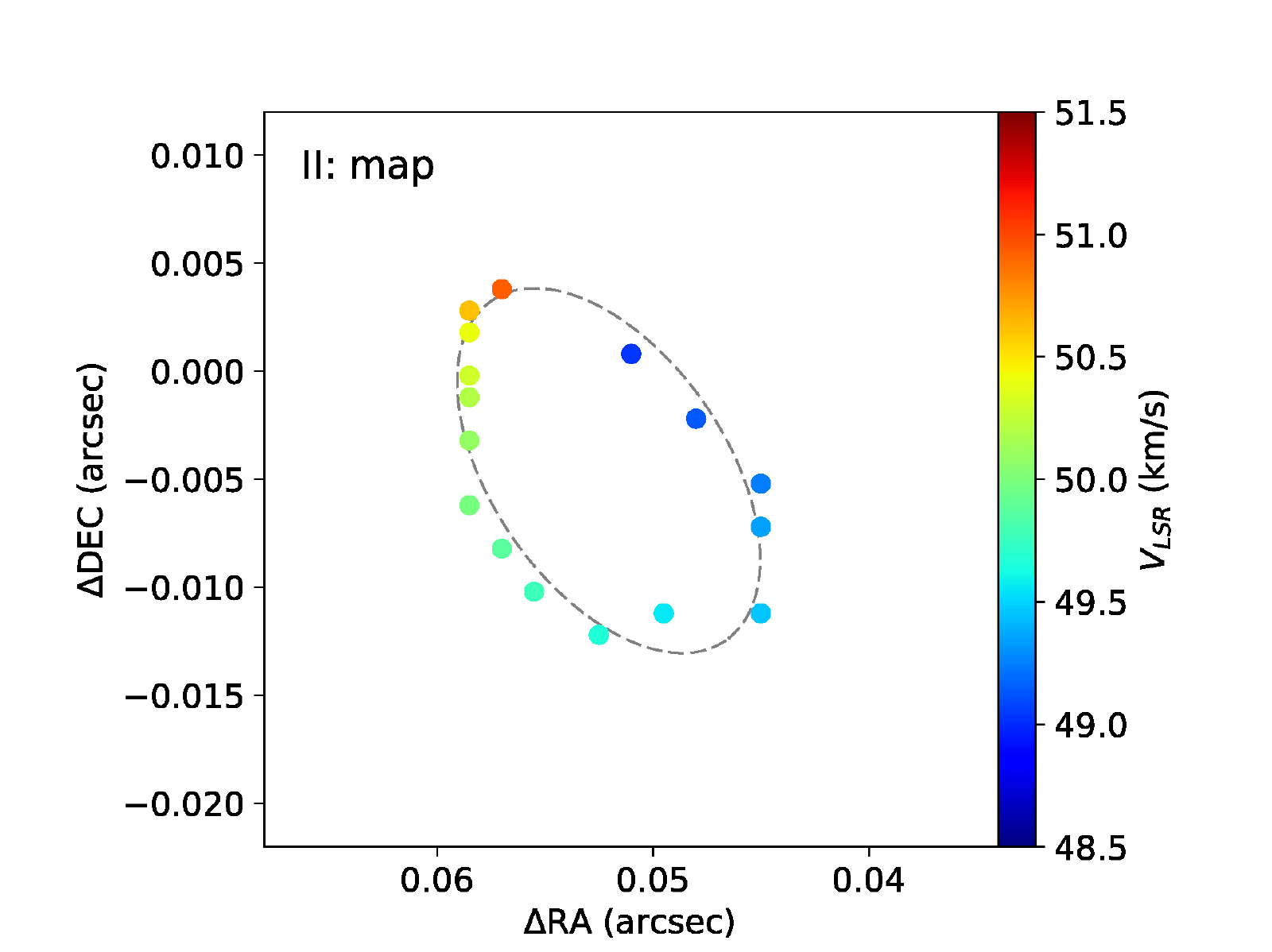}{0.35\textwidth}{}
					\fig{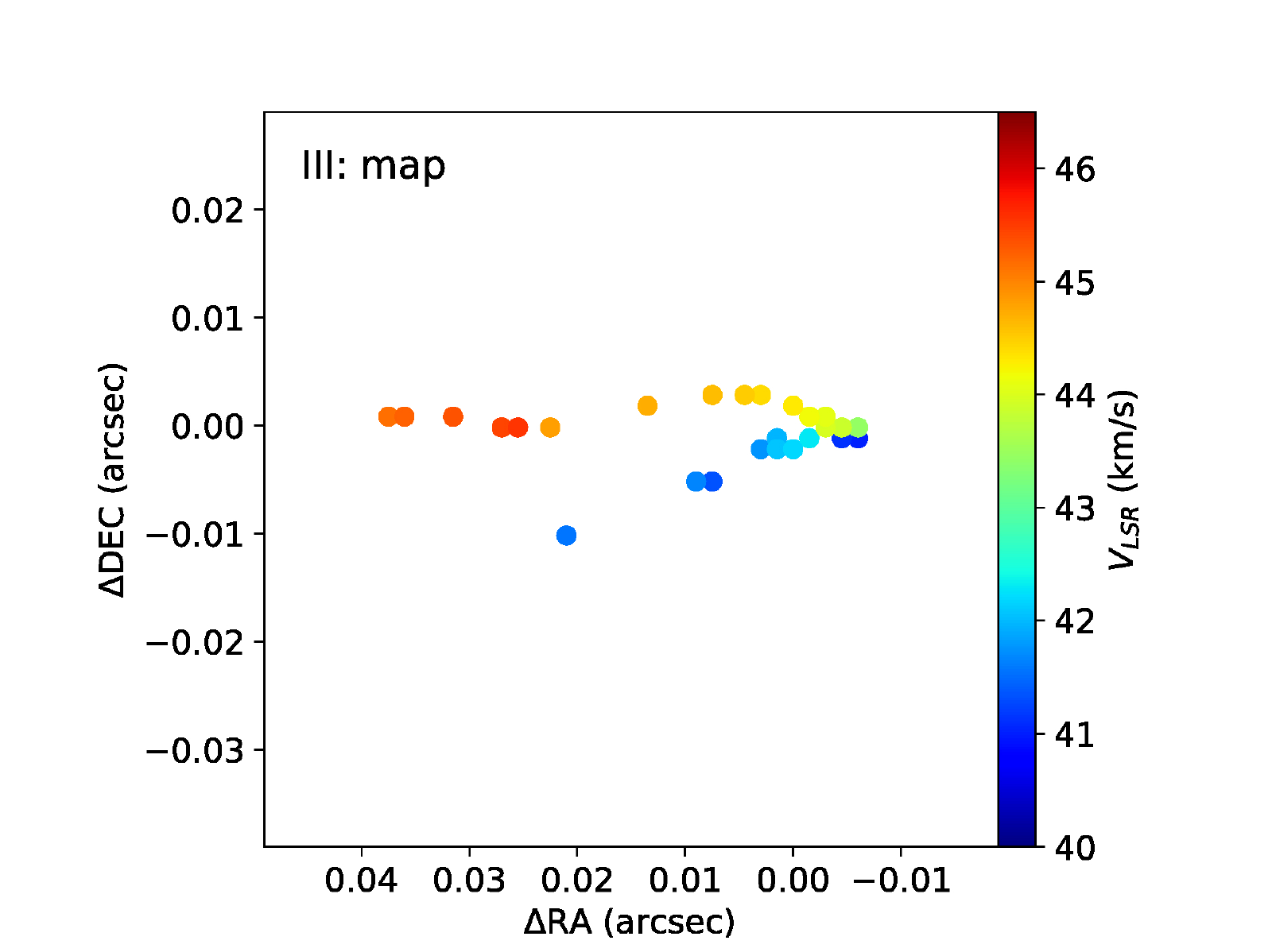}{0.35\textwidth}{}
          }
\gridline{\fig{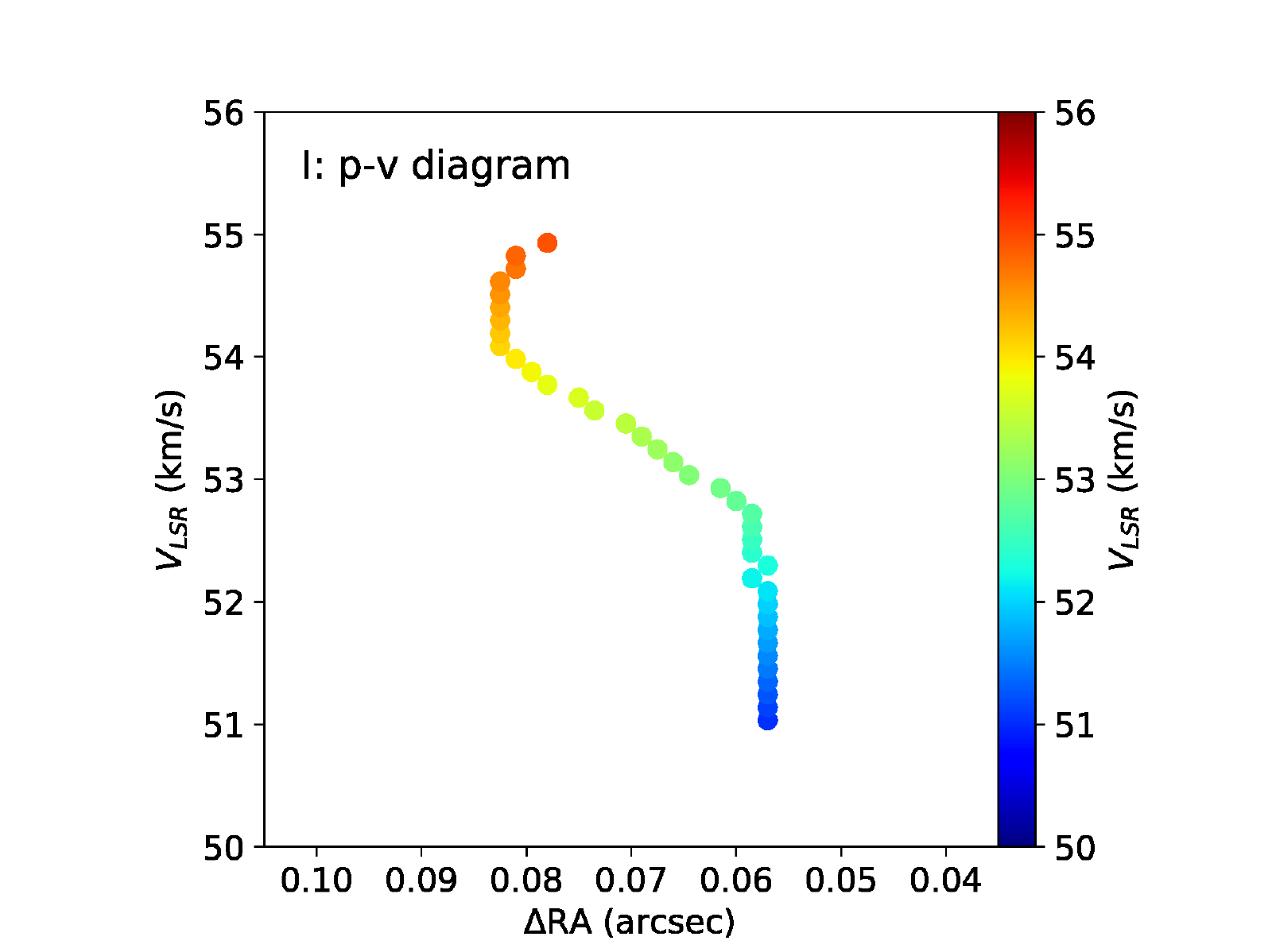}{0.35\textwidth}{(a)}
					\fig{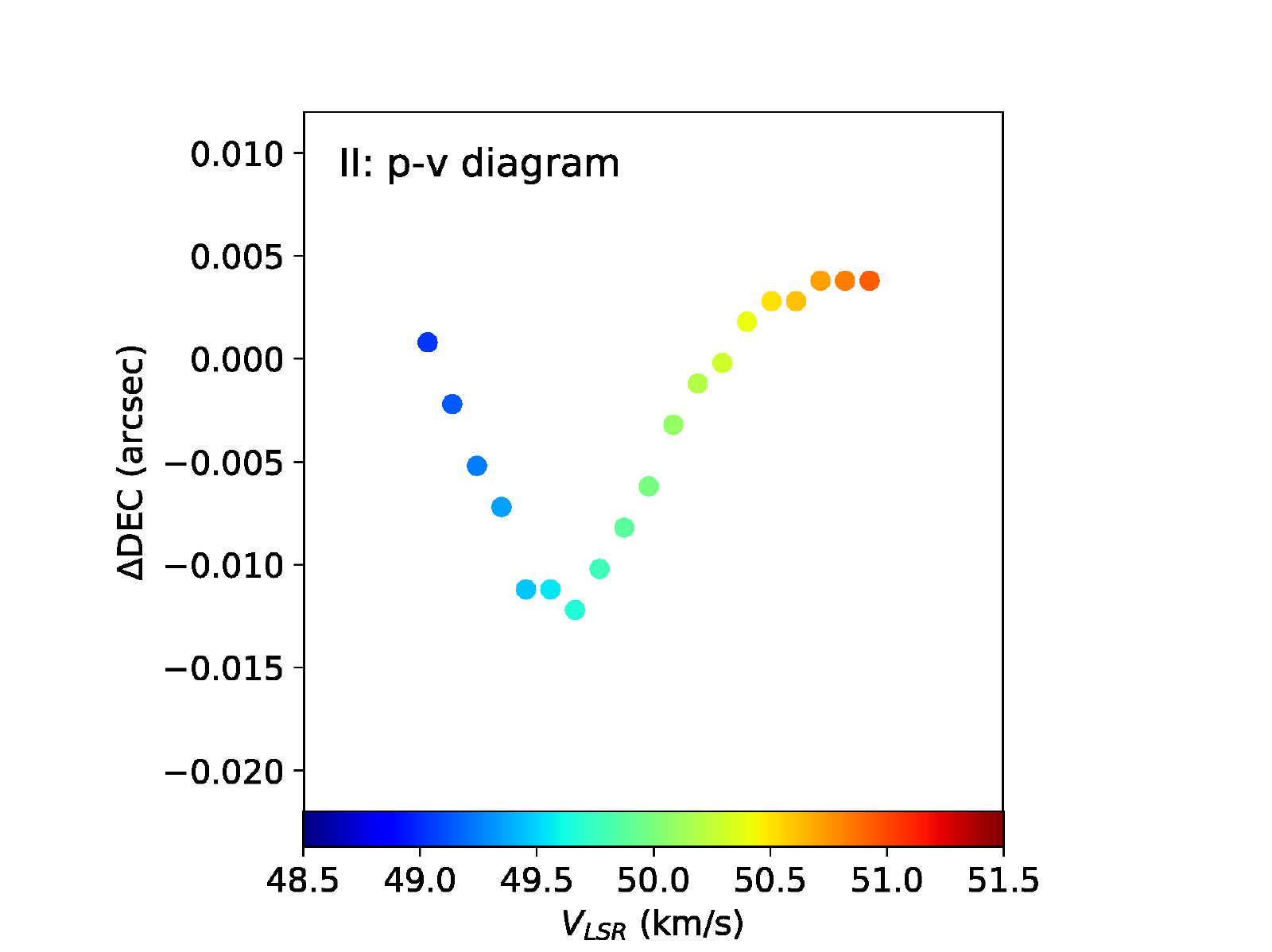}{0.35\textwidth}{(b)}
					\fig{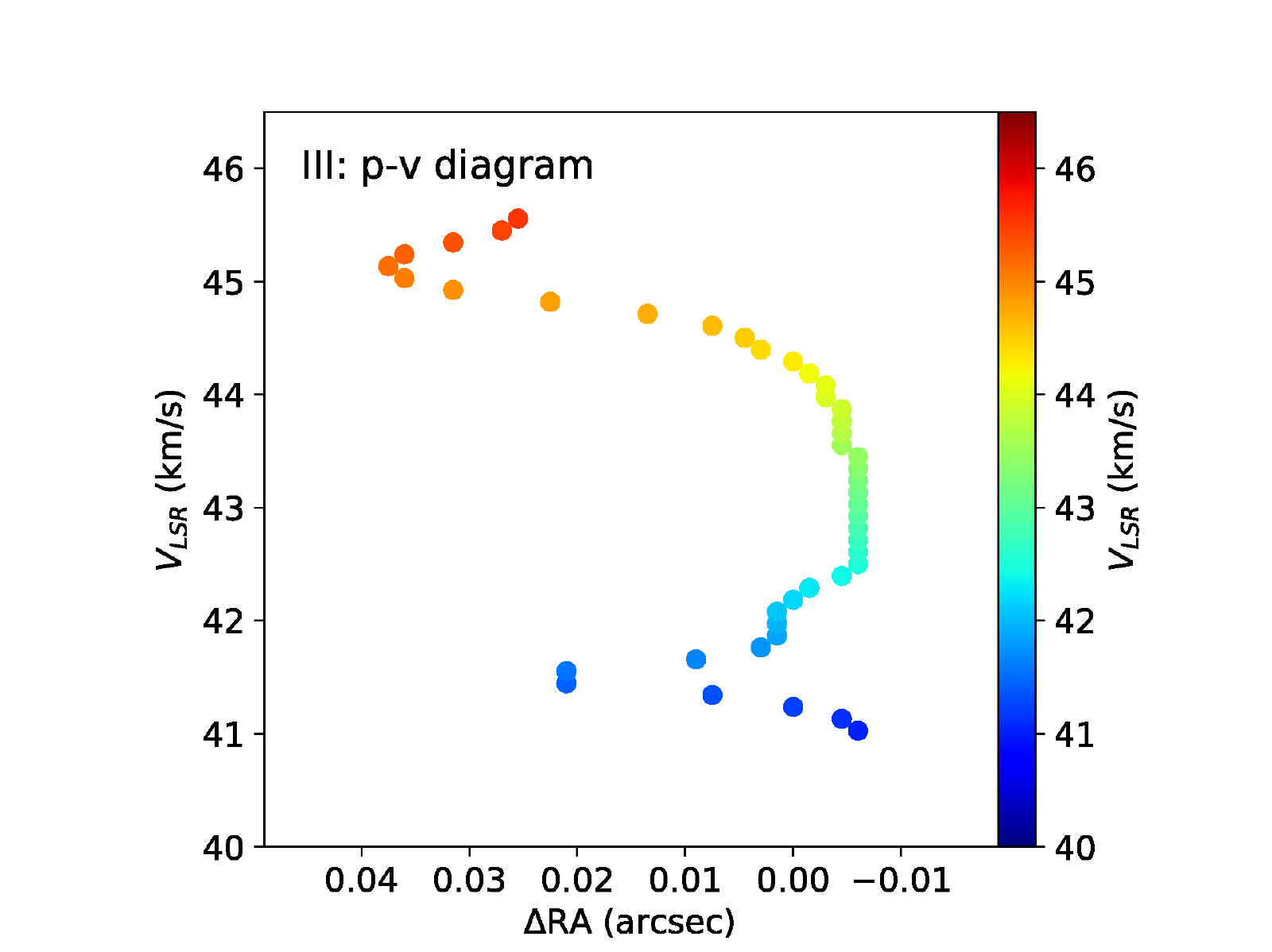}{0.35\textwidth}{(c)}
				  }
\caption{Close-up of H$_2$O maser group G2 in the JVLA 1 region: (a) cluster I, (b) cluster II, (c) cluster III containing bursting maser feature (position (0,0)). Plot is color-coded by radial velocity (see colorbars for color scale).  Positional offsets are relative to the strongest 22 GHz H$_2$O maser spot. \label{fig:G2}}
\end{figure*}

Cluster II comprises maser spots with velocities of 49-51 km~s$^{-1}$ and traces an ellipse-like structure with a size of $\sim$20 mas (Figure \ref{fig:G2}(b)).
A least-squares fit of the cluster by an ellipse (see Figure \ref{fig:G2}(b)) yields a semi-major axis of 9.54 ($\pm$0.52) mas and a semi-minor axis of 5.42 ($\pm$0.48) mas, with an inclination angle of $\sim$55$^{\circ}$.
Clusters I and III are located symmetrically around cluster II (see Figure \ref{fig:G2}): 
the red (cluster I) consists of maser spots with velocities 52-55 km~s$^{-1}$ and is located to the east, while the blue (cluster III) with velocities 41-46 km~s$^{-1}$ is detected to the west. It is notable that both clusters show velocity ranges of about 6~km~s$^{-1}$ (see Figures \ref{fig:G2} (a) and (c)).
Cluster III shows a small ($\sim$40 mas) V-shaped structure with the bursting maser located at its apex.
Structures of similar morphology are reported in the the literature, e.g. see \cite{2016MNRAS.460..283B}, and are proposed to be associated with the shocked gas at the interface between a jet and the ambient medium. 
The flux densities of maser components forming clusters I and III are about an order of magnitude higher than those of cluster II.
Cluster IV exhibits blue velocities of 36-41 km~s$^{-1}$ and is detected to the west of clusters I-III (see Figure \ref{fig:h2ojvla1}). The cluster extent is $\sim$0.06$^{\prime\prime}$ and does not show a particularly ordered structure.

\subsubsection{CH$_3$OH Masers}

\begin{figure*}
\gridline{\fig{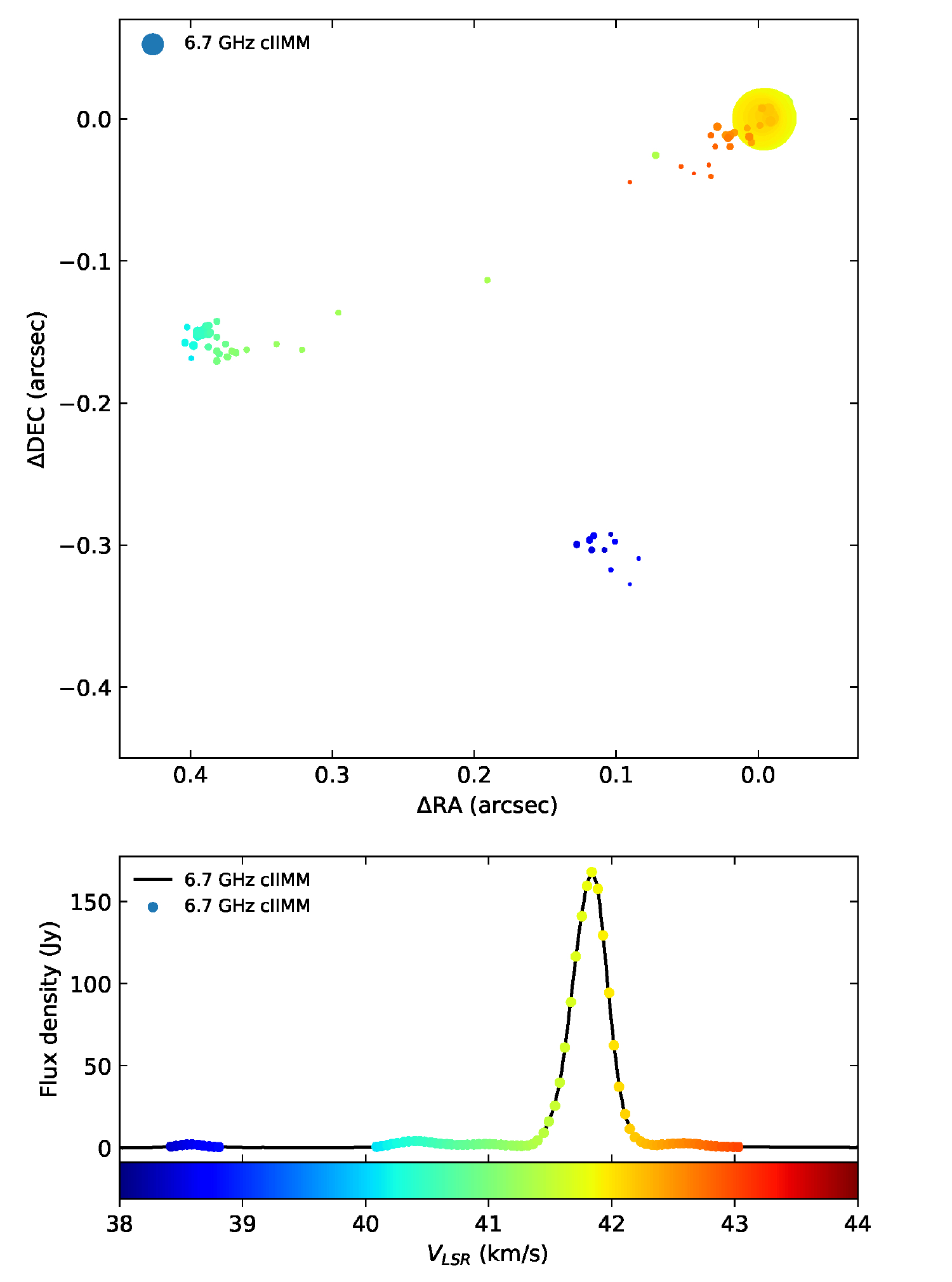}{0.4\textwidth}{(a)}
          \fig{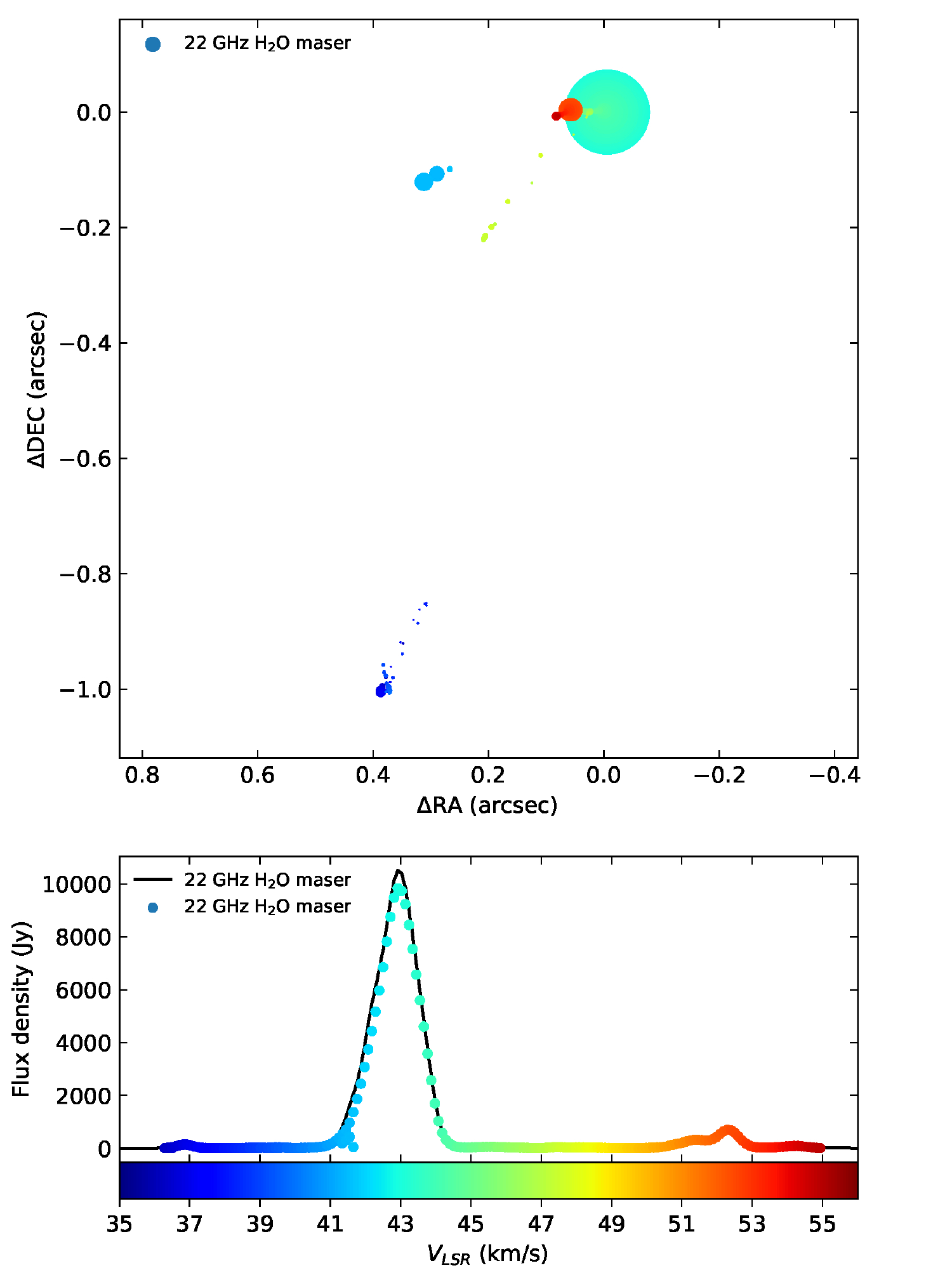}{0.4\textwidth}{(b)}
          }
\gridline{\fig{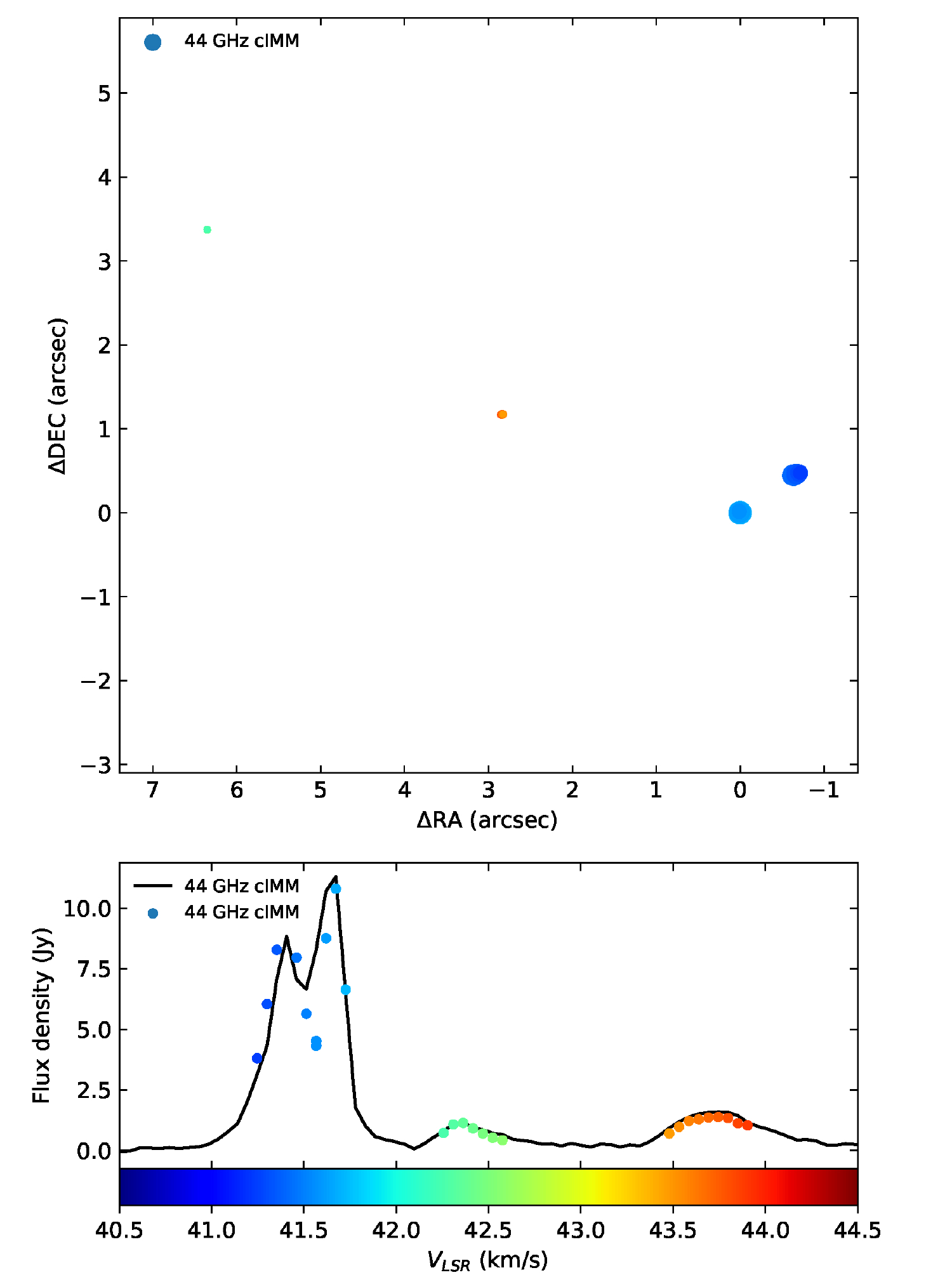}{0.4\textwidth}{(c)}
				  }
\caption{Maser data at (a) 6.7 GHz, (b) 22 GHz, and (c) 44 GHz for G25.65+1.05: the upper panel -- a map of the maser spots and the lower panel -- the maser source spectrum. Plots are color-coded by radial velocity (see colorbar for color scale). The diameter of each spot is proportional to the flux. Positional offsets are relative to the strongest maser spot at each frequency.\label{fig:maps}}
\end{figure*}

\begin{figure}
\plotone{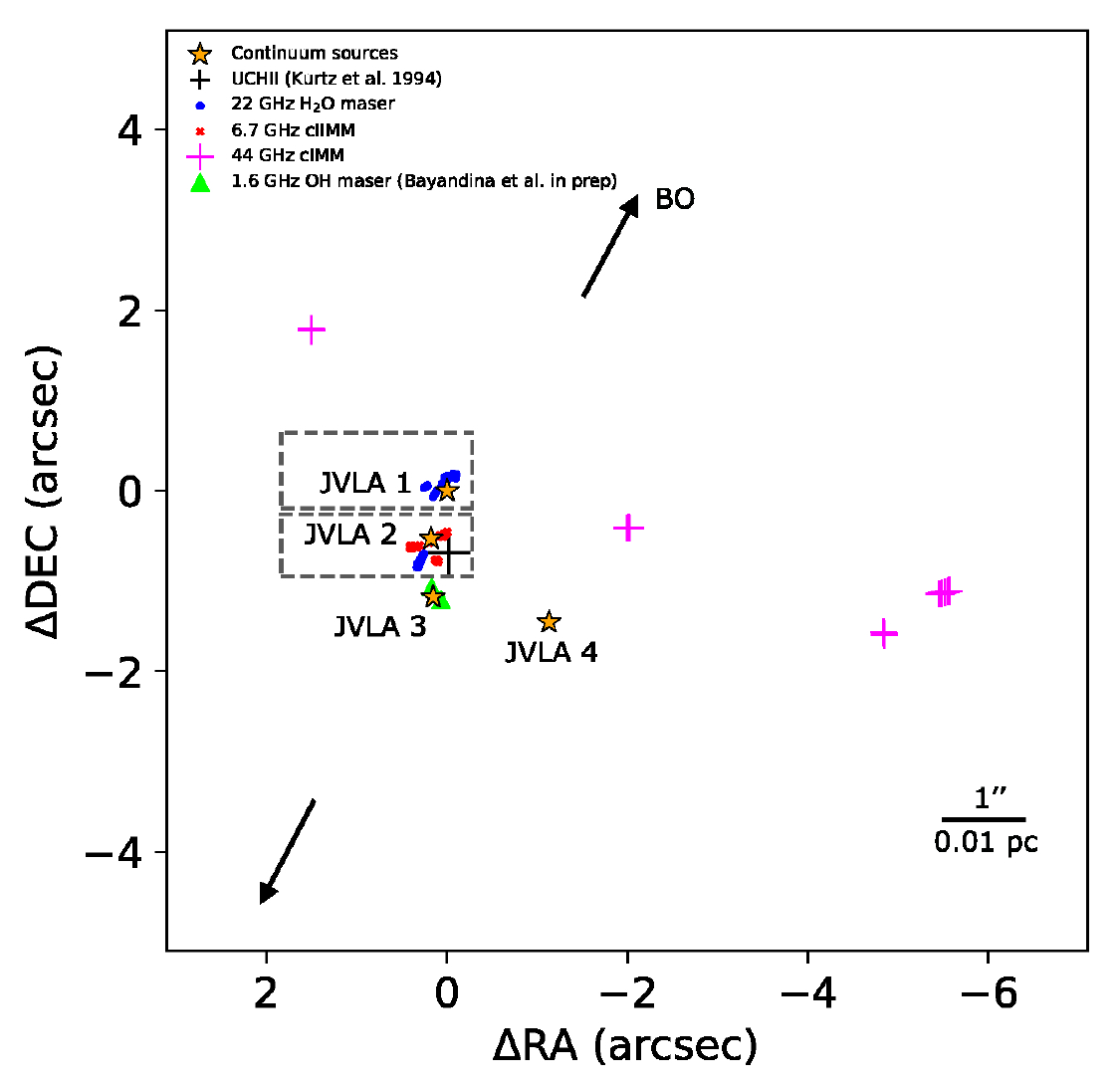}
\caption{Objects detected toward G25.65+1.05 with the JVLA: continuum sources (see Table 2) -- orange stars, 22 GHz H$_2$O masers -- blue circles, 6.7 GHz cIIMMs -- red ``x'' crosses, 44 GHz cIMMs -- magenta crosses, 1665 MHz OH masers (pre-burst C-configuration JVLA observations of 2013, see \cite{2018ApJS..00..00B}) -- green triangles. Black cross indicates the position of UCHII region detected at 3.6 cm in \cite{1994ApJS..91..659K}. Black arrows represent the direction and the position angle (but not the actual position) of bipolar outflow from  \cite{2013AA..557..94S}. Positional offsets are relative to the JVLA 1 continuum source. The physical scale label (in pc) assumes the distance to the source of 2.08 kpc (the BeSSeL Survey Bayesian Distance Calculator). \label{fig:summap}}
\end{figure}

\begin{figure}
\plotone{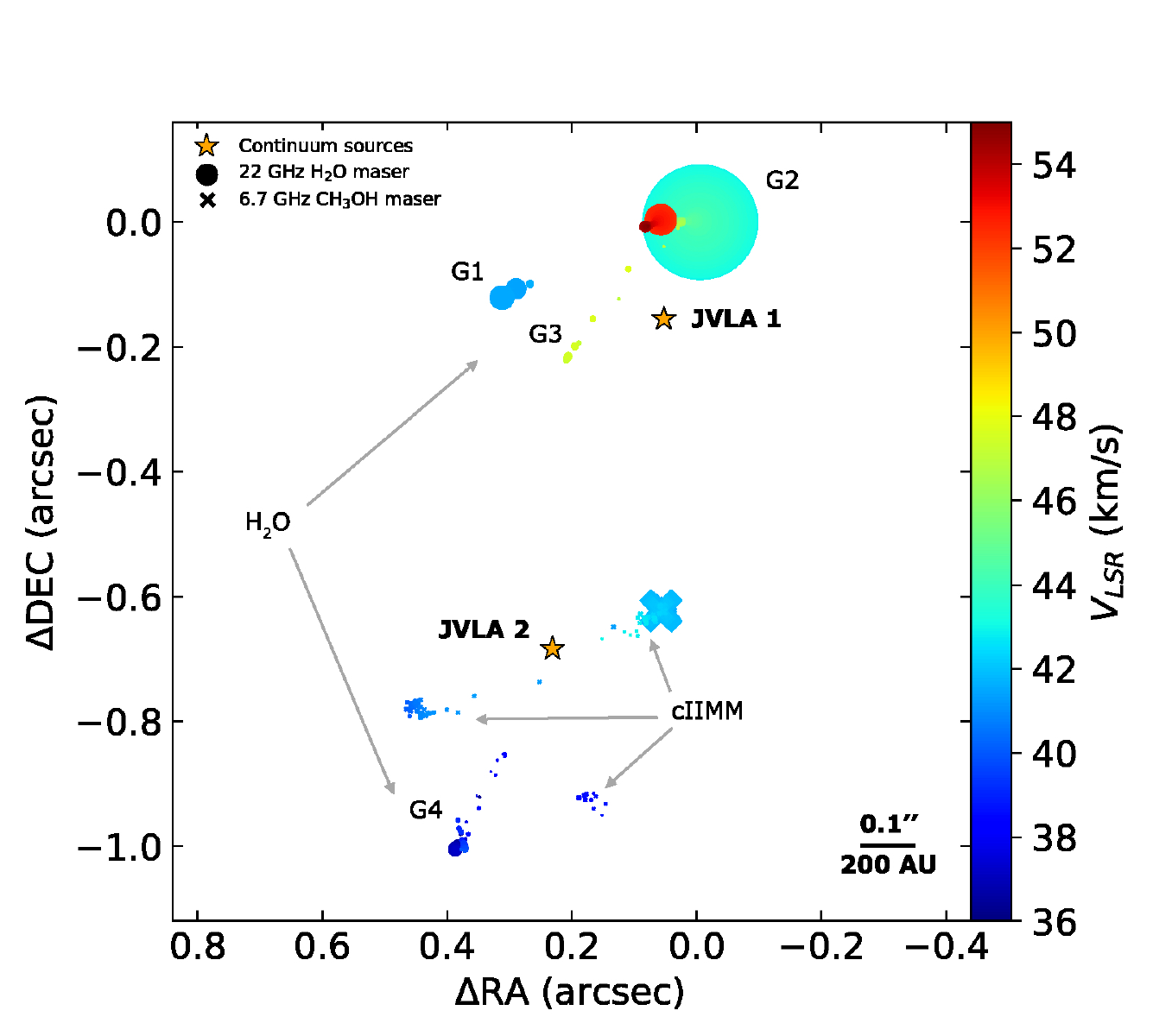}
\caption{Distribution of 22 GHz H$_2$O (marked by circles) and 6.7 GHz CH$_3$OH (marked by ``x'' crosses) maser spots detected in vicinity of JVLA 1 and 2 sources. The diameter of each spot is proportional to the flux. For 22 GHz H$_2$O maser groups labeling see Table \ref{tab:T22GHZ}. Plot is color-coded by radial velocity (see colorbar for color scale). Positional offsets are relative to the strongest 22 GHz H$_2$O maser spot. The physical scale label (in AU) assumes the distance to the source of 2.08 kpc (the BeSSeL Survey Bayesian Distance Calculator). \label{fig:j1+j2}}
\end{figure}

6.7~GHz CH$_3$OH maser emission is detected in the velocity range $\sim$38-44 km~s$^{-1}$, with the peak at 41.84 km~s$^{-1}$ (Figure~\ref{fig:maps}). 
Detected maser lines are blue-shifted fromthe source systemic velocity. Maser spots are distributed over an area of size $\sim$0.4$^{\prime\prime}$ (Figure \ref{fig:maps}) and spatially associated with the continuum source JVLA 2 (Figures \ref{fig:summap} and \ref{fig:j1+j2}). 

There are three loci of 6.7~GHz CH$_3$OH maser emission. The northern two loci are elongated in the NW-SE direction with a size $\sim$500 mas and located symmetrically around JVLA 2. Maser spots exhibit velocities of 40-43 km~s$^{-1}$ and display the largest cIIMM flux densities in the region.
The southern locus is compact, with a size of $\sim$50 mas (Figure \ref{fig:j1+j2}) and formed by weak maser spots with velocity $\sim$38.6 km~s$^{-1}$.
The northern and southern loci are separated from each other by $\sim$200~mas.

No 12~GHz class II CH$_3$OH maser emission was found above the 3$\sigma$ level of 0.16 Jy of these observations. Their non-detection is relevant to the discussion in Section 5. 

44~GHz CH$_3$OH maser emission was detected in the velocity range $\sim$41-44 km~s$^{-1}$, with the peak at 41.67 km~s$^{-1}$ (Figure~\ref{fig:maps}). Maser features are distributed along the NE-SW direction over a range $\sim$8$^{\prime\prime}$ (Figures \ref{fig:maps} and \ref{fig:summap}). Four spatial clusters of 44 GHz cIMM are present, with the NE-SW line located towards the NW of the continuum sources.

\section{Discussion}

\subsection{Nature of the Continuum Sources} \label{Nature of the continuum sources}

In previous observations, made with lower angular resolution, continuum emission detected in the region was not resolved into separate sources.
A single continuum peak at 3.6 cm (8.4 GHz) was detected in two different B-configuration VLA observations: \cite{1994ApJS..91..659K} and \cite{1995MNRAS..276..1024J}. Both studies reported a flux density of $\sim$3.8 mJy, although the peak emission coordinates were slightly different -- \cite{1994ApJS..91..659K} places the peak closer to JVLA 2, while \cite{1995MNRAS..276..1024J} locate it closer to JVLA 1.
The higher resolution, multi-frequency observations reported here allow for more detailed analyses of the continuum sources in this region.

The negative spectral indices of JVLA 1 and 2 (and possibly of JVLA 4)
suggest that these are all relatively young objects, which are not
yet photo-ionizing their surroundings.  Rather, the emission appears
to be synchrotron radiation. Synchrotron jets are known to arise in 
young stellar objects (e.g., see the review by \cite{2018A&ARv..26....3A}). 
Such a mechanism appears to be at work in JVLA 1 and 2, possibly
related to shocked gas near these embedded sources.
Synchrotron emission can also arise near massive binary stars (e.g., \cite{2012ApJ...755..152R}).
The relatively flat spectrum of JVLA 3, and the detection of OH masers in its vicinity, 
suggest that it may be free-free emission from an optically thin HII region,
in which case it is likely to be the oldest object of the four. 

JVLA 1 and 2 appear to be in different evolutionary stages, based on their association with water and methanol masers (see
Figures \ref{fig:summap} and \ref{fig:j1+j2}).  
ALMA archive data (Project Code 2012.1.00826.S) provide a 350 GHz image showing a point source at the position of JVLA 2. An unresolved bright IR source detected in K-band images \citep{2002AA.394..225Z, 2010MNRAS...404..661V} is also coincident with JVLA 2. 
Moreover, its association with cIIMM strongly suggests that this source is an embedded, accreting, massive young stellar object (MYSO).

\begin{figure*}
\gridline{\fig{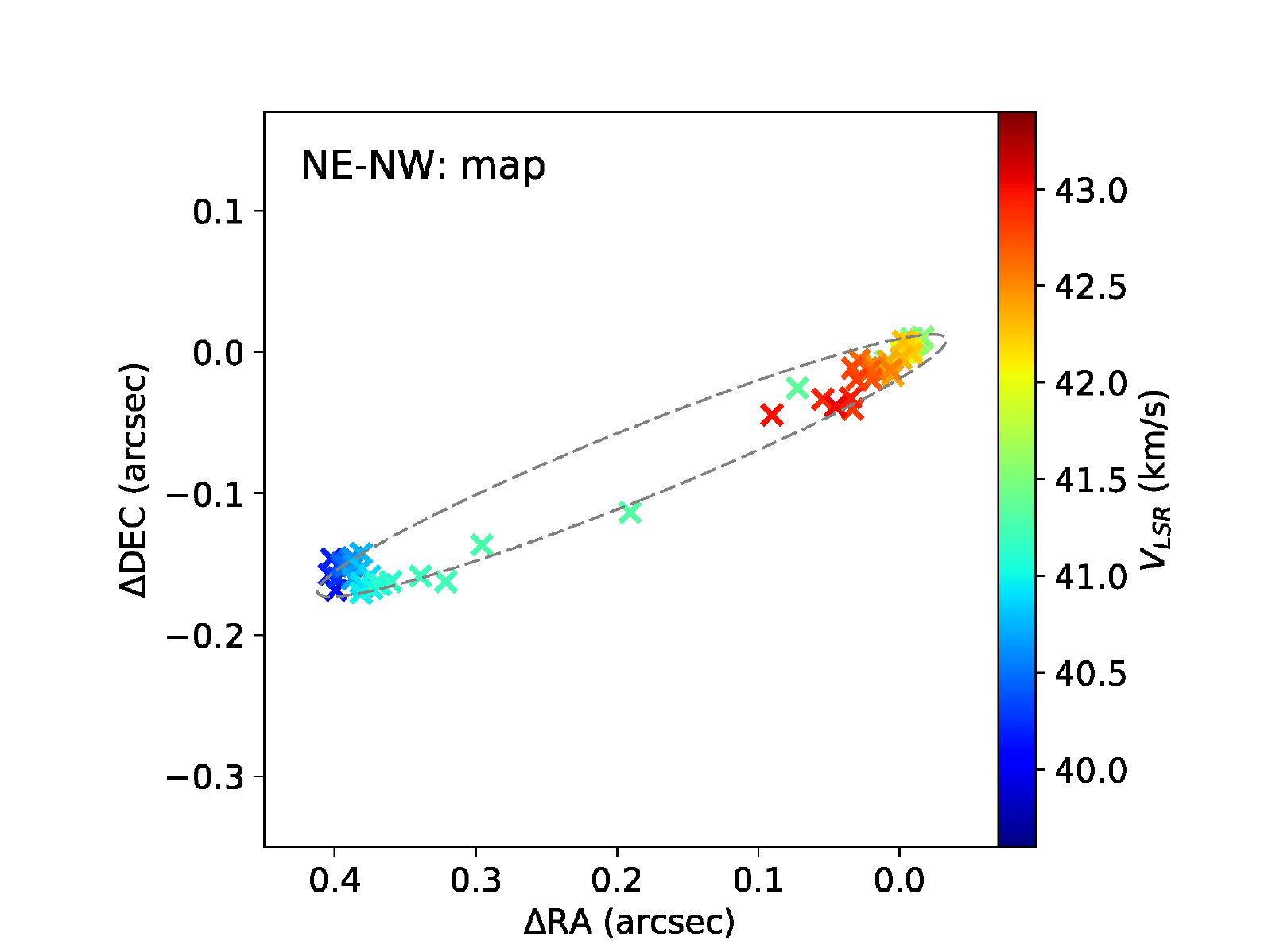}{0.75\textwidth}{}
          }
\gridline{\fig{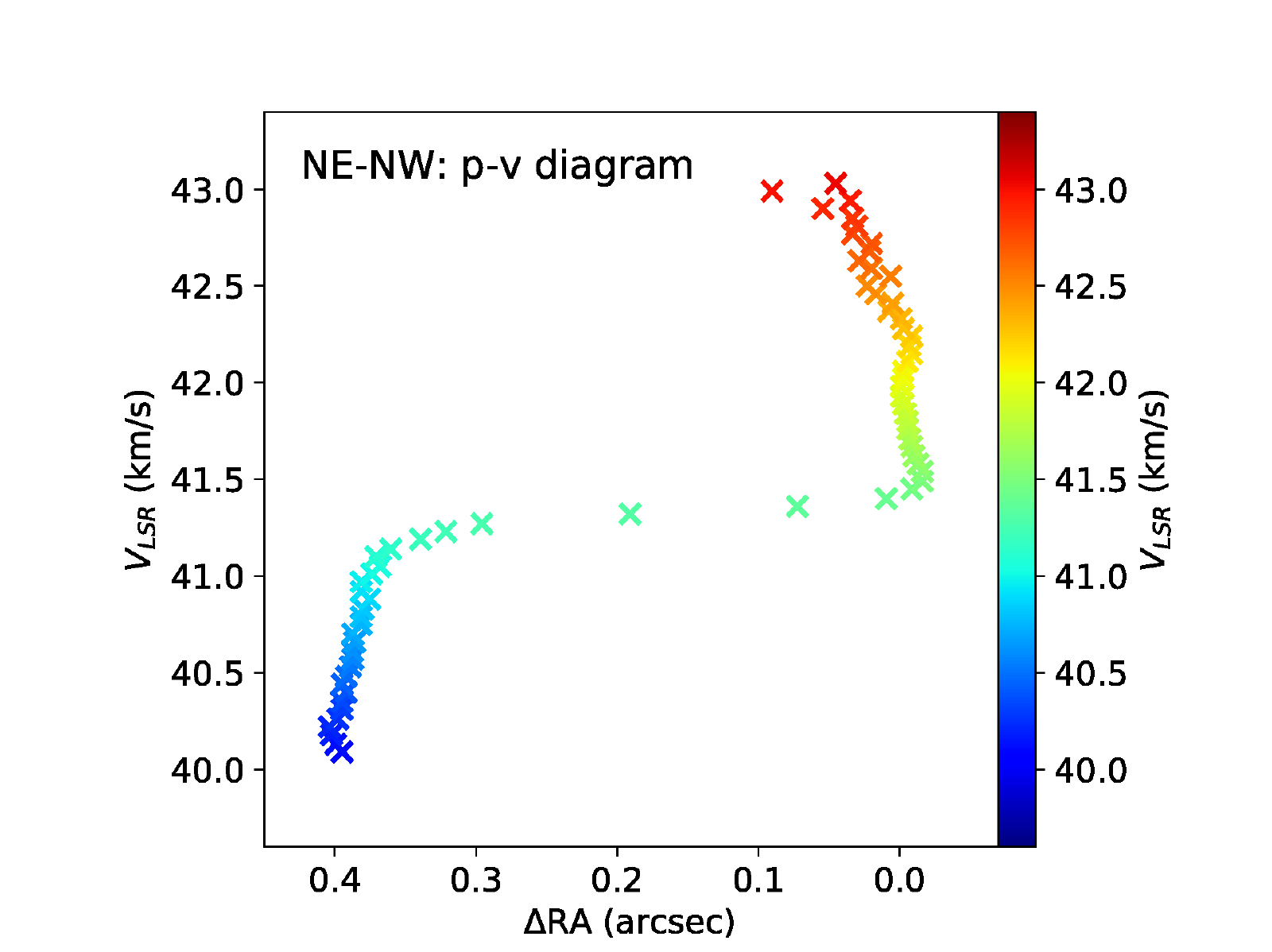}{0.75\textwidth}{}
				  }
\caption{6.7 GHz CH$_3$OH maser clusters NE and NW: upper panel -- map and a least squares fitting of ellipse, bottom panel -- p-v diagram. Plot is color-coded by radial velocity (see colorbar for color scale). \label{fig:67Npv}}
\end{figure*}

The 6.7 GHz methanol masers in JVLA 2 seem to trace an edge-on disk. A least-squares ellipse fit to the data (Figure \ref{fig:67Npv}) gives a semi-major axis of $\sim$240 mas and semi-minor axis of $\sim$25 mas, with an inclination angle of $\sim$68$^{\circ}$. Thus, assuming a kinematic distance of D = 2.08 kpc (BeSSeL, \url{http://bessel.vlbi-astrometry.org}), the linear size of the putative disk is $\sim$1000 AU which is similar to the typical sizes of circumstellar disks traced by 6.7 GHz CH$_3$OH masers (e.g., \cite{1998ApJ..508..275N}). The position-velocity diagram for the NE and NW clusters (Figure \ref{fig:67Npv}) is in general agreement with the theoretical calculations for protostellar discs \citep{1991MNRAS.251..707R}.
An alternative explanation is a bipolar outflow.  Assuming a 45$^\circ$
orientation with respect to the sky, the 4 km~s$^{-1}$ velocity range over a
distance of $\sim$1000 AU suggests a dynamic age of $\sim$1000 yr for such an outflow.

The 6.7~GHz class II CH$_3$OH maser is the only maser species for which we have pre-burst interferometric images.
The ATCA 6.7~GHz image of the region showed the chain of four CH$_3$OH maser spots A, B, C, and D linearly distributed over $\sim$1.5$\arcsec$ in the NS direction \citep{1998MNRAS.301..640W}: spot A is undetected in the current JVLA observations, while our northern maser group and southern maser group can be associated with spots B and C/D, respectively. Comparison of the most recent compact array 6.7~GHz map \citep{2016ApJ...833..18H} and our 6.7~GHz data does not show significant changes in the CH$_3$OH maser spot distribution, with only a shift by $\sim$50~mas ($\sim$100 AU) of the southern maser cluster position.
In VLBI observations of \cite{2014PASJ.66....31F} only the NW maser cluster was detected, while in EVN observations of \cite{2015AA..578..102S} positions of both the NE and NW maser clusters were determined. The southern cluster was not detected in either VLBI study, which may indicate extended structure.

The nature of JVLA 1 is less certain. Its lack of cIIMM emission, declining spectrum and exclusive association with water masers (Figures \ref{fig:summap} and \ref{fig:j1+j2}) may indicate that the continuum emission traces a shocked region of synchrotron emission, possibly a jet driven from the JVLA 2 MYSO. On the other hand, JVLA 1 may itself be a young stellar object with a jet. An accurate source distance would assist in distinguishing between these two scenarios by providing a distance for the presumed source-ejecta separation between JVLA 1 and 2.

Null detections of the 12.2 GHz maser line were reported by \cite{1994MNRAS.269..257G} and \cite{2010MNRAS.401.2219B} with detection limits of $\sim$10 Jy and $\sim$0.5 Jy, respectively. 
Our observations also do not detect maser emission at a 3$\sigma$ level of 0.16 Jy.
\cite{2010MNRAS.401.2219B} suggests that 12.2 GHz CH$_3$OH masers are associated with a later evolutionary phase of massive star formation. Therefore, the non-detection of this maser species in G25.65+1.05 may be indicative of an early evolutionary stage for the MYSO(s) in G25.65+1.05.

\subsection{Outflows/Episodic Ejection Traced by Masers}

No interferometric observation of the 44 GHz class I masers in G25.65+1.05 has previously been reported in the literature. 

\begin{figure}
\plotone{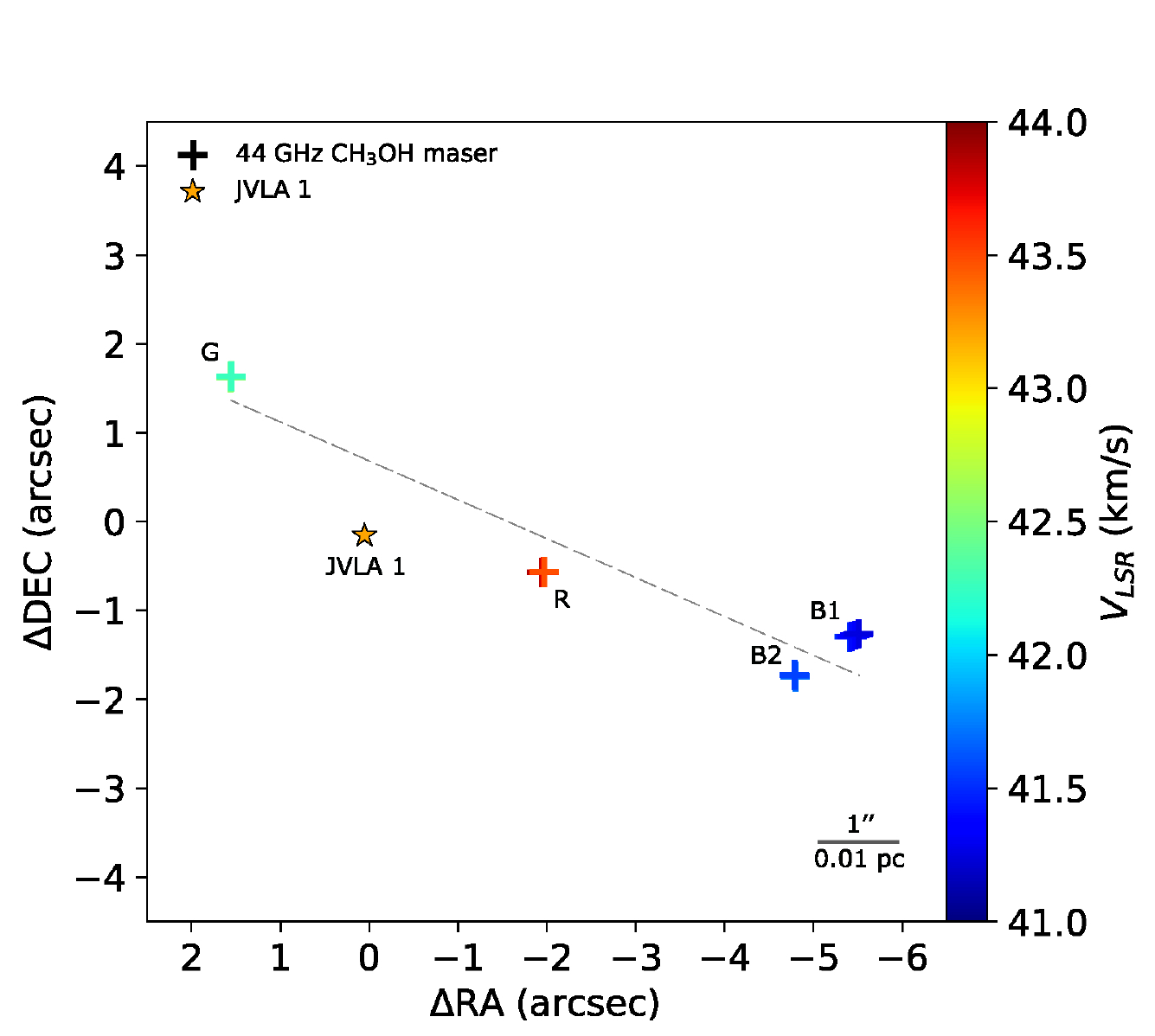}
\caption{Distribution of 44 GHz CH$_3$OH maser spots. For 44 GHz CH$_3$OH  maser groups labeling see Table \ref{tab:T44GHZ}. The dashed line represents the least squares fit of maser positions. Plot is color-coded by radial velocity (see colorbar for color scale). Positional offsets are relative to the JVLA 1 continuum source. The physical scale label (in pc) assumes the distance to the source of 2.08 kpc (the BeSSeL Survey Bayesian Distance Calculator). \label{fig:44}}
\end{figure}

\begin{figure}
\plotone{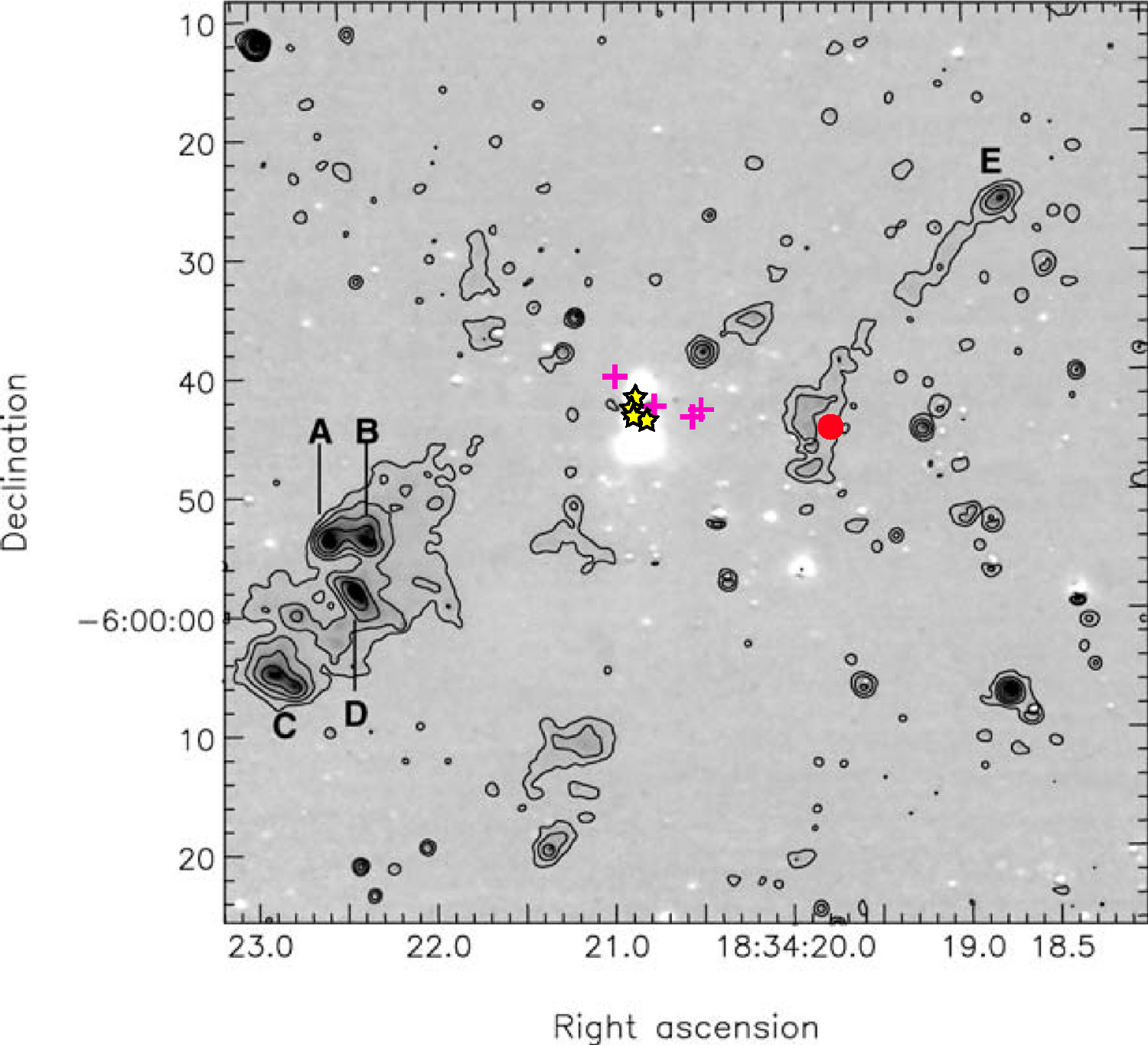}
\caption{Objects detected toward G25.65+1.05 with the JVLA are overplotted on continuum-subtracted $\nu$ = 1-0 S(1) H$_2$ image from Fig.2 in \cite{2006MNRAS..367..238T}: continuum sources (see Table 2) -- yellow stars, 44 GHz cIMMs -- magenta crosses. Position of IRAS 18316-0602 is indicated by red circle. For the levels see \cite{2006MNRAS..367..238T}. \label{fig:h2}}
\end{figure}

\begin{figure}
\plotone{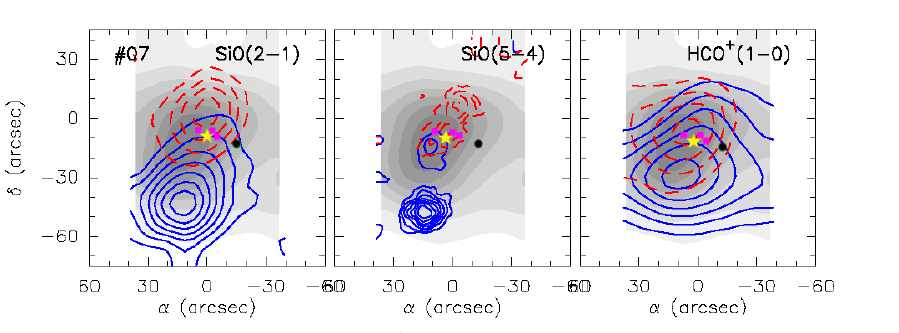}
\caption{Objects detected toward G25.65+1.05 with the JVLA are overplotted on SiO (2-1), SiO (5-4), and HCO$^+$ (1-0) outflow maps from Fig.2 in \cite{2013AA..557..94S}: average position of continuum sources (see Table 2) -- yellow star, 44 GHz cIMMs -- magenta squares. Position of IRAS 18316-0602 is indicated by black circle. The gray scale corresponds to the N$_2$H$^+$ (1−0) integrated emission. Blue-solid and red-dashed contours represent blue- and red-shifted integrated wing emission, respectively. For the levels see Table 2 in \cite{2013AA..557..94S}. \label{fig:BO}}
\end{figure}

The two brighter 44 GHz cIMM blue features at velocities 41.4 and 41.7 km~s$^{-1}$ (B1 and B2; hereafter cIMM features are named in accordance with their velocity --- see Table \ref{tab:T44GHZ} and Figure \ref{fig:44}) are separated by  $\sim$6$^{\prime\prime}$ ($\sim$0.06~pc) from the continuum source JVLA 1, associated with the bursting H$_2$O maser (see Figure \ref{fig:44}). Weaker maser features G (green) at the velocity 42.4 km~s$^{-1}$, and R (red) at the velocity 43.7 km~s$^{-1}$ are located almost symmetrically about JVLA 1 at $\sim$2$^{\prime\prime}$, i.e. 0.02 pc, (Figure \ref{fig:44}). Maser feature R spatially coincides with the peak of 1-0 S(1) H$_2$ line emission (Figure \ref{fig:h2}), while no H$_2$ emission was detected at the positions of the ``green'' and ``blue'' cIMM features \citep{2006MNRAS..367..238T}.  Figures \ref{fig:h2} and \ref{fig:BO} show that cIMM features fall along a straight line perpendicular to the axis of NW-SE oriented large-scale bipolar outflow detected in 1-0 S(1) H$_2$ \citep{2006MNRAS..367..238T} and SiO (2-1), SiO (5-4), HCO$^+$ (1-0) lines \citep{2013AA..557..94S}. 

\begin{figure}
\plotone{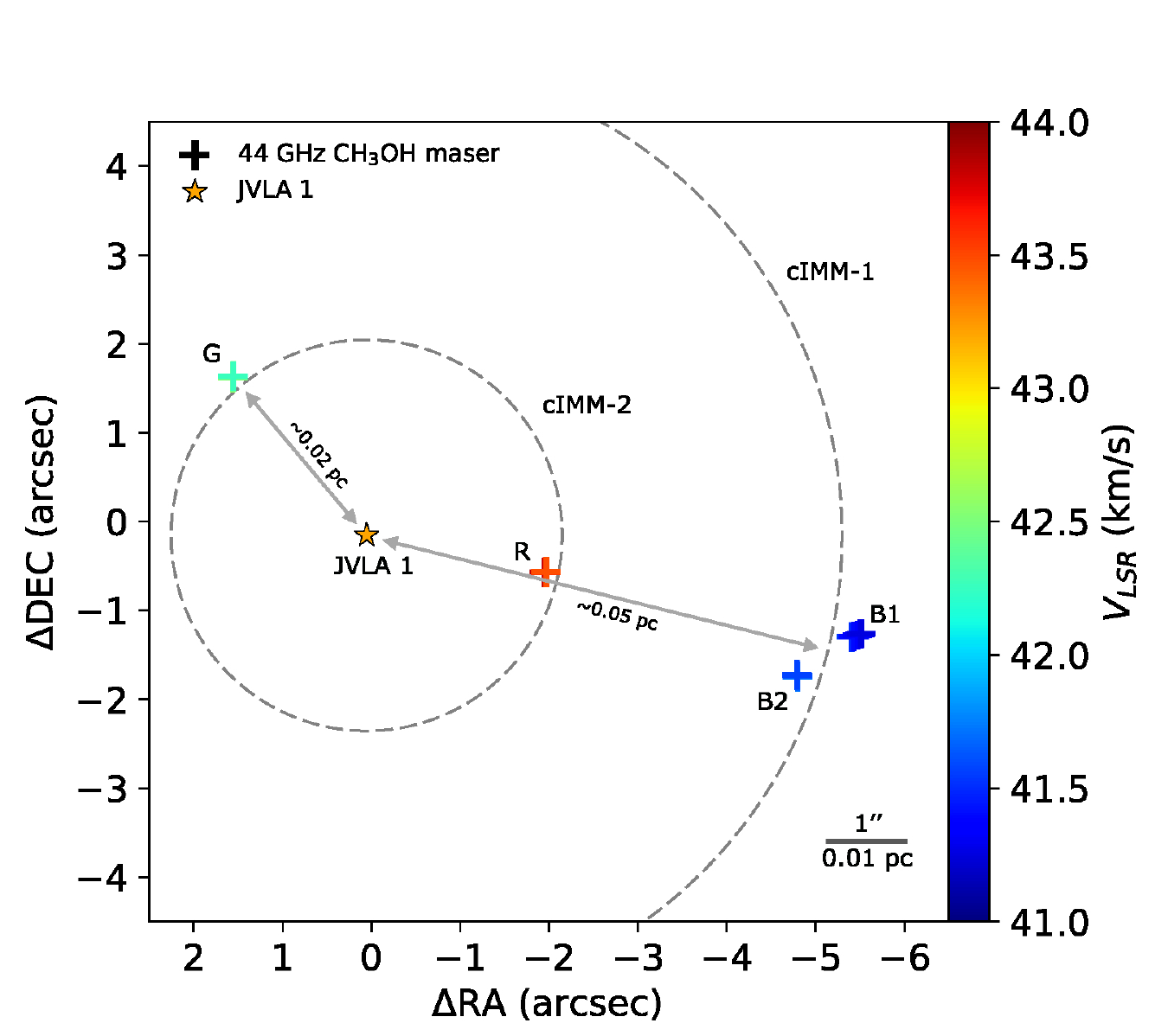}
\caption{Possible model: episodic ejection traced by cIMMs. Two episodes of possible ejection:
1) cIMM-1: traced by B1 and B2 features, radius is $\sim$0.05 pc; 2) cIMM-2: traced by G and R features, radius is $\sim$0.02 pc.
Plot is color-coded by radial velocity (see colorbar for color scale). Positional offsets are relative to the JVLA 1 continuum source. The physical scale label (in pc) assumes the distance to the source of 2.08 kpc (the BeSSeL Survey Bayesian Distance Calculator). \label{fig:44ej}}
\end{figure}

\begin{figure}
\plotone{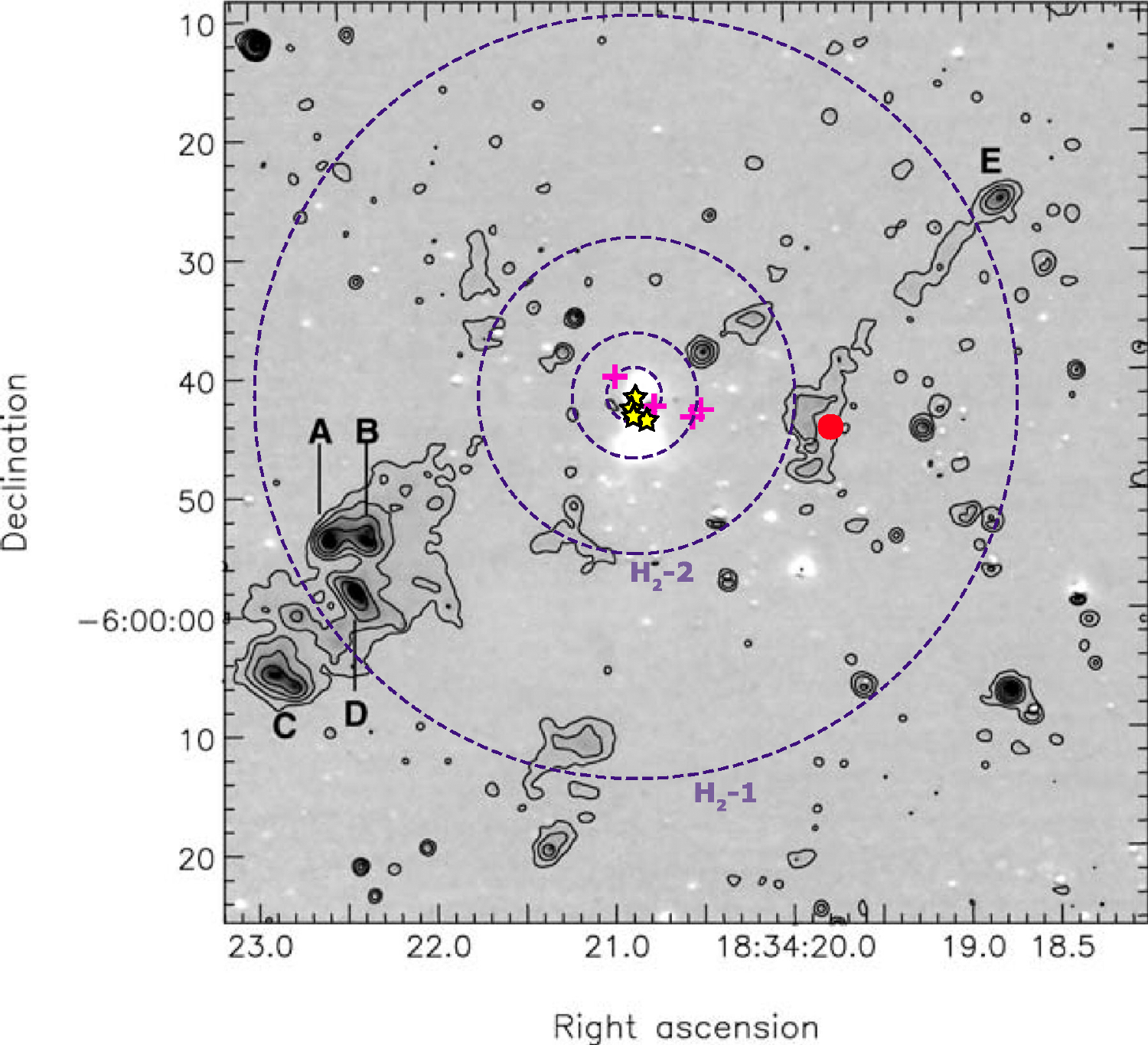}
\caption{Possible model: episodic ejection traced by H$_2$ emission. Possibly two more episodes of ejection are traced by H$_2$ emission: 1) H$_2$-1: traced by H$_2$ knots A, B, C, D, E -- radius is $\sim$0.3 pc; 2) H$_2$-2: traced by weak H$_2$ knots -- radius is $\sim$0.13 pc. \label{fig:H2ej}}
\end{figure}

Various scenarios can be proposed to explain the locations of the 44
GHz cIMM components.  Given their roughly linear orientation, they
might trace a compact bipolar outflow roughly aligned with the
large-scale outflow revealed by molecular line observations described
in Section 1.   Another plausible explanation is episodic ejections
from JVLA 1.  As shown in Figure \ref{fig:44ej}, the separation between JVLA 1
and the blue masers is about 2$\times$ larger than that between 
JVLA 1 and the red and green masers.  In this scenario, the former
(blue) masers would be older, with a 5.2$^{\prime\prime}$ (0.05~pc) separation 
from JVLA 1, while the latter (red and green) masers would be younger,
with a 0.2$^{\prime\prime}$ (0.02~pc) separation from JVLA 1. 
Two, even earlier ejection events, may be traced by H$_2$ emission
(see Figure \ref{fig:H2ej}).  The older ejection event, traced by H$_2$ knots
A -- E, with a radius of $\sim$ 0.3 pc, while a younger event would
be traced by the weaker H$_2$ emission with a radius of $\sim$0.13 pc.

\subsection{Possible Model of the Bursting Source} 

Maser bursts have recently been a topic of great interest for studies of accretion and ejection in the context of star formation. In particular, the recognition that massive stars may undergo episodic accretion bursts similar to low mass stars (for example the FUOrs and ExOrs, \cite{2014PPVI...387}), has recently led to several important new insights on the topic \citep{2017ApJ...837L..29H, 2016Nature..13..276C}. Furthermore, the realization that such bursts are generally followed by ejection events has lead astronomers to look for the ``smoking gun'' of episodic accretion in evidence of episodic ejections \citep{2016MNRAS.460..283B, 2017MNRAS.467.2367B}. However, direct evidence of such accretion and ejection events is scarce, partly due to the rarity of such events and also due to the difficulty in obtaining the required continuum mapping monitoring programs with highly-subscribed instruments. Masers, such as the 6.7 GHz methanol and 22 GHz water masers, trace regions near the sites of accretion and ejection, respectively \citep{1996IAUS..178..163M}, and thus provide a viable method to search for new accretion or ejection activity in large sample of massive stars through maser monitoring observations.

In fact, examples of such rapid changes have already been reported in the literature. Following a 6.7 GHz maser flare in S255IR \citep{2015ATel.8286....1F, 2016Nature..13..276C} a $\sim$3 mag brightening in the $H$ and ${Ks}$ infrared bands was detected, which was attributed to a disk-mediated accretion burst onto a massive young stellar object. 
Similarly, the increased luminosity detected in NGC 6334 I and accompanied by strong maser flares was attributed to a sudden accretion event. Observations by \cite{2017ApJ...837L..29H} show that the free-free emission (at 1.5 cm) from NGC6334I has fallen by a factor of 4; such a decrease would arise quite naturally if the accretion event reduced the ionizing photon flux from the young stellar object.  At the high densities of these massive star formation regions, the recombination timescale for the ionized gas can be of the order of a few weeks, hence the free-free continuum emission can track the same accretion events, but showing a decrease while the IR luminosity increases.
Both these star forming regions showed significant changes in their maser emission during the aforementioned events. A byproduct of these findings was increased interest in the burst behavior of maser emission, interpreted in the context of star formation as described above. While single-dish monitoring results can imply such processes, imaging is required to associate spectral temporal changes to physical processes seen in star forming regions. 

The current literature proposes three main routes for producing maser bursts; the increase of continuum emission entering the maser (S255, NGC6334, see above), changes in the pumping conditions \citep{2004PASJ...56L..15H}, and the overlapping of maser clouds along the line of sight to the observer, as was reported by \cite{2005ApJ..634..459S, 2014PASJ...66..106H} in Orion KL.

G25.65+1.05 has recently enjoyed intensive observational studies because of the repeated, intense 22 GHz water maser flares, similar in intensity to those of Orion KL and W49N.  Our VLA observations were made about two months after the water maser outburst that occurred in September 2017 \citep{2017ATel10728....1V, 2019MNRAS.482L..90V}. The burst lasted for several days and reached a peak flux density of about 65~kJy, according to single-dish monitoring.  At the time of the VLA observations, the correlated H$_2$O maser flux density remained at about 10 kJy.  

Our JVLA images reveal for the first time the location of the bursting maser in G25.65+1.05, with the additional context of the distributions of other maser species and investigation of the nature of the continuum sources (see Section \ref{Nature of the continuum sources}). The bursting maser is found to reside at the periphery of a continuum source JVLA 1, which may be either a young protostellar object with a jet, or a region of shocked gas possibly ejected from the JVLA 2 MYSO. As such, the maser is probably associated with protostellar ejections rather than accretion events.

Our interpretation of the water masers associated with the burst feature in JVLA 1 is as follows. Two maser sheets propagating in a shock are delineated by the large V-shape formed by the G1-G2-G3 groups. At their intersect in G2, the fine scale distribution traces the intersect down to a smaller, similarly orientated v-shape that comprises G2 cluster III (Figure \ref{fig:G2} (c)). Consequently, the bursting maser is observed at the exact point where the two maser sheets intersect in the sky-plane of the observer, i.e. where the path length of maser emission significantly increased along the line of sight. 

Providing that the two sheets have a non-zero relative proper motion it can be expected that the intersect point will shift in location along the lengths of the maser sheets. This could explain the repeating nature of the bursts in G25.65+1.05. Measurements of the relative proper motions via VLBI observations would be required to test this scenario and could lead to the possibility to detect future burst events.

Considering the location and morphology of the water masers in G25.65+1.05, we consider the most likely burst scenario to be an increase of maser path length along the line of sight to the observer; i.e., a repeat of the Orion KL scenario \citep{2005ApJ..634..459S}.
The fast appearance and fast decay of the burst in G25.65+1.05 \citep{2019MNRAS.482L..90V} supports this view, since the other maser burst scenarios would require the presence of physical changes on intra-day timescales, to match and explain the same timescales of maser variability.

Further study of the burst process is being conducted by our group to clarify the case for G25.65+1.05, and to extend the sample to other bursting masers in current and future campaigns. In this effort we propose define a catalog of Bursting Maser Objects (further abbreviated as ``BMO'') based on the format used for other transient astronomical events, for example novas. For the burst event in G25.65+1.05 located in the Serpens constellation we suggest the name ``BMO-Serp''.

\section{Conclusions}

This work may be summarized as follows:

\begin{enumerate}
	\item Spectral line and continuum observations of G25.65+1.05 were conducted with JVLA B-configuration in C, Ku, K, and Q frequency bands during the post- H$_2$O maser burst epoch of November-December 2017. 
	\item Continuum emission above a 5$\sigma$ detection level is found in all four frequency bands -- positions of four continuum sources (JVLA 1, 2, 3, and 4) were determined.
	\item Spectral line emission is detected in all frequency bands except Ku-band -- no 12~GHz class II CH$_3$OH maser emission was found.
	\item Milliarcsecond accuracy J2000 coordinates, radial velocities, and flux densities were obtained for 22 GHz H$_2$O and 6.7 and 44 GHz CH$_3$OH masers. Spot maps and spectra for each maser are presented. 
	\item A map of 22 GHz H$_2$O maser spots was obtained for the first time ever, and the absolute position of the 22 GHz H$_2$O bursting feature was determined. 
	\item Three groups of 6.7 GHz class II CH$_3$OH masers associated with continuum source JVLA 2 are detected. Comparison of pre- and post-burst maps of 6.7 GHz CH$_3$OH maser emission didn't reveal significant changes in maser spot distribution. 
	\item A map of 44 GHz class I CH$_3$OH maser spots was obtained for the first time.  We detected  four 44 GHz maser features oriented in a linear structure that is roughly aligned with a large-scale bipolar outflow operating in the region.
	\item Based on the presence of 6.7 GHz masers and the ALMA detection at 350 GHz,
we propose that JVLA 2 is a massive, accreting protostar.  JVLA 1 may be a
marginally older object (because of the absence of cIIMM) but both objects
appear to be relatively young, and likely host shocked gas, which is 
responsible for their cm-wave emission.  A definitive interpretation of
JVLA 3 and 4 must await multi-frequency observations at higher resolution
and sensitivity.
	\item We report the location and morphology of the 22 GHz H$_2$O
maser emission, and note the absence of change in other maser
species in the field.  This indicates that the H$_2$O maser burst most
likely was caused by the increase of maser path length along the
line of sight to the observer due to intersection of two maser sheets.
\end{enumerate}

\acknowledgments
We are grateful to the National Radio Astronomy
Observatory of the USA for the opportunity to observe
with the JVLA and to the NRAO New Mexico staff for their assistance
in carrying out the observations.

We thank the Maser Monitoring Organization (M2O), a loosely organized collaboration and network of maser monitoring telescopes, for the unobstructed sharing of information on the source observations made and for fruitful discussion. Special thanks to A. E. Volvach (RT-22 Simeiz), M. Olech (RT-32 Toru\'n), and G. C. MacLeod (RT-26 HartRAO) for providing single-dish data on the date of our JVLA observations.

This work
was partially supported by Program 28 of the Russian
Academy of Sciences, ``The Cosmos: Fundamental
Processes and Their Interconnections''. RB acknowledges support through the EACOA Fellowship from the East Asian Core Observatories Association.




\newpage
\startlongtable
\begin{deluxetable}{cccccccccc}
\tablewidth{0pt}
\tablecaption{Observation parameters \label{tab:obs}}
\tablehead{
\colhead{Band} & \colhead{Obs. Date} & \colhead{Integ. Time} &
\multicolumn{3}{c}{Continuum} &
\multicolumn{4}{c}{Spectral Line} \\ \cmidrule(lr){4-6} \cmidrule(lr){7-10}
\colhead{} & \colhead{} & \colhead{} &
\colhead{Synth. Beam} & \colhead{PA} & \colhead{$\sigma$} &
\colhead{Vel. res.} & \colhead{Synth. Beam} & \colhead{PA} & \colhead{$\sigma$} \\
\colhead{} & \colhead{} & \colhead{(min)} &
\colhead{(arcsec)} & \colhead{($^\circ$)} & \colhead{($\mu$Jy/beam)} &
\colhead{(km~s$^{-1}$)} & \colhead{(arcsec)} & \colhead{($^\circ$)} & \colhead{(mJy/beam)} 
}
\startdata
C  (6.0 GHz)  & 2017-Dec-09 & 10 & 1.82$\times$1.15 & 23.01 & 8  & 0.04 & 1.24$\times$0.86 & 19.78 & 35 \\
Ku (15 GHz)   & 2017-Dec-09 & 12 & 0.70$\times$0.47 & 20.51 & 10 & 0.05 & 0.62$\times$0.44 & 19.16 & 53 \\
K  (22 GHz)   & 2017-Dec-09 & 12 & 0.48$\times$0.33 & 15.43 & 15 & 0.11 & 0.35$\times$0.25 & 11.25 & 60 \\
Q  (45 GHz)   & 2017-Nov-02 & 10 & 0.33$\times$0.19 & 44.27 & 41 & 0.05 & 0.24$\times$0.14 & 33.08 & 36 \\
\enddata
\end{deluxetable}

\startlongtable
\begin{deluxetable*}{ccccc}
\tablewidth{400pt}
\tablecaption{List of detected continuum sources \label{tab:cont}}
%\tablewidth{0pt}
\tablehead{
\colhead{Freq.} &   
\colhead{RA(J2000)} & \colhead{DEC(J2000)} & \colhead{Integrated} & \colhead{Peak}  \\
\colhead{} &  
\colhead{} & \colhead{} & 
\colhead{flux} & \colhead{flux}  \\
\colhead{(GHz)} &  
\colhead{($^h$~$^m$~$^s$)} & \colhead{($^\circ$~$\arcmin$~$\arcsec$)} & 
\colhead{(mJy)} & \colhead{(mJy/beam)} 
}
%\colnumbers
\startdata
6\tablenotemark{a}  &  18:34:20.9000$\pm$0.0003\tablenotemark{b} & -05:59:42.033$\pm$0.009 & 3.94$\pm$0.05 & 2.66$\pm$0.02  \\ \hline
\multicolumn{5}{c}{JVLA 1} \\ \hline
15  &   18:34:20.9002$\pm$0.0001 & -05:59:41.698$\pm$0.001 & 2.53$\pm$0.01 & 1.97$\pm$0.01  \\
22  &   18:34:20.9003$\pm$0.0001 & -05:59:41.682$\pm$0.003 & 2.52$\pm$0.04 & 1.65$\pm$0.02  \\
45  &   18:34:20.8903$\pm$0.0014 & -05.59.41.739$\pm$0.032 & 1.59$\pm$0.26 & 0.47$\pm$0.06  \\ \hline
\multicolumn{5}{c}{JVLA 2} \\ \hline
15  &   18:34:20.9133$\pm$0.0001 & -05:59:42.264$\pm$0.002 & 0.79$\pm$0.01 & 0.76$\pm$0.01  \\
22  &   18:34:20.9140$\pm$0.0001 & -05:59:42.266$\pm$0.004 & 1.12$\pm$0.03 & 0.95$\pm$0.02  \\
45  &   18:34:20.9078$\pm$0.0008 & -05:59:42.375$\pm$0.012 & 0.60$\pm$0.09 & 0.70$\pm$0.05  \\ \hline
\multicolumn{5}{c}{JVLA 3} \\ \hline
15  &   18:34:20.9122$\pm$0.0001 & -05:59:42.876$\pm$0.003 & 0.66$\pm$0.01 & 0.54$\pm$0.01  \\
22  &   18:34:20.9113$\pm$0.0004 & -05:59:42.923$\pm$0.007 & 0.71$\pm$0.04 & 0.49$\pm$0.02  \\
45  &   18:34:20.8932$\pm$0.0031 & -05:59:42.892$\pm$0.076 & 0.71$\pm$0.26 & 0.21$\pm$0.06  \\ \hline
\multicolumn{5}{c}{JVLA 4} \\ \hline
15  &   18:34:20.8244$\pm$0.0002 & -05:59:43.175$\pm$0.004 & 0.43$\pm$0.01 & 0.37$\pm$0.01  \\
22  &   18:34:20.8258$\pm$0.0006 & -05:59:43.151$\pm$0.013 & 0.43$\pm$0.03 & 0.31$\pm$0.02  \\ 
\enddata
\tablenotetext{a}{Sources are unresolved at this frequency.} 
\tablenotetext{b}{~Statistical errors of the fit are listed. A systematic error in the position of the sources could be of order 0.15$\arcsec$, due to position uncertainty of the phase reference source J1832-1035.} 
\end{deluxetable*}

\newpage
\startlongtable
\begin{deluxetable}{lccccccc}
\tablewidth{0pt}
\tablecaption{22 GHz H$_2$O maser parameters \label{tab:T22GHZ}}
\tablehead{
\colhead{RA(J2000)} & \colhead{DEC(J2000)} &
\colhead{Integrated} & \colhead{Peak} &
\colhead{V$_{LSR}$} & \colhead{Group\tablenotemark{a}} & \colhead{Cluster\tablenotemark{b}} \\
\colhead{} & \colhead{} &
\colhead{flux} & \colhead{flux} &
\colhead{} & \colhead{} & \colhead{} \\
\colhead{($^h$~$^m$~$^s$)} & \colhead{($^\circ$~$\arcmin$~$\arcsec$)} &
\colhead{(Jy)} & \colhead{(Jy/beam)} &
\colhead{(km~s$^{-1}$)} & \colhead{} & \colhead{}
}
\startdata
18:34:20.9187$\pm$0.0005 & -05:59:41.682$\pm$0.013 & 210$\pm$30 & 190$\pm$10 & 41.34 & G1 & \\
18:34:20.9173$\pm$0.0007 & -05:59:41.668$\pm$0.013 & 440$\pm$50 & 310$\pm$20 & 41.45 & & \\
18:34:20.9188$\pm$0.0005 & -05:59:41.682$\pm$0.011 & 490$\pm$50 & 440$\pm$30 & 41.55 & & \\
18:34:20.9158$\pm$0.0011 & -05:59:41.660$\pm$0.017 & 780$\pm$110& 460$\pm$40 & 41.66 & & \\
\hline
18:34:20.9018$\pm$0.0001 & -05:59:41.557$\pm$0.002 & 230$\pm$10   & 220.47$\pm$1.97 & 51.03 & G2 & I \\
18:34:20.9018$\pm$0.0001 & -05:59:41.557$\pm$0.002 & 280$\pm$10   & 260.73$\pm$2.23 & 51.14 & & \\
18:34:20.9018$\pm$0.0001 & -05:59:41.557$\pm$0.001 & 320$\pm$10   & 299.10$\pm$2.47 & 51.24 & & \\
18:34:20.9018$\pm$0.0001 & -05:59:41.557$\pm$0.001 & 340$\pm$10   & 323.80$\pm$2.63 & 51.35 & & \\
18:34:20.9018$\pm$0.0001 & -05:59:41.557$\pm$0.001 & 350$\pm$10   & 327.50$\pm$2.67 & 51.45 & & \\
18:34:20.9018$\pm$0.0001 & -05:59:41.557$\pm$0.001 & 330$\pm$10   & 314.57$\pm$2.57 & 51.56 & & \\
18:34:20.9018$\pm$0.0001 & -05:59:41.557$\pm$0.001 & 320$\pm$10   & 301.37$\pm$2.50 & 51.66 & & \\
18:34:20.9018$\pm$0.0001 & -05:59:41.557$\pm$0.001 & 330$\pm$10   & 306.87$\pm$2.53 & 51.77 & & \\
18:34:20.9018$\pm$0.0001 & -05:59:41.557$\pm$0.001 & 370$\pm$10   & 350$\pm$10 & 51.88 & & \\
18:34:20.9018$\pm$0.0001 & -05:59:41.557$\pm$0.001 & 460$\pm$10   & 430$\pm$10 & 51.98 & & \\
18:34:20.9018$\pm$0.0001 & -05:59:41.557$\pm$0.001 & 580$\pm$10   & 540$\pm$10 & 52.09 & & \\
18:34:20.9019$\pm$0.0001 & -05:59:41.557$\pm$0.001 & 690$\pm$10   & 650$\pm$10 & 52.19 & & \\
18:34:20.9018$\pm$0.0001 & -05:59:41.557$\pm$0.001 & 760$\pm$10   & 710$\pm$10 & 52.30 & & \\
18:34:20.9019$\pm$0.0001 & -05:59:41.557$\pm$0.001 & 730$\pm$10   & 690$\pm$10 & 52.40 & & \\
18:34:20.9019$\pm$0.0001 & -05:59:41.557$\pm$0.001 & 630$\pm$10   & 590$\pm$10 & 52.51 & & \\
18:34:20.9019$\pm$0.0001 & -05:59:41.557$\pm$0.001 & 470$\pm$10   & 440$\pm$10 & 52.61 & & \\
18:34:20.9019$\pm$0.0001 & -05:59:41.558$\pm$0.002 & 310$\pm$10   & 293.07$\pm$2.43 & 52.72 & & \\
18:34:20.9020$\pm$0.0001 & -05:59:41.558$\pm$0.002 & 188.50$\pm$3.03   & 176.90$\pm$1.67 & 52.82 & & \\
18:34:20.9021$\pm$0.0001 & -05:59:41.559$\pm$0.002 & 108.53$\pm$1.90   & 101.77$\pm$1.03 & 52.93 & & \\
18:34:20.9023$\pm$0.0001 & -05:59:41.560$\pm$0.002 & 63.37$\pm$1.17   & 59.57$\pm$0.63 & 53.03 & & \\
18:34:20.9024$\pm$0.0001 & -05:59:41.561$\pm$0.002 & 39.67$\pm$0.80   & 37.47$\pm$0.43 & 53.14 & & \\
18:34:20.9025$\pm$0.0001 & -05:59:41.563$\pm$0.002 & 28.60$\pm$0.67   & 27.23$\pm$0.37 & 53.24 & & \\
18:34:20.9026$\pm$0.0001 & -05:59:41.563$\pm$0.002 & 24.67$\pm$0.60   & 23.57$\pm$0.33 & 53.35 & & \\
18:34:20.9027$\pm$0.0001 & -05:59:41.564$\pm$0.002 & 25.60$\pm$0.60   & 24.53$\pm$0.33 & 53.46 & & \\
18:34:20.9029$\pm$0.0001 & -05:59:41.565$\pm$0.002 & 30.27$\pm$0.67   & 28.80$\pm$0.37 & 53.56 & & \\
18:34:20.9030$\pm$0.0001 & -05:59:41.566$\pm$0.002 & 38.13$\pm$0.77   & 36.20$\pm$0.43 & 53.67 & & \\
18:34:20.9032$\pm$0.0001 & -05:59:41.566$\pm$0.002 & 49.37$\pm$0.93   & 46.73$\pm$0.50 & 53.77 & & \\
18:34:20.9033$\pm$0.0001 & -05:59:41.567$\pm$0.002 & 63.50$\pm$1.13   & 59.90$\pm$0.63 & 53.88 & & \\
18:34:20.9034$\pm$0.0001 & -05:59:41.568$\pm$0.002 & 78.87$\pm$1.37   & 74.37$\pm$0.73 & 53.98 & & \\
18:34:20.9035$\pm$0.0001 & -05:59:41.568$\pm$0.002 & 92.17$\pm$1.57   & 86.93$\pm$0.87 & 54.09 & & \\
18:34:20.9035$\pm$0.0001 & -05:59:41.568$\pm$0.002 & 99.80$\pm$1.70   & 94.07$\pm$0.93 & 54.19 & & \\
18:34:20.9035$\pm$0.0001 & -05:59:41.568$\pm$0.002 & 98.47$\pm$1.67   & 92.70$\pm$0.90 & 54.30 & & \\
18:34:20.9035$\pm$0.0001 & -05:59:41.568$\pm$0.002 & 86.47$\pm$1.47   & 81.67$\pm$0.80 & 54.40 & & \\
18:34:20.9035$\pm$0.0001 & -05:59:41.568$\pm$0.002 & 67.00$\pm$1.17   & 63.27$\pm$0.63 & 54.51 & & \\
18:34:20.9035$\pm$0.0001 & -05:59:41.568$\pm$0.002 & 44.97$\pm$0.83   & 42.63$\pm$0.47 & 54.61 & & \\
18:34:20.9034$\pm$0.0001 & -05:59:41.568$\pm$0.002 & 25.97$\pm$0.60   & 24.90$\pm$0.33 & 54.72 & & \\
18:34:20.9034$\pm$0.0001 & -05:59:41.568$\pm$0.004 & 13.07$\pm$0.50   & 12.84$\pm$0.27 & 54.82 & & \\
18:34:20.9032$\pm$0.0002 & -05:59:41.567$\pm$0.006 & 5.93$\pm$0.40   & 6.06$\pm$0.23 & 54.93 & & \\
18:34:20.9014$\pm$0.0002 & -05:59:41.560$\pm$0.005 & 7.73$\pm$0.43   & 7.42$\pm$0.24 & 49.03 & & II \\
18:34:20.9012$\pm$0.0003 & -05:59:41.563$\pm$0.006 & 5.77$\pm$0.40   & 5.52$\pm$0.21 & 49.14 & & \\
18:34:20.9010$\pm$0.0003 & -05:59:41.566$\pm$0.006 & 4.92$\pm$0.33   & 4.51$\pm$0.18 & 49.24 & & \\
18:34:20.9010$\pm$0.0003 & -05:59:41.568$\pm$0.006 & 4.25$\pm$0.28   & 3.85$\pm$0.15 & 49.35 & & \\
18:34:20.9010$\pm$0.0003 & -05:59:41.572$\pm$0.006 & 4.37$\pm$0.29   & 3.75$\pm$0.15 & 49.45 & & \\
18:34:20.9013$\pm$0.0003 & -05:59:41.572$\pm$0.007 & 4.59$\pm$0.31   & 3.90$\pm$0.16 & 49.56 & & \\
18:34:20.9015$\pm$0.0003 & -05:59:41.573$\pm$0.007 & 5.47$\pm$0.37   & 4.73$\pm$0.19 & 49.66 & & \\
18:34:20.9017$\pm$0.0003 & -05:59:41.571$\pm$0.007 & 6.47$\pm$0.43   & 5.80$\pm$0.23 & 49.77 & & \\
18:34:20.9018$\pm$0.0002 & -05:59:41.569$\pm$0.005 & 8.37$\pm$0.47   & 7.76$\pm$0.26 & 49.87 & & \\
18:34:20.9019$\pm$0.0002 & -05:59:41.567$\pm$0.004 & 11.13$\pm$0.50   & 10.47$\pm$0.27 & 49.98 & & \\
18:34:20.9019$\pm$0.0001 & -05:59:41.564$\pm$0.003 & 15.50$\pm$0.50   & 14.64$\pm$0.28 & 50.08 & & \\
18:34:20.9019$\pm$0.0001 & -05:59:41.562$\pm$0.003 & 22.17$\pm$0.57   & 21.00$\pm$0.31 & 50.19 & & \\
18:34:20.9019$\pm$0.0001 & -05:59:41.561$\pm$0.002 & 32.70$\pm$0.70   & 30.97$\pm$0.37 & 50.30 & & \\
18:34:20.9019$\pm$0.0001 & -05:59:41.559$\pm$0.002 & 48.87$\pm$0.93   & 46.13$\pm$0.50 & 50.40 & & \\
18:34:20.9019$\pm$0.0001 & -05:59:41.558$\pm$0.002 & 71.53$\pm$1.27   & 67.37$\pm$0.70 & 50.51 & & \\
18:34:20.9019$\pm$0.0001 & -05:59:41.558$\pm$0.002 & 99.70$\pm$1.73   & 93.83$\pm$0.93 & 50.61 & & \\
18:34:20.9018$\pm$0.0001 & -05:59:41.557$\pm$0.002 & 130.90$\pm$2.23   & 123.13$\pm$1.20 & 50.72 & & \\
18:34:20.9018$\pm$0.0001 & -05:59:41.557$\pm$0.002 & 163.23$\pm$2.70   & 153.53$\pm$1.47 & 50.82 & & \\
18:34:20.9018$\pm$0.0001 & -05:59:41.557$\pm$0.002 & 196.80$\pm$3.13   & 184.93$\pm$1.70 & 50.93 & & \\
18:34:20.8977$\pm$0.0001 & -05:59:41.562$\pm$0.002 & 296.67$\pm$0.01   & 262.47$\pm$2.50 & 41.13 & & III \\
18:34:20.8980$\pm$0.0001 & -05:59:41.563$\pm$0.002  & 460.00$\pm$6.67   & 386.67$\pm$3.33 & 41.24 & & \\
18:34:20.8985$\pm$0.0001 & -05:59:41.566$\pm$0.002  & 716.67$\pm$13.33  & 536.67$\pm$6.67 & 41.34 & & \\
18:34:20.8994$\pm$0.0001 & -05:59:41.571$\pm$0.002  & 1100.00$\pm$23.33  & 706.67$\pm$10.00 & 41.45 & & \\
18:34:20.8994$\pm$0.0001 & -05:59:41.571$\pm$0.002  & 1500.00$\pm$30.00  & 966.67$\pm$13.33 & 41.55 & & \\
18:34:20.8986$\pm$0.0001 & -05:59:41.566$\pm$0.002  & 1840.00$\pm$33.33  & 1370.00$\pm$16.67 & 41.66 & & \\
18:34:20.8982$\pm$0.0001 & -05:59:41.563$\pm$0.002  & 2280.00$\pm$33.33  & 1866.67$\pm$16.67 & 41.76 & & \\
18:34:20.8981$\pm$0.0001 & -05:59:41.562$\pm$0.001  & 2883.33$\pm$40.00  & 2443.33$\pm$20.00 & 41.87 & & \\
18:34:20.8981$\pm$0.0001 & -05:59:41.562$\pm$0.001  & 3636.67$\pm$46.67  & 3073.33$\pm$23.33 & 41.97 & & \\
18:34:20.8981$\pm$0.0001 & -05:59:41.563$\pm$0.001  & 4443.33$\pm$60.00  & 3743.33$\pm$30.00 & 42.08 & & \\
18:34:20.8980$\pm$0.0001 & -05:59:41.563$\pm$0.001  & 5200.00$\pm$66.67  & 4436.67$\pm$33.33 & 42.19 & & \\
18:34:20.8979$\pm$0.0001 & -05:59:41.562$\pm$0.001  & 5886.67$\pm$73.33  & 5173.33$\pm$36.67 & 42.29 & & \\
18:34:20.8977$\pm$0.0001 & -05:59:41.561$\pm$0.001  & 6600.00$\pm$80.00  & 5976.67$\pm$43.33 & 42.40 & & \\
18:34:20.8976$\pm$0.0001 & -05:59:41.561$\pm$0.001  & 7420.00$\pm$86.67  & 6853.33$\pm$46.67 & 42.50 & & \\
18:34:20.8976$\pm$0.00004 & -05:59:41.561$\pm$0.001 & 8390.00$\pm$96.67  & 7823.33$\pm$53.33 & 42.61 & & \\
18:34:20.8976$\pm$0.00004 & -05:59:41.561$\pm$0.001 & 9360.00$\pm$106.67 & 8756.67$\pm$56.67 & 42.71 & & \\
18:34:20.8976$\pm$0.00004 & -05:59:41.561$\pm$0.001 & 10143.33$\pm$116.67 & 9490.00$\pm$63.33 & 42.82 & & \\
18:34:20.8976$\pm$0.00004 & -05:59:41.561$\pm$0.001 & 10520.00$\pm$120.00 & 9843.33$\pm$63.33 & 42.92 & & \\
18:34:20.8976$\pm$0.00004 & -05:59:41.561$\pm$0.001 & 10416.67$\pm$120.00 & 9743.33$\pm$63.33 & 43.03 & & \\
18:34:20.8976$\pm$0.00004 & -05:59:41.561$\pm$0.001 & 9880.00$\pm$113.33 & 9243.33$\pm$60.00 & 43.13 & & \\
18:34:20.8976$\pm$0.00004 & -05:59:41.561$\pm$0.001 & 9033.33$\pm$103.33 & 8453.33$\pm$56.67 & 43.24 & & \\
18:34:20.8976$\pm$0.00004 & -05:59:41.561$\pm$0.001 & 8060.00$\pm$93.33  & 7536.67$\pm$50.00 & 43.34 & & \\ 
18:34:20.8976$\pm$0.00004 & -05:59:41.561$\pm$0.001 & 7036.67$\pm$80.00  & 6576.67$\pm$43.33 & 43.45 & & \\
18:34:20.8977$\pm$0.00004 & -05:59:41.561$\pm$0.001 & 5993.33$\pm$70.00  & 5600.00$\pm$36.67 & 43.55 & & \\
18:34:20.8977$\pm$0.00004 & -05:59:41.561$\pm$0.001 & 4930.00$\pm$56.67  & 4603.33$\pm$30.00 & 43.66 & & \\
18:34:20.8977$\pm$0.00004 & -05:59:41.561$\pm$0.001 & 3836.67$\pm$43.33  & 3580.00$\pm$23.33 & 43.76 & & \\
18:34:20.8977$\pm$0.0001 & -05:59:41.561$\pm$0.001  & 2763.33$\pm$33.33  & 2576.67$\pm$16.67 & 43.87 & & \\
18:34:20.8978$\pm$0.0001 & -05:59:41.561$\pm$0.001  & 1826.67$\pm$23.33  & 1700.00$\pm$13.33 & 43.98 & & \\
18:34:20.8978$\pm$0.0001 & -05:59:41.560$\pm$0.001  & 1110.00$\pm$13.33  & 1033.33$\pm$6.67 & 44.08 & & \\
18:34:20.8979$\pm$0.0001 & -05:59:41.560$\pm$0.001  & 633.33$\pm$6.67   & 586.67$\pm$3.33 & 44.19 & & \\
18:34:20.8980$\pm$0.0001 & -05:59:41.559$\pm$0.001  & 346.67$\pm$0.01   & 322.27$\pm$2.67 & 44.29 & & \\
18:34:20.8982$\pm$0.0001 & -05:59:41.558$\pm$0.002  & 188.17$\pm$3.07   & 174.63$\pm$1.67 & 44.40 & & \\
18:34:20.8983$\pm$0.0001 & -05:59:41.558$\pm$0.002  & 104.77$\pm$1.83   & 96.70$\pm$1.00 & 44.50 & & \\
18:34:20.8985$\pm$0.0001 & -05:59:41.558$\pm$0.002  & 62.50$\pm$1.17   & 57.33$\pm$0.63 & 44.61 & & \\
18:34:20.8989$\pm$0.0001 & -05:59:41.559$\pm$0.002  & 42.57$\pm$0.83   & 38.43$\pm$0.43 & 44.71 & & \\
18:34:20.8995$\pm$0.0001 & -05:59:41.561$\pm$0.002  & 34.63$\pm$0.73   & 30.93$\pm$0.40 & 44.82 & & \\
18:34:20.9001$\pm$0.0001 & -05:59:41.560$\pm$0.002  & 34.20$\pm$0.73   & 30.30$\pm$0.37 & 44.92 & & \\
18:34:20.9004$\pm$0.0001 & -05:59:41.560$\pm$0.002  & 38.27$\pm$0.80   & 33.80$\pm$0.40 & 45.03 & & \\
18:34:20.9005$\pm$0.0001 & -05:59:41.560$\pm$0.002  & 45.47$\pm$0.90   & 39.63$\pm$0.47 & 45.13 & & \\
18:34:20.9004$\pm$0.0001 & -05:59:41.560$\pm$0.002  & 54.57$\pm$1.07   & 46.67$\pm$0.53 & 45.24 & & \\
18:34:20.9001$\pm$0.0001 & -05:59:41.560$\pm$0.002  & 63.67$\pm$1.20   & 52.97$\pm$0.60 & 45.34 & & \\
18:34:20.8998$\pm$0.0001 & -05:59:41.561$\pm$0.002  & 71.13$\pm$1.33   & 57.93$\pm$0.67 & 45.45 & & \\
18:34:20.8997$\pm$0.0001 & -05:59:41.561$\pm$0.002  & 72.53$\pm$1.37   & 58.83$\pm$0.67 & 45.56 & & \\
18:34:20.8957$\pm$0.0003 & -05:59:41.539$\pm$0.009  & 4.73$\pm$0.43   & 4.61$\pm$0.23 & 36.29 &  & IV \\ 
18:34:20.8954$\pm$0.0004 & -05:59:41.540$\pm$0.011  & 5.83$\pm$0.63   & 5.60$\pm$0.33 & 36.39 & & \\
18:34:20.8948$\pm$0.0006 & -05:59:41.536$\pm$0.015  & 7.10$\pm$1.07   & 6.73$\pm$0.60 & 36.50 & & \\
18:34:20.8949$\pm$0.0005 & -05:59:41.543$\pm$0.012  & 4.30$\pm$0.53   & 4.20$\pm$0.31 & 37.34 & & \\
18:34:20.8951$\pm$0.0005 & -05:59:41.544$\pm$0.010  & 3.67$\pm$0.40   & 3.53$\pm$0.22 & 37.44 & & \\
18:34:20.8954$\pm$0.0004 & -05:59:41.547$\pm$0.010  & 3.37$\pm$0.37   & 3.23$\pm$0.20 & 37.55 & & \\
18:34:20.8957$\pm$0.0004 & -05:59:41.548$\pm$0.009  & 2.95$\pm$0.27   & 2.85$\pm$0.15 & 37.66 & & \\
18:34:20.8961$\pm$0.0003 & -05:59:41.551$\pm$0.009  & 2.83$\pm$0.25   & 2.69$\pm$0.14 & 37.76 & & \\
18:34:20.8960$\pm$0.0004 & -05:59:41.560$\pm$0.010  & 2.81$\pm$0.26   & 2.58$\pm$0.14 & 37.87 & & \\
18:34:20.8957$\pm$0.0004 & -05:59:41.570$\pm$0.011  & 3.02$\pm$0.32   & 2.68$\pm$0.17 & 37.97 & & \\
18:34:20.8954$\pm$0.0005 & -05:59:41.577$\pm$0.014  & 3.23$\pm$0.40   & 2.69$\pm$0.20 & 38.08 & & \\
18:34:20.8952$\pm$0.0006 & -05:59:41.579$\pm$0.015  & 3.67$\pm$0.47   & 3.02$\pm$0.23 & 38.18 & & \\
18:34:20.8954$\pm$0.0005 & -05:59:41.574$\pm$0.013  & 4.10$\pm$0.50   & 3.46$\pm$0.25 & 38.29 & & \\
18:34:20.8957$\pm$0.0004 & -05:59:41.564$\pm$0.011  & 4.87$\pm$0.50   & 4.28$\pm$0.27 & 38.39 & & \\
18:34:20.8961$\pm$0.0004 & -05:59:41.551$\pm$0.009  & 5.87$\pm$0.53   & 5.41$\pm$0.29 & 38.50 & & \\
18:34:20.8965$\pm$0.0003 & -05:59:41.539$\pm$0.007  & 8.07$\pm$0.60   & 7.65$\pm$0.33 & 38.60 & & \\
18:34:20.8965$\pm$0.0003 & -05:59:41.532$\pm$0.006  & 10.17$\pm$0.67   & 9.67$\pm$0.37 & 38.71 & & \\
18:34:20.8967$\pm$0.0002 & -05:59:41.527$\pm$0.005  & 12.90$\pm$0.73   & 12.23$\pm$0.40 & 38.81 & & \\
18:34:20.8969$\pm$0.0002 & -05:59:41.525$\pm$0.005  & 14.43$\pm$0.77   & 13.67$\pm$0.40 & 38.92 & & \\
18:34:20.8969$\pm$0.0002 & -05:59:41.525$\pm$0.005  & 14.80$\pm$0.77   & 14.03$\pm$0.43 & 39.02 & & \\
18:34:20.8969$\pm$0.0002 & -05:59:41.525$\pm$0.005  & 13.40$\pm$0.73   & 12.70$\pm$0.40 & 39.13 & & \\
18:34:20.8968$\pm$0.0003 & -05:59:41.526$\pm$0.006  & 11.60$\pm$0.77   & 10.97$\pm$0.43 & 39.24 & & \\
18:34:20.8966$\pm$0.0004 & -05:59:41.528$\pm$0.008  & 8.87$\pm$0.80   & 8.47$\pm$0.43 & 39.34 & & \\
18:34:20.8962$\pm$0.0005 & -05:59:41.531$\pm$0.011  & 7.00$\pm$0.87   & 6.63$\pm$0.47 & 39.45 & & \\
18:34:20.8956$\pm$0.0006 & -05:59:41.535$\pm$0.014  & 5.57$\pm$0.83   & 5.20$\pm$0.47 & 39.55 & & \\
18:34:20.8960$\pm$0.0005 & -05:59:41.555$\pm$0.011  & 4.90$\pm$0.57   & 4.75$\pm$0.32 & 39.77 & & \\
18:34:20.8966$\pm$0.0003 & -05:59:41.562$\pm$0.008  & 5.63$\pm$0.50   & 5.60$\pm$0.28 & 39.87 & & \\
18:34:20.8970$\pm$0.0003 & -05:59:41.563$\pm$0.006  & 6.73$\pm$0.47   & 6.83$\pm$0.26 & 39.97 & & \\
18:34:20.8973$\pm$0.0002 & -05:59:41.562$\pm$0.006  & 8.87$\pm$0.57   & 8.84$\pm$0.32 & 40.08 & & \\
18:34:20.8974$\pm$0.0002 & -05:59:41.561$\pm$0.005  & 11.73$\pm$0.53   & 11.51$\pm$0.31 & 40.18 & & \\
18:34:20.8974$\pm$0.0001 & -05:59:41.562$\pm$0.003  & 16.30$\pm$0.57   & 15.74$\pm$0.31 & 40.29 & & \\
18:34:20.8974$\pm$0.0001 & -05:59:41.562$\pm$0.003  & 21.13$\pm$0.63   & 20.20$\pm$0.33 & 40.39 & & \\
18:34:20.8974$\pm$0.0001 & -05:59:41.561$\pm$0.003  & 28.63$\pm$0.70   & 27.30$\pm$0.40 & 40.50 & & \\
18:34:20.8975$\pm$0.0001 & -05:59:41.561$\pm$0.002  & 39.07$\pm$0.80   & 36.90$\pm$0.43 & 40.60 & & \\
18:34:20.8975$\pm$0.0001 & -05:59:41.561$\pm$0.002  & 54.97$\pm$1.03   & 51.70$\pm$0.57 & 40.71 & & \\
18:34:20.8975$\pm$0.0001 & -05:59:41.561$\pm$0.002  & 80.20$\pm$1.47   & 74.93$\pm$0.80 & 40.82 & & \\
18:34:20.8976$\pm$0.0001 & -05:59:41.561$\pm$0.002  & 122.17$\pm$2.17   & 113.27$\pm$1.17 & 40.92 & & \\
18:34:20.8976$\pm$0.0001 & -05:59:41.562$\pm$0.002  & 189.73$\pm$3.17   & 172.90$\pm$1.70 & 41.03 & & \\
\hline
18:34:20.8996$\pm$0.0001 & -05:59:41.560$\pm$0.002 & 68.90$\pm$1.30 & 56.17$\pm$0.63 & 45.66 & G3 & \\
18:34:20.8996$\pm$0.0001 & -05:59:41.560$\pm$0.002 & 61.03$\pm$1.17 & 50.40$\pm$0.60 & 45.77 & & \\
18:34:20.8996$\pm$0.0001 & -05:59:41.560$\pm$0.002 & 52.10$\pm$1.03 & 43.97$\pm$0.53 & 45.87 & & \\
18:34:20.8996$\pm$0.0001 & -05:59:41.560$\pm$0.002 & 43.27$\pm$0.87 & 37.30$\pm$0.47 & 45.98 & & \\
18:34:20.8996$\pm$0.0001 & -05:59:41.560$\pm$0.002 & 36.30$\pm$0.73 & 31.97$\pm$0.40 & 46.08 & & \\
18:34:20.8995$\pm$0.0001 & -05:59:41.560$\pm$0.002 & 30.53$\pm$0.67 & 27.30$\pm$0.37 & 46.19 & & \\
18:34:20.8995$\pm$0.0001 & -05:59:41.561$\pm$0.002 & 25.77$\pm$0.63 & 23.27$\pm$0.33 & 46.29 & & \\
18:34:20.8995$\pm$0.0001 & -05:59:41.562$\pm$0.003 & 21.00$\pm$0.57 & 19.06$\pm$0.30 & 46.40 & & \\
18:34:20.8996$\pm$0.0001 & -05:59:41.562$\pm$0.003 & 16.83$\pm$0.53 & 15.29$\pm$0.28 & 46.50 & & \\
18:34:20.8998$\pm$0.0002 & -05:59:41.562$\pm$0.004 & 13.03$\pm$0.50 & 11.87$\pm$0.27 & 46.61 & & \\
18:34:20.8999$\pm$0.0002 & -05:59:41.565$\pm$0.004 & 10.57$\pm$0.50 & 9.47$\pm$0.26 & 46.71 & & \\
18:34:20.9000$\pm$0.0003 & -05:59:41.570$\pm$0.006 & 8.93$\pm$0.50  & 7.50$\pm$0.25 & 46.82 & & \\
18:34:20.9015$\pm$0.0004 & -05:59:41.600$\pm$0.008 & 10.00$\pm$0.63 & 6.05$\pm$0.26 & 46.92 & & \\
18:34:20.9063$\pm$0.0006 & -05:59:41.684$\pm$0.012 & 14.73$\pm$0.97 & 6.85$\pm$0.32 & 47.03 & & \\
18:34:20.9106$\pm$0.0002 & -05:59:41.755$\pm$0.005 & 24.50$\pm$0.93 & 15.20$\pm$0.37 & 47.14 & & \\
18:34:20.9118$\pm$0.0001 & -05:59:41.778$\pm$0.003 & 37.63$\pm$0.93 & 28.30$\pm$0.43 & 47.24 & & \\
18:34:20.9119$\pm$0.0001 & -05:59:41.781$\pm$0.003 & 51.23$\pm$1.10 & 39.80$\pm$0.53 & 47.35 & & \\
18:34:20.9117$\pm$0.0001 & -05:59:41.776$\pm$0.002 & 60.07$\pm$1.23 & 45.03$\pm$0.57 & 47.45 & & \\
18:34:20.9110$\pm$0.0001 & -05:59:41.760$\pm$0.003 & 62.30$\pm$1.33 & 41.20$\pm$0.57 & 47.56 & & \\
18:34:20.9091$\pm$0.0002 & -05:59:41.716$\pm$0.004 & 56.77$\pm$1.33 & 30.90$\pm$0.50 & 47.66 & & \\
18:34:20.9053$\pm$0.0002 & -05:59:41.636$\pm$0.005 & 44.97$\pm$1.23 & 24.33$\pm$0.47 & 47.77 & & \\
18:34:20.9021$\pm$0.0001 & -05:59:41.571$\pm$0.003 & 34.70$\pm$0.87 & 27.50$\pm$0.43 & 47.87 & & \\
18:34:20.9017$\pm$0.0001 & -05:59:41.560$\pm$0.002 & 34.17$\pm$0.73 & 30.93$\pm$0.37 & 47.98 & & \\
18:34:20.9016$\pm$0.0001 & -05:59:41.558$\pm$0.002 & 33.97$\pm$0.70 & 31.77$\pm$0.37 & 48.08 & & \\
18:34:20.9016$\pm$0.0001 & -05:59:41.557$\pm$0.002 & 32.80$\pm$0.67 & 30.87$\pm$0.37 & 48.19 & & \\
18:34:20.9016$\pm$0.0001 & -05:59:41.557$\pm$0.002 & 30.47$\pm$0.67 & 28.80$\pm$0.37 & 48.29 & & \\
18:34:20.9016$\pm$0.0001 & -05:59:41.557$\pm$0.002 & 27.70$\pm$0.63 & 26.23$\pm$0.33 & 48.40 & & \\
18:34:20.9016$\pm$0.0001 & -05:59:41.557$\pm$0.002 & 24.50$\pm$0.57 & 23.19$\pm$0.32 & 48.50 & & \\
18:34:20.9015$\pm$0.0001 & -05:59:41.558$\pm$0.003 & 20.93$\pm$0.53 & 19.85$\pm$0.30 & 48.61 & & \\
18:34:20.9015$\pm$0.0001 & -05:59:41.558$\pm$0.003 & 17.03$\pm$0.50 & 16.27$\pm$0.28 & 48.72 & & \\
18:34:20.9015$\pm$0.0001 & -05:59:41.559$\pm$0.003 & 13.70$\pm$0.47 & 13.08$\pm$0.26 & 48.82 & & \\
18:34:20.9014$\pm$0.0002 & -05:59:41.559$\pm$0.005 & 10.10$\pm$0.47 & 9.73$\pm$0.26 & 48.93 & & \\
\hline
18:34:20.9212$\pm$0.0006 & -05:59:42.482$\pm$0.016 & 10.20$\pm$0.83 &  5.16$\pm$0.29 & 36.29 & G4 & \\
18:34:20.9232$\pm$0.0002 & -05:59:42.540$\pm$0.006 & 23.93$\pm$1.13 &  16.53$\pm$0.50 & 36.39 & & \\
18:34:20.9236$\pm$0.0001 & -05:59:42.556$\pm$0.004 & 52.57$\pm$1.77 &  41.97$\pm$0.87 & 36.50 & & \\
18:34:20.9237$\pm$0.0001 & -05:59:42.561$\pm$0.003 & 94.83$\pm$2.70 &  79.90$\pm$1.37 & 36.60 & & \\
18:34:20.9238$\pm$0.0001 & -05:59:42.564$\pm$0.003 & 0.14$\pm$0.01 &   117.13$\pm$1.83 & 36.71 & & \\
18:34:20.9238$\pm$0.0001 & -05:59:42.565$\pm$0.003 & 0.16$\pm$0.01 &   138.63$\pm$2.10 & 36.81 & & \\
18:34:20.9238$\pm$0.0001 & -05:59:42.566$\pm$0.003 & 0.16$\pm$0.01 &   136.70$\pm$2.03 & 36.92 & & \\
18:34:20.9238$\pm$0.0001 & -05:59:42.567$\pm$0.003 & 126.87$\pm$3.17 & 112.20$\pm$1.67 & 37.02 & & \\
18:34:20.9237$\pm$0.0001 & -05:59:42.567$\pm$0.003 & 85.03$\pm$2.17 &  74.90$\pm$1.13 & 37.13 & & \\
18:34:20.9236$\pm$0.0001 & -05:59:42.567$\pm$0.003 & 46.70$\pm$1.33 &  40.37$\pm$0.67 & 37.23 & & \\
18:34:20.9232$\pm$0.0001 & -05:59:42.565$\pm$0.004 & 22.93$\pm$0.80 &  19.23$\pm$0.40 & 37.34 & & \\
18:34:20.9229$\pm$0.0002 & -05:59:42.562$\pm$0.006 & 11.93$\pm$0.60 &  9.67$\pm$0.29 & 37.44 & & \\
18:34:20.9227$\pm$0.0003 & -05:59:42.549$\pm$0.009 & 7.20$\pm$0.53 &   5.42$\pm$0.25 & 37.55 & & \\
18:34:20.9226$\pm$0.0005 & -05:59:42.522$\pm$0.012 & 5.87$\pm$0.47 &   3.65$\pm$0.19 & 37.66 & & \\
18:34:20.9215$\pm$0.0006 & -05:59:42.480$\pm$0.015 & 5.63$\pm$0.47 &   2.92$\pm$0.16 & 37.76 & & \\
18:34:20.9200$\pm$0.0007 & -05:59:42.441$\pm$0.016 & 6.63$\pm$0.50 &   3.15$\pm$0.17 & 37.87 & & \\
18:34:20.9193$\pm$0.0007 & -05:59:42.423$\pm$0.015 & 8.40$\pm$0.63 &   4.02$\pm$0.21 & 37.97 & & \\
18:34:20.9187$\pm$0.0006 & -05:59:42.413$\pm$0.014 & 10.77$\pm$0.77 &  5.11$\pm$0.25 & 38.08 & & \\
18:34:20.9185$\pm$0.0006 & -05:59:42.412$\pm$0.014 & 12.93$\pm$0.90 &  6.19$\pm$0.30 & 38.18 & & \\
18:34:20.9185$\pm$0.0006 & -05:59:42.416$\pm$0.014 & 14.67$\pm$1.00 &  6.96$\pm$0.33 & 38.29 & & \\
18:34:20.9195$\pm$0.0005 & -05:59:42.447$\pm$0.013 & 16.37$\pm$1.03 &  7.83$\pm$0.33 & 38.39 & & \\
18:34:20.9213$\pm$0.0004 & -05:59:42.500$\pm$0.010 & 18.63$\pm$1.10 &  9.93$\pm$0.40 & 38.50 & & \\
18:34:20.9224$\pm$0.0002 & -05:59:42.541$\pm$0.007 & 21.93$\pm$1.07 &  14.50$\pm$0.47 & 38.60 & & \\
18:34:20.9228$\pm$0.0002 & -05:59:42.557$\pm$0.005 & 25.97$\pm$1.03 &  19.50$\pm$0.50 & 38.71 & & \\
18:34:20.9229$\pm$0.0002 & -05:59:42.559$\pm$0.005 & 28.80$\pm$1.17 &  21.90$\pm$0.53 & 38.81 & & \\
18:34:20.9231$\pm$0.0002 & -05:59:42.551$\pm$0.007 & 29.30$\pm$1.37 &  20.90$\pm$0.60 & 38.92 & & \\
18:34:20.9234$\pm$0.0003 & -05:59:42.532$\pm$0.010 & 29.27$\pm$1.63 &  18.13$\pm$0.67 & 39.02 & & \\
18:34:20.9235$\pm$0.0003 & -05:59:42.519$\pm$0.012 & 30.43$\pm$1.80 &  17.47$\pm$0.67 & 39.13 & & \\
18:34:20.9232$\pm$0.0003 & -05:59:42.538$\pm$0.009 & 32.47$\pm$1.73 &  20.73$\pm$0.70 & 39.24 & & \\
18:34:20.9229$\pm$0.0002 & -05:59:42.556$\pm$0.006 & 36.00$\pm$1.50 &  26.80$\pm$0.70 & 39.34 & & \\
18:34:20.9228$\pm$0.0001 & -05:59:42.562$\pm$0.004 & 40.37$\pm$1.30 &  32.47$\pm$0.63 & 39.45 & & \\
18:34:20.9228$\pm$0.0001 & -05:59:42.563$\pm$0.003 & 40.67$\pm$1.17 &  33.87$\pm$0.57 & 39.55 & & \\
18:34:20.9227$\pm$0.0001 & -05:59:42.564$\pm$0.003 & 34.27$\pm$0.97 &  29.07$\pm$0.50 & 39.66 & & \\
18:34:20.9227$\pm$0.0001 & -05:59:42.566$\pm$0.004 & 23.83$\pm$0.73 &  20.53$\pm$0.37 & 39.76 & & \\
18:34:20.9228$\pm$0.0002 & -05:59:42.568$\pm$0.005 & 13.87$\pm$0.60 &  12.16$\pm$0.31 & 39.87 & & \\
18:34:20.9232$\pm$0.0003 & -05:59:42.566$\pm$0.007 & 7.37$\pm$0.53 &   6.60$\pm$0.29 & 39.97 & & \\
\enddata
\tablenotetext{a}{Conditional large-scale grouping of 22 GHz H$_2$O maser spots detected in the field.} 
\tablenotetext{b}{~22 GHz H$_2$O maser clusters detected in the vicinity of G2 group.} 
\end{deluxetable}

\newpage
\startlongtable
\begin{deluxetable}{lccccccc}
\tablewidth{0pt}
\tablecaption{6.7 GHz CH$_3$OH maser parameters \label{tab:T67GHZ}}
\tablehead{
\colhead{RA(J2000)} & \colhead{DEC(J2000)} &
\colhead{Integrated} & \colhead{Peak} &
\colhead{V$_{LSR}$} & \colhead{Cluster\tablenotemark{a}} \\
\colhead{} & \colhead{} &
\colhead{flux} & \colhead{flux} &
\colhead{} & \colhead{} \\
\colhead{($^h$~$^m$~$^s$)} & \colhead{($^\circ$~$\arcmin$~$\arcsec$)} &
\colhead{(Jy)} & \colhead{(Jy/beam)} &
\colhead{(km~s$^{-1}$)} & \colhead{}
}
\startdata
18:34:20.9090$\pm$0.0006 & -05:59:42.476$\pm$0.014 & 0.79$\pm$0.03  & 0.90$\pm$0.02  & 38.42 & S \\ 
18:34:20.9093$\pm$0.0004 & -05:59:42.487$\pm$0.010 & 1.22$\pm$0.04  & 1.30$\pm$0.02  & 38.46 & \\
18:34:20.9099$\pm$0.0003 & -05:59:42.487$\pm$0.008 & 1.65$\pm$0.04  & 1.75$\pm$0.03  & 38.51 & \\
18:34:20.9106$\pm$0.0003 & -05:59:42.483$\pm$0.007 & 1.97$\pm$0.04  & 2.03$\pm$0.02  & 38.55 & \\
18:34:20.9100$\pm$0.0003 & -05:59:42.480$\pm$0.008 & 1.93$\pm$0.04  & 2.07$\pm$0.03  & 38.59 & \\
18:34:20.9098$\pm$0.0003 & -05:59:42.477$\pm$0.008 & 1.71$\pm$0.04  & 1.83$\pm$0.03  & 38.64 & \\
18:34:20.9088$\pm$0.0004 & -05:59:42.481$\pm$0.010 & 1.36$\pm$0.04  & 1.44$\pm$0.03  & 38.68 & \\
18:34:20.9090$\pm$0.0005 & -05:59:42.501$\pm$0.014 & 0.99$\pm$0.04  & 1.02$\pm$0.02  & 38.73 & \\
18:34:20.9077$\pm$0.0007 & -05:59:42.493$\pm$0.018 & 0.70$\pm$0.04  & 0.74$\pm$0.02  & 38.77 & \\
18:34:20.9081$\pm$0.0009 & -05:59:42.511$\pm$0.022 & 0.48$\pm$0.03  & 0.56$\pm$0.02  & 38.81 & \\
\hline
18:34:20.9284$\pm$0.0007 & -05:59:42.333$\pm$0.018 & 0.70$\pm$0.04  & 0.74$\pm$0.02  & 40.09 & NE \\ 
18:34:20.9287$\pm$0.0004 & -05:59:42.352$\pm$0.011 & 1.12$\pm$0.04  & 1.18$\pm$0.02  & 40.13 & \\
18:34:20.9289$\pm$0.0003 & -05:59:42.330$\pm$0.009 & 1.67$\pm$0.04  & 1.73$\pm$0.03  & 40.18 & \\
18:34:20.9290$\pm$0.0002 & -05:59:42.341$\pm$0.006 & 2.30$\pm$0.04  & 2.40$\pm$0.03  & 40.22 & \\
18:34:20.9286$\pm$0.0002 & -05:59:42.343$\pm$0.005 & 2.95$\pm$0.04  & 3.05$\pm$0.03  & 40.26 & \\
18:34:20.9284$\pm$0.0002 & -05:59:42.333$\pm$0.005 & 3.44$\pm$0.05  & 3.60$\pm$0.03  & 40.31 & \\
18:34:20.9284$\pm$0.0002 & -05:59:42.336$\pm$0.004 & 3.82$\pm$0.05  & 3.93$\pm$0.03  & 40.35 & \\
18:34:20.9282$\pm$0.0002 & -05:59:42.333$\pm$0.004 & 3.95$\pm$0.05  & 4.06$\pm$0.03  & 40.40 & \\
18:34:20.9284$\pm$0.0002 & -05:59:42.334$\pm$0.004 & 3.96$\pm$0.05  & 4.08$\pm$0.03  & 40.44 & \\
18:34:20.9282$\pm$0.0002 & -05:59:42.335$\pm$0.004 & 3.70$\pm$0.05  & 3.81$\pm$0.03  & 40.48 & \\
18:34:20.9280$\pm$0.0002 & -05:59:42.330$\pm$0.005 & 3.29$\pm$0.05  & 3.43$\pm$0.03  & 40.53 & \\
18:34:20.9279$\pm$0.0002 & -05:59:42.335$\pm$0.005 & 2.88$\pm$0.05  & 2.99$\pm$0.03  & 40.57 & \\
18:34:20.9279$\pm$0.0002 & -05:59:42.329$\pm$0.006 & 2.51$\pm$0.04  & 2.59$\pm$0.03  & 40.61 & \\
18:34:20.9278$\pm$0.0003 & -05:59:42.334$\pm$0.006 & 2.17$\pm$0.04  & 2.26$\pm$0.03  & 40.66 & \\
18:34:20.9279$\pm$0.0003 & -05:59:42.344$\pm$0.007 & 1.93$\pm$0.04  & 2.03$\pm$0.03  & 40.70 & \\
18:34:20.9275$\pm$0.0003 & -05:59:42.326$\pm$0.007 & 1.82$\pm$0.04  & 1.87$\pm$0.02  & 40.75 & \\
18:34:20.9275$\pm$0.0003 & -05:59:42.337$\pm$0.008 & 1.82$\pm$0.04  & 1.87$\pm$0.03  & 40.79 & \\
18:34:20.9274$\pm$0.0003 & -05:59:42.349$\pm$0.007 & 1.93$\pm$0.04  & 1.93$\pm$0.03  & 40.83 & \\
18:34:20.9271$\pm$0.0003 & -05:59:42.342$\pm$0.007 & 2.02$\pm$0.04  & 2.04$\pm$0.03  & 40.88 & \\
18:34:20.9275$\pm$0.0003 & -05:59:42.347$\pm$0.006 & 2.21$\pm$0.04  & 2.22$\pm$0.03  & 40.92 & \\
18:34:20.9275$\pm$0.0003 & -05:59:42.354$\pm$0.006 & 2.32$\pm$0.04  & 2.31$\pm$0.03  & 40.97 & \\
18:34:20.9270$\pm$0.0003 & -05:59:42.351$\pm$0.006 & 2.23$\pm$0.04  & 2.22$\pm$0.03  & 41.01 & \\
18:34:20.9266$\pm$0.0003 & -05:59:42.348$\pm$0.006 & 2.13$\pm$0.04  & 2.10$\pm$0.02  & 41.05 & \\
18:34:20.9268$\pm$0.0003 & -05:59:42.347$\pm$0.007 & 1.94$\pm$0.04  & 1.92$\pm$0.02  & 41.10 & \\
18:34:20.9261$\pm$0.0004 & -05:59:42.346$\pm$0.008 & 1.71$\pm$0.04  & 1.63$\pm$0.03  & 41.14 & \\
18:34:20.9247$\pm$0.0004 & -05:59:42.342$\pm$0.009 & 1.58$\pm$0.04  & 1.50$\pm$0.02  & 41.19 & \\
18:34:20.9235$\pm$0.0004 & -05:59:42.346$\pm$0.009 & 1.47$\pm$0.04  & 1.39$\pm$0.02  & 41.23 & \\
18:34:20.9218$\pm$0.0005 & -05:59:42.320$\pm$0.010 & 1.48$\pm$0.04  & 1.33$\pm$0.03  & 41.27 & \\
18:34:20.9148$\pm$0.0005 & -05:59:42.297$\pm$0.009 & 1.64$\pm$0.05  & 1.47$\pm$0.03  & 41.32 & \\
\hline
18:34:20.9069$\pm$0.0003 & -05:59:42.209$\pm$0.006 & 2.54$\pm$0.05  & 2.27$\pm$0.03  & 41.36 & NW \\ 
18:34:20.9027$\pm$0.0002 & -05:59:42.190$\pm$0.004 & 4.83$\pm$0.06  & 4.71$\pm$0.03  & 41.40 & \\
18:34:20.9015$\pm$0.0001 & -05:59:42.174$\pm$0.003 & 9.21$\pm$0.08  & 9.29$\pm$0.04  & 41.45 & \\
18:34:20.9010$\pm$0.0001 & -05:59:42.173$\pm$0.003 & 15.86$\pm$0.12  & 16.08$\pm$0.07  & 41.49 & \\
18:34:20.9010$\pm$0.0001 & -05:59:42.174$\pm$0.002 & 25.23$\pm$0.18  & 25.64$\pm$0.10  & 41.54 & \\
18:34:20.9012$\pm$0.0001 & -05:59:42.177$\pm$0.002 & 39.20$\pm$0.25  & 39.90$\pm$0.14  & 41.58 & \\
18:34:20.9014$\pm$0.0001 & -05:59:42.179$\pm$0.002 & 60.23$\pm$0.36  & 61.20$\pm$0.20  & 41.62 & \\
18:34:20.9015$\pm$0.0001 & -05:59:42.180$\pm$0.002 & 87.39$\pm$0.48  & 88.81$\pm$0.29  & 41.67 & \\
18:34:20.9016$\pm$0.0001 & -05:59:42.181$\pm$0.002 & 114.70$\pm$0.65  & 116.55$\pm$0.36  & 41.71 & \\
18:34:20.9017$\pm$0.0001 & -05:59:42.183$\pm$0.002 & 138.91$\pm$0.74  & 141.15$\pm$0.43  & 41.76 & \\
18:34:20.9017$\pm$0.0001 & -05:59:42.183$\pm$0.002 & 156.96$\pm$0.87  & 159.48$\pm$0.48  & 41.80 & \\
18:34:20.9018$\pm$0.0001 & -05:59:42.184$\pm$0.002 & 165.22$\pm$0.87  & 168.00$\pm$0.52  & 41.84 & \\
18:34:20.9019$\pm$0.0001 & -05:59:42.183$\pm$0.002 & 155.22$\pm$0.83  & 157.65$\pm$0.48  & 41.89 & \\
18:34:20.9020$\pm$0.0001 & -05:59:42.184$\pm$0.002 & 127.43$\pm$0.70  & 129.46$\pm$0.40  & 41.93 & \\
18:34:20.9020$\pm$0.0001 & -05:59:42.183$\pm$0.002 & 93.00$\pm$0.52  & 94.53$\pm$0.30  & 41.98 & \\
18:34:20.9019$\pm$0.0001 & -05:59:42.182$\pm$0.002 & 61.29$\pm$0.37  & 62.40$\pm$0.21  & 42.02 & \\
18:34:20.9019$\pm$0.0001 & -05:59:42.183$\pm$0.002 & 36.60$\pm$0.24  & 37.30$\pm$0.14  & 42.06 & \\
18:34:20.9017$\pm$0.0001 & -05:59:42.181$\pm$0.003 & 20.26$\pm$0.16  & 20.72$\pm$0.09  & 42.11 & \\
18:34:20.9015$\pm$0.0001 & -05:59:42.183$\pm$0.003 & 11.29$\pm$0.09  & 11.57$\pm$0.05  & 42.15 & \\
18:34:20.9016$\pm$0.0001 & -05:59:42.177$\pm$0.003 & 6.40$\pm$0.06  & 6.60$\pm$0.03  & 42.19 & \\
18:34:20.9015$\pm$0.0002 & -05:59:42.185$\pm$0.005 & 3.72$\pm$0.05  & 3.84$\pm$0.03  & 42.24 & \\
18:34:20.9019$\pm$0.0002 & -05:59:42.176$\pm$0.006 & 2.34$\pm$0.04  & 2.41$\pm$0.03  & 42.28 & \\
18:34:20.9020$\pm$0.0003 & -05:59:42.188$\pm$0.007 & 1.77$\pm$0.04  & 1.88$\pm$0.03  & 42.33 & \\
18:34:20.9026$\pm$0.0003 & -05:59:42.190$\pm$0.008 & 1.71$\pm$0.04  & 1.75$\pm$0.02  & 42.37 & \\
18:34:20.9024$\pm$0.0003 & -05:59:42.200$\pm$0.007 & 1.92$\pm$0.04  & 2.02$\pm$0.03  & 42.41 & \\
18:34:20.9032$\pm$0.0002 & -05:59:42.193$\pm$0.006 & 2.21$\pm$0.04  & 2.32$\pm$0.02  & 42.46 & \\
18:34:20.9036$\pm$0.0002 & -05:59:42.195$\pm$0.006 & 2.54$\pm$0.04  & 2.66$\pm$0.03  & 42.50 & \\
18:34:20.9025$\pm$0.0002 & -05:59:42.196$\pm$0.005 & 2.73$\pm$0.04  & 2.83$\pm$0.03  & 42.55 & \\
18:34:20.9034$\pm$0.0002 & -05:59:42.195$\pm$0.006 & 2.81$\pm$0.05  & 2.93$\pm$0.03  & 42.59 & \\
18:34:20.9040$\pm$0.0002 & -05:59:42.189$\pm$0.006 & 2.68$\pm$0.04  & 2.77$\pm$0.03  & 42.63 & \\
18:34:20.9035$\pm$0.0002 & -05:59:42.197$\pm$0.006 & 2.30$\pm$0.04  & 2.45$\pm$0.03  & 42.68 & \\
18:34:20.9034$\pm$0.0003 & -05:59:42.203$\pm$0.008 & 1.91$\pm$0.04  & 1.98$\pm$0.03  & 42.72 & \\
18:34:20.9043$\pm$0.0003 & -05:59:42.195$\pm$0.009 & 1.52$\pm$0.04  & 1.60$\pm$0.02  & 42.77 & \\
18:34:20.9041$\pm$0.0004 & -05:59:42.203$\pm$0.010 & 1.22$\pm$0.03  & 1.28$\pm$0.02  & 42.81 & \\
18:34:20.9043$\pm$0.0005 & -05:59:42.224$\pm$0.012 & 0.97$\pm$0.04  & 1.04$\pm$0.02  & 42.85 & \\
18:34:20.9057$\pm$0.0006 & -05:59:42.217$\pm$0.014 & 0.84$\pm$0.04  & 0.87$\pm$0.02  & 42.90 & \\
18:34:20.9044$\pm$0.0007 & -05:59:42.216$\pm$0.017 & 0.70$\pm$0.04  & 0.75$\pm$0.02  & 42.94 & \\
18:34:20.9081$\pm$0.0008 & -05:59:42.228$\pm$0.019 & 0.59$\pm$0.03  & 0.65$\pm$0.02  & 42.99 & \\
18:34:20.9051$\pm$0.0010 & -05:59:42.222$\pm$0.023 & 0.48$\pm$0.03  & 0.55$\pm$0.02  & 43.03 & \\
\enddata
\tablenotetext{a}{Conditional large-scale grouping of 6.7 GHz CH$_3$OH maser spots detected in the field.} 
\end{deluxetable}

\newpage
\startlongtable
\begin{deluxetable}{lccccccc}
\tablewidth{0pt}
\tablecaption{44 GHz CH$_3$OH maser parameters \label{tab:T44GHZ}}
\tablehead{
\colhead{RA(J2000)} & \colhead{DEC(J2000)} &
\colhead{Integrated} & \colhead{Peak} &
\colhead{V$_{LSR}$} & \colhead{Cluster\tablenotemark{a}} \\
\colhead{} & \colhead{} &
\colhead{flux} & \colhead{flux} &
\colhead{} & \colhead{} \\
\colhead{($^h$~$^m$~$^s$)} & \colhead{($^\circ$~$\arcmin$~$\arcsec$)} &
\colhead{(Jy)} & \colhead{(Jy/beam)} &
\colhead{(km~s$^{-1}$)} & \colhead{}
}
\startdata
18:34:20.7684$\pm$0.0003 & -05:59:42.132$\pm$0.007 & 1.03$\pm$0.11 & 0.97$\pm$0.06 & 43.90 & R \\
18:34:20.7680$\pm$0.0003 & -05:59:42.131$\pm$0.007 & 1.12$\pm$0.12 & 1.08$\pm$0.07 & 43.85 & \\
18:34:20.7676$\pm$0.0003 & -05:59:42.131$\pm$0.006 & 1.34$\pm$0.13 & 1.29$\pm$0.07 & 43.80 & \\
18:34:20.7672$\pm$0.0003 & -05:59:42.125$\pm$0.007 & 1.38$\pm$0.14 & 1.30$\pm$0.08 & 43.75 & \\
18:34:20.7670$\pm$0.0003 & -05:59:42.128$\pm$0.007 & 1.36$\pm$0.14 & 1.32$\pm$0.08 & 43.69 & \\
18:34:20.7667$\pm$0.0003 & -05:59:42.129$\pm$0.006 & 1.28$\pm$0.13 & 1.27$\pm$0.07 & 43.64 & \\
18:34:20.7665$\pm$0.0003 & -05:59:42.127$\pm$0.007 & 1.22$\pm$0.12 & 1.15$\pm$0.07 & 43.59 & \\
18:34:20.7665$\pm$0.0003 & -05:59:42.129$\pm$0.007 & 0.98$\pm$0.11 & 0.94$\pm$0.06 & 43.53 & \\
18:34:20.7667$\pm$0.0004 & -05:59:42.127$\pm$0.007 & 0.70$\pm$0.08 & 0.71$\pm$0.05 & 43.48 & \\
\hline
18:34:21.0018$\pm$0.0005 & -05:59:39.942$\pm$0.009 & 0.42$\pm$0.06 & 0.40$\pm$0.03 & 42.58 & G \\
18:34:21.0015$\pm$0.0004 & -05:59:39.937$\pm$0.008 & 0.52$\pm$0.07 & 0.50$\pm$0.04 & 42.52 & \\
18:34:21.0017$\pm$0.0004 & -05:59:39.921$\pm$0.009 & 0.70$\pm$0.09 & 0.63$\pm$0.04 & 42.47 & \\
18:34:21.0018$\pm$0.0003 & -05:59:39.922$\pm$0.007 & 0.91$\pm$0.10 & 0.87$\pm$0.06 & 42.42 & \\
18:34:21.0015$\pm$0.0003 & -05:59:39.929$\pm$0.007 & 1.14$\pm$0.12 & 1.07$\pm$0.06 & 42.36 & \\
18:34:21.0017$\pm$0.0003 & -05:59:39.923$\pm$0.006 & 1.08$\pm$0.10 & 1.05$\pm$0.06 & 42.31 & \\
18:34:21.0017$\pm$0.0003 & -05:59:39.925$\pm$0.007 & 0.72$\pm$0.08 & 0.72$\pm$0.04 & 42.26 & \\
\hline
18:34:20.5784$\pm$0.0002 & -05:59:43.310$\pm$0.004 & 6.64$\pm$0.40 & 5.77$\pm$0.22 & 41.73 & B2 \\
18:34:20.5782$\pm$0.0002 & -05:59:43.300$\pm$0.004 & 10.8$\pm$0.60 & 9.76$\pm$0.32 & 41.67 & \\
18:34:20.5785$\pm$0.0004 & -05:59:43.295$\pm$0.007 & 8.76$\pm$1.04 & 8.32$\pm$0.56 & 41.62 & \\
18:34:20.5789$\pm$0.0005 & -05:59:43.283$\pm$0.009 & 4.52$\pm$0.72 & 4.27$\pm$0.38 & 41.57 & \\
\hline                                                                                      
18:34:20.5377$\pm$0.0007 & -05:59:42.857$\pm$0.015 & 4.32$\pm$0.96 & 3.72$\pm$0.48  & 41.57 & B1 \\
18:34:20.5374$\pm$0.0005 & -05:59:42.859$\pm$0.009 & 5.64$\pm$0.84 & 4.72$\pm$0.40  & 41.51 & \\
18:34:20.5367$\pm$0.0007 & -05:59:42.856$\pm$0.011 & 7.96$\pm$1.20 & 4.52$\pm$0.48  & 41.46 & \\
18:34:20.5358$\pm$0.0009 & -05:59:42.851$\pm$0.011 & 9.36$\pm$1.40 & 4.44$\pm$0.48  & 41.41 & \\
18:34:20.5334$\pm$0.0008 & -05:59:42.838$\pm$0.010 & 8.28$\pm$1.12 & 3.83$\pm$0.37  & 41.36 & \\
18:34:20.5312$\pm$0.0007 & -05:59:42.827$\pm$0.010 & 6.04$\pm$0.84 & 3.05$\pm$0.30  & 41.30 & \\
18:34:20.5304$\pm$0.0006 & -05:59:42.824$\pm$0.009 & 3.80$\pm$0.52 & 2.54$\pm$0.22  & 41.25 & \\
\enddata
\tablenotetext{a}{Conditional large-scale grouping of 44 GHz CH$_3$OH maser spots detected in the field, spots are labeled according to their velocity: ``B1'' and ``B2'' -- blue spots, ``G'' -- green, and ``R'' -- red.} 
\end{deluxetable}

%#1

%#2

%#3

%#4

%#5

%#6

%#7

%#8
\begin{figure}
\plotone{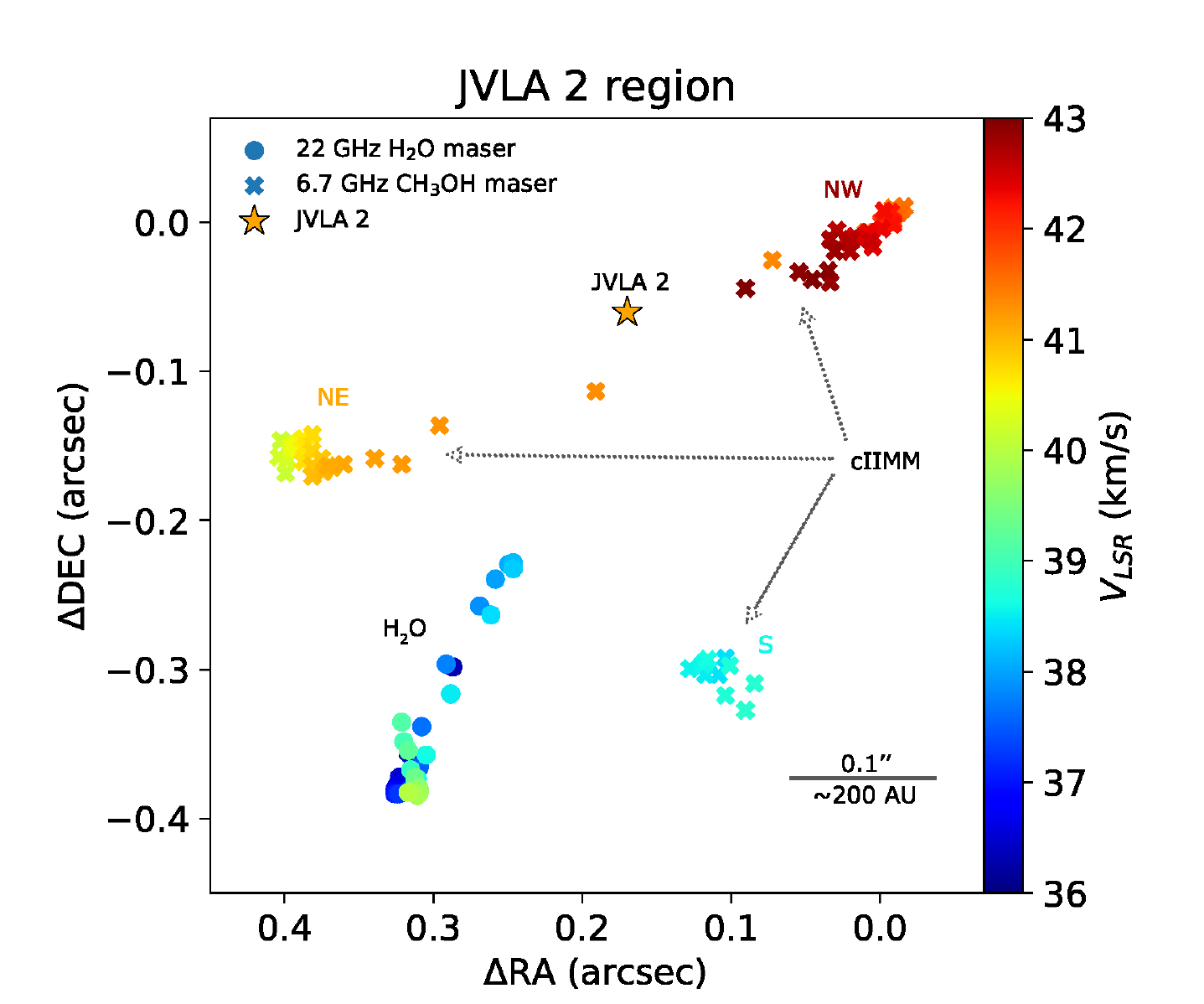}
\caption{Distribution of 6.7 GHz CH$_3$OH and 22 GHz H$_2$O maser spots detected in JVLA 2 region. For 6.7 GHz CH$_3$OH maser clusters labeling see Table \ref{tab:T67GHZ}.  Plot is color-coded by radial velocity (see colorbar for color scale). Positional offsets are relative to the strongest 6.7 GHz  CH$_3$OH maser spot. The physical scale label (in AU) assumes the distance to the source of 2.08 kpc (the BeSSeL Survey Bayesian Distance Calculator). \label{fig:jvla2}}
\end{figure}

%#9

%#10

%#11

%#12

%#13

%#14


\newpage
\begin{thebibliography}{}

\bibitem[Anglada et al.(2018)]{2018A&ARv..26....3A} Anglada, G., Rodr{\'\i}guez, L. F., Carrasco-Gonz$\acute{a}$lez, C. \ 2018, \aapr, 26, 3
\bibitem[Araya et al.(2010)]{2010ApJ..717..133A} Araya, E. D., Hofner, P., Goss, W. M., Kurtz, S., Richards, A. M. S., Linz, H., Olmi, L.,  Sewilo, M. \ 2010, \apj, 717L, 133
\bibitem[Ashimbaeva et al.(2017)]{2017ATel11042....1A} Ashimbaeva, N. T., Platonov, M. A., Rudnitskij, G. M., Tolmachev, A. M. \ 2017, ATel \#11042
\bibitem[Audard et al.(2014)]{2014PPVI...387} Audard, M., Abraham, P., Dunham, M. M., et al. \ 2014, in Protostars and Planets VI, eds. H. Beuther, R. S. Klessen, C. P. Dullemond, \& T. Henning (Tucson: University of Arizona Press), 387
\bibitem[Batrla et al.(1987)]{1987Natur.326...49B} Batrla, W., Matthews, H., Menten, K., Walmsley, C. \ 1987, \nat, 326, 49
\bibitem[Bayandina et al.(2019)]{2018ApJS..00..00B} Bayandina, O. S., Val'tts, I. E., Kurtz, S. E., Shakhvorostova, N. N. \ 2019, in preparation
\bibitem[Brand et al.(1994)]{1994AAS..103..541B} Brand, J., Cesaroni, R., Caselli, P., Catarzi, M., Codella, C., Comoretto, G., Curioni, G. P., Curioni, P., Di Franco, S., Felli, M., Giovanardi, C., Olmi, L., Palagi, F., Palla, F., Panella, D., Pareschi, G., Rossi, E., Speroni, N., Tofani, G. \ 1994, \aaps, 103, 541
\bibitem[Breen et al.(2010)]{2010MNRAS.401.2219B} Breen, S. L., Ellingsen, S. P., Caswell, J. L., Lewis, B. E. \ 2010, \mnras, 401, 2219
\bibitem[Bronfman et al.(1996)]{1996AAS..115..81B} Bronfman, L., Nyman, L.-A., May, J. \ 1996, \aaps, 115, 81
\bibitem[Burns et al.(2016)]{2016MNRAS.460..283B} Burns, R. A., Handa, T., Nagayama, T., Sunada, K., Omodaka, T. \ 2016, \mnras, 460, 283
\bibitem[Burns et al.(2017)]{2017MNRAS.467.2367B} Burns, R. A., Handa, T., Imai, H., Nagayama, T., Omodaka, T., Hirota, T., Motogi, K., van Langevelde, H. J., Baan, W. A. \ 2017,  \mnras, 467, 2367
\bibitem[Caratti o Garatti et al.(2016)]{2016Nature..13..276C} Caratti o Garatti, A.,  Stecklum, B.,  Garcia Lopez, R.,  Eisloffel, J.,  Ray, T. P.,  Sanna, A.,  Cesaroni, R.,  Walmsley, C. M.,  Oudmaijer, R. D.,  de Wit, W. J.,  Moscadelli, L.,  Greiner, J., Krabbe, A.,   Fischer, C.,  Klein, R.,  Ibanez, J. M. \ 2016, Nature Physics, 13, 276
\bibitem[Clark(1980)]{1980AA....89..377C} Clark, B. G. \ 1980, \aap, 89, 377
\bibitem[Dartois et al.(2000)]{2000AA..361..1095D} Dartois, E., Gerin, M., d'Hendecourt, L. \ 2000, \aap, 361, 1095
\bibitem[Deguchi et al.(1989)]{1989ApJ..340L..17D} Deguchi, S. \& Watson, W.D, \ 1989, \apj, 340, L17
\bibitem[Fontani et al.(2010)]{2010AA..517..56F} Fontani, F., Cesaroni, R., Furuya, R. S. \ 2010, \aap, 517, 56
\bibitem[Fujisawa et al.(2015)]{2015ATel.8286....1F} Fujisawa, K., Yonekura, Y., Sugiyama, K., Horiuchi, H., Hayashi, T., Hachisuka, K., Matsumoto, N., Niinuma, K. \ 2015, ATel \#8286
\bibitem[Fujisawa et al.(2014)]{2014PASJ.66....31F} Fujisawa, K., Sugiyama,  K., Motogi, K., Hachisuka, K., Yonekura, Y., Sawada-Satoh, S., Matsumoto, N., Sorai, K., Momose, M., Saito, Y.,  Takaba, H., Ogawa, H., Kimura, K., Niinuma, K., Hirano, D., Omodaka, T., Kobayashi, H., Kawaguchi, N., Shibata, K. M., Honma, M., Hirota, T., Murata, Y., Doi, A., Mochizuki, N., Shen, Z., Chen, X., Xia, B., Li, B., Kim, K.-T. \ 2014, \pasj, 66, 31
\bibitem[Garay et al.(1996)]{1996ApJ.459..193G} Garay, G., Ramirez, S., Rodriguez, L. F., Curiel, S., Torrelles, J. M. \ 1996, \apj, 459 193
\bibitem[Gaylard et al.(1994)]{1994MNRAS.269..257G} Gaylard, M. J., MacLeod, G. C., van der Walt, D. J. \ 1994, \mnras, 269, 257
\bibitem[Green \& McClure-Griffiths(2011)]{2011MNRAS.417.2500G} Green, J. A., McClure-Griffiths, N. M. \ 2011, \mnras, 417, 2500
\bibitem[Hirota et al.(2011)]{2011ApJ...739L..59H} Hirota, T., Tsuboi, M., Fujisawa, K., Honma, M., Kawaguchi, N., Kim, M. K., Kobayashi, H., Imai, H., Omodaka, T., Shibata, K. M., Shimoikura, T., Yonekura, Y. \ 2011, \apj, 739, L59
\bibitem[Hirota et al.(2014)]{2014PASJ...66..106H}  Hirota, T., Tsuboi, M., Kurono, Y., Fujisawa, K., Honma, M.,  Kim, M. K.,  Imai, H., Yonekura, Y. \ 2014, \pasj, 66, 106
\bibitem[Honma et al.(2004)]{2004PASJ...56L..15H} Honma, M., Yoon, K. C., Bushimata, T., Fujii, T., Hirota, T., Horiai, K., Imai, H., Inomata, N., Ishitsuka, J., Iwadate, K., Jike, T., Kameya, O., Kamohara, R., Kan-Ya, Y., Kawaguchi, N., Kobayashi, H. et al. \ 2004, \pasj, 56, 15
\bibitem[Hu et al.(2016)]{2016ApJ...833..18H} Hu, B., Menten, K. M., Wu, Y.,  Bartkiewicz, A., Rygl, K., Reid, M. J., Urquhart, J. S.,   Zheng, X. \ 2016, \apj, 833, 18
\bibitem[Hunter et al.(2017)]{2017ApJ...837L..29H} Hunter, T. R., Brogan, C. L., MacLeod, G., Cyganowski, C. J., Chandler, C. J., Chibueze, J. O., Friesen, R., Indebetouw, R., Thesner, C., Young, K. H. \ 2017, \apj, 837, L29
\bibitem[Jenness et al.(1995)]{1995MNRAS..276..1024J} Jenness, T., Scott, P. F., Padman, R. \ 1995, \mnras, 276, 1024
\bibitem[Kurtz et al.(1994)]{1994ApJS..91..659K} Kurtz, S., Churchwell, E., Wood, D. O. S. \ 1994, \apjs, 91, 659
\bibitem[Lee et al.(2001)]{2001ApJ..557..429L} Lee C.-F., Stone J. M., Ostriker E. C., Mundy L. G. \ 2001, \apj, 557, 429
\bibitem[Lekht et al.(2018)]{2018ARep...62..213L} Lekht, E. E., Pashchenko, M. I., Rudnitskij, G. M.,  Tolmachev, A. M. \ 2018, ARep, 62, 213
\bibitem[L{\'o}pez-Sepulcre et al.(2011)]{2011AA..526L..2L} L{\'o}pez-Sepulcre, A., Walmsley, C. M., Cesaroni, R., Codella, C., Schuller, F., Bronfman, L., Carey, S. J., Menten, K. M., Molinari, S., Noriega-Crespo, A. \ 2011, \aap, 526, L2
\bibitem[McCutcheon et al.(1995)]{1995AJ....110.1762M} McCutcheon, W. H., Sato, T., Purton, C. R., Matthews, H. E., Dewdney, P. E. \ 1995, AJ, 110, 1762
\bibitem[Menten(1996)]{1996IAUS..178..163M} Menten, K. \ 1996, in Molecules in Astrophysics: Probes \& Processes, ed. E. F. van Dishoeck, IAU Symp., 178, 163
\bibitem[Menten(1991)]{1991ASPC..16..119M} Menten, K. \ 1991, ASPCS, 16, 119
\bibitem[Menten(2012)]{2012IAUS..287..506M} Menten, K. M. \ 2012, IAUS287, Cosmic Masers -- from OH to H$_0$, ed. by R. S. Booth, E. M. L. Humphreys and W. H. T. Vlemmings, 506
\bibitem[Molinari et al.(1996)]{1996AA..308..573M} Molinari, S., Brand, J., Cesaroni, R., Palla, F. \ 1996, \aap, 308, 573
\bibitem[Moscadelli et al.(2017)]{2017AA..600..8M} Moscadelli, L., Sanna, A., Goddi, C.,  Walmsley, M. C., Cesaroni, R.,  Caratti o Garatti, A.,  Stecklum, B., Menten, K. M., Kraus, A. \ 2017, \aap, 600, L8
\bibitem[Norris et al.(1998)]{1998ApJ..508..275N} Norris, R. P., Byleveld, S. E., Diamond, P. J., Ellingsen, S. P., Ferris, R. H., Gough, R. G., Kesteven, M. J., McCulloch, P. M., Phillips, C. J., Reynolds, J. E., Tzioumis, A. K., Takahashi, Y., Troup, E. R., Wellington,  K. J., \ 1998, \apj, 508, 275
\bibitem[Palla et al.(1991)]{1991AA..246..249P} Palla, F., Brand, J., Comoretto, G., Felli, M., Cesaroni, R. \ 1991, \aap, 246, 249
\bibitem[Reid et al.(2016)]{2016ApJ...823...77R} Reid, M. J., Dame, T. M., Menten, K. M., Brunthaler, A. \ 2016, \apj, 823, 77
\bibitem[Richer \& Padman(1991)]{1991MNRAS.251..707R} Richer, J. S. \& Padman, R. \ 1991, \mnras, 251, 707
\bibitem[Rodr{\'\i}guez et al.(2012)]{2012ApJ...755..152R}
Rodr{\'\i}guez, L. F., Gonz$\acute{a}$lez, R. F., Montes, G., Asiri, H. M., Raga, A. C., Cant$\acute{o}$, J. \ 2012, \apj, 755, 152
\bibitem[S$\acute{a}$nchez-Monge et al. (2013)]{2013AA..557..94S} S$\acute{a}$nchez-Monge, A., L$\acute{o}$pez-Sepulcre, A., Cesaroni, R., Walmsley, C. M., Codella, C., Beltr$\acute{a}$n, M. T., Pestalozzi, M., Molinari, S. \ 2013, \aap, 557, A94
\bibitem[Shepherd \& Churchwell(1996)]{1996ApJ..457..267S} Shepherd, D. S., Churchwell, E. \ 1996, \apj, 457, 267
\bibitem[Shimoikura et al.(2005)]{2005ApJ..634..459S} Shimoikura, T., Kobayashi, H., Omodaka, T., Diamond, P. J., Matveyenko, L. I., Fujisawa, K. \ 2005, \apj, 634, 459
\bibitem[Slysh et al.(1999)]{1999AAS..134..115S} Slysh, V. I., Val'tts, I. E., Kalenskii, S. V., Voronkov, M. A., Palagi, F., Tofani, G., Catarzi, M. \ 1999, \aaps, 134, 115
\bibitem[Sobolev et al.(2017)]{2017ATel10788....1S} Sobolev, A. M., Bisyarina, A. P., Tatarnikov, A. M., Antokhin, I., Volvach, A. E. \ 2017, ATel \#10788
\bibitem[Sugiyama et al.(2017)]{2017ATel10757....1S} Sugiyama, K., Saito, Y., Akitaya, H., Yonekura, Y., Momose, M. \ 2017, ATel \#10757
\bibitem[Sunada et al.(2007)]{2007PASJ..59..1185S} Sunada, K., Nakazato, T., Ikeda, N., Hongo, S., Kitamura, Y., Yang, J. \ 2007, \pasj, 59, 1185 
\bibitem[Surcis et al.(2015)]{2015AA..578..102S} Surcis, G., Vlemmings, W. H. T., van Langevelde,  H. J., Hutawarakorn Kramer, B., Bartkiewicz, A.,  Blasi, M. G. \ 2015, \aap, 578, A102
\bibitem[Szymczak et al.(2000)]{2000AAS.143..269S} Szymczak, M., Hrynek, G., Kus, A. J. \ 2000, \aaps, 143, 269
\bibitem[Szymczak et al.(2016)]{2016MNRAS.459L..56S} Szymczak, M., Olech, M., Wolak, P., Bartkiewicz, A., Gawronski, M. \ 2016, \mnras, 459, L56
\bibitem[Todd \& Howat(2006)]{2006MNRAS..367..238T}  Todd, S. P. \&  Howat, S. K. R. \ 2006, \mnras, 367, 238
\bibitem[Thompson et al.(2006)]{2006AA..453..1003T} Thompson, M. A., Hatchell, J., Walsh, A. J., Macdonald, G. H., Millar, T. J. \ 2006, \aap, 453, 1003
\bibitem[van der Walt et al.(1995)]{1995AAS...110..81V} van der Walt, D. J., Gaylard, M. J., MacLeod, G. C. \ 1995, \aaps, 110, 81
\bibitem[van der Walt et al.(2016)]{2016AA...588A..47V} van der Walt, D. J., Maswanganye, J. P., Etoka, S., Goedhart, S., van den Heever, S. P. \ 2016, \aap, 588, 47
\bibitem[Varricatt et al.(2010)]{2010MNRAS...404..661V} Varricatt, W. P., Davis, C. J., Ramsay, S., Todd, S. P. \ 2010, \mnras, 404, 661
\bibitem[Volvach et al.(2017a)]{2017ATel10728....1V}	Volvach, A. E., Volvach, L. N., MacLeod, G., Lekht, E. E., Rudnitskij, G. M., Tolmachev, A. M. \ 2017, ATel \#10728
\bibitem[Volvach et al.(2017b)]{2017ATel10853....1V} Volvach, A. E., Volvach, L. N., MacLeod, G., Bayandina, O. S., Shakhvorostova, N. N., Val'tts, I. E.  \ 2017, ATel \#10853
\bibitem[Volvach et al.(2019)]{2019MNRAS.482L..90V}	Volvach,  L. N., Volvach,  A. E., Larionov,  M. G., MacLeod,  G. C., van den Heever, S. P., Wolak, P., \& Olech,  M. \ 2019, \mnras, 482, L90
\bibitem[Walsh et al.(1997)]{1997MNRAS.291..261W} Walsh, A. J., Hyland, A. R., Robinson, G., Burton, M. G. \ 1997, \mnras, 291, 261
\bibitem[Walsh et al.(1998)]{1998MNRAS.301..640W} Walsh, A. J., Burton, M. G., Hyland, A. R., Robinson, G. \ 1998, \mnras, 301, 640
\bibitem[Zavagno et al.(2002)]{2002AA.394..225Z} 	Zavagno, A., Deharveng, L., Nadeau, D., Caplan, J. \ 2002, \aap, 394, 225
\end{thebibliography}
\end{document}